\documentclass[a4paper,oneside]{amsart}
\usepackage{latexsym}
\usepackage{amsmath}
\usepackage{amssymb}
\usepackage{amsfonts}
\usepackage{float}
\usepackage{esint}
\usepackage[utf8]{inputenc}
\usepackage[T1]{fontenc}   
\usepackage{graphicx}
\usepackage{color}
\allowdisplaybreaks %needed to allow page breaking in the middle of multiline equations
\definecolor{Blue}{rgb}{0.,0.,1.}
\definecolor{Red}{rgb}{1.,0.,0.}
\definecolor{Green}{rgb}{0.,1.,0.}
\let\origmaketitle\maketitle
\def\maketitle{
  \begingroup
  \def\uppercasenonmath##1{} % this disables uppercasing title
  \let\MakeUppercase\relax % this disables uppercasing authors
	\origmaketitle
  \endgroup
	}

\newcounter{smallarabics}
\newenvironment{arabicenumerate}
{\begin{list}{{\normalfont\textrm{(\arabic{smallarabics})}}}
  {\usecounter{smallarabics}\setlength{\itemindent}{0cm}
   \setlength{\leftmargin}{5ex}\setlength{\labelwidth}{4ex}
   \setlength{\topsep}{0.75\parsep}\setlength{\partopsep}{0ex}
   \setlength{\itemsep}{0ex}}}
{\end{list}}

\newcounter{smallroman}
\newenvironment{romanenumerate}
{\begin{list}{{\normalfont\textrm{(\roman{smallroman})}}}
  {\usecounter{smallroman}\setlength{\itemindent}{0cm}
   \setlength{\leftmargin}{5ex}\setlength{\labelwidth}{4ex}
   \setlength{\topsep}{0.75\parsep}\setlength{\partopsep}{0ex}
   \setlength{\itemsep}{0ex}}}
{\end{list}}

\newcommand{\ben}{\begin{arabicenumerate}}  
\newcommand{\een}{\end{arabicenumerate}}

\def\init{\setcounter{equation}{0}}

\newtheorem{theoreme}{Theorem }[section]
\newtheorem{proposition}[theoreme]{Proposition}
\newtheorem{hypothesis}[theoreme]{Hypothesis}
\newtheorem{lemma}[theoreme]{Lemma}
\newtheorem{definition}[theoreme]{Definition}

\newtheorem{remark}[theoreme]{Remark}
\newtheorem{example}[theoreme]{Example}
\newcommand{\beq}{\begin{equation}}
\newcommand{\eeq}{\end{equation}}

\newcommand{\bex}{\begin{example}}
\newcommand{\eex}{\end{example}}
\def\bel{\begin{lemma}}
\def\eel{\end{lemma}}
\def\bet{\begin{theoreme}}
\def\eet{\end{theoreme}}
\def\bed{\begin{definition}}
\def\eed{\end{definition}}
\def\ber{\begin{remark}}
\def\eer{\end{remark}}

\def\rr{{\mathbb R}}
\def\zz{{\mathbb Z}}
\def\cc{{\mathbb C}}
\def\nn{{\mathbb N}}

\def\part{{\rm par}}

\def\Re{{\rm Re}}

\def\bar{\overline}

\def\cinf{C^\infty}
\def\c0inf{C_0^\infty}

\def\proof{
\noindent{\bf Proof.}\ \ }

\def\sh{{\rm sh}}
\def\ch{{\rm ch}}

\def\cV{{\mathcal V}}

\def\cD{{\mathcal D}}
\def\cU{{\mathcal U}}

\def\cC{{\mathcal C}}
\def\cW{{\mathcal W}}

\def\i{{\rm i}}

\def\qed{$\Box$\medskip}

\def \p{ \partial}

\def\12{\frac{1}{2}}
\def\14{\frac{1}{4}}

\def\e{{\rm e}}

\def\bbbone{{\mathchoice {\rm 1\mskip-4mu l} {\rm 1\mskip-4mu l}
{\rm 1\mskip-4.5mu l} {\rm 1\mskip-5mu l}}}
\def\one{\bbbone}
\def\cH{{\mathcal H}}

\def\ii{{\rm j}}

\def\coinf{C_0^\infty}

\def\cK{{\mathcal K}}

\def \p{ \partial}

\def\12{\frac{1}{2}}

\def\e{{\rm e}}

\def\cH{{\mathcal H}}

\def\Diff{{\rm Diff}}

\def\bep{\begin{proposition}}
\def\eep{\end{proposition}}

\def\Op{{\rm Op}}
\def\Opw{{\rm Op}^{\rm w}}

\newcommand{\mat}[4]{\left(\begin{array}{cc}#1 &#2  \\ #3 &#4 \end{array}\right)}

\newcommand{\col}[2]{\left(\begin{array}{c}#1 \\#2\end{array} \right)}

\def\CARal{{\rm C\hskip 0.25 em \hbox{\raise 1.72 ex 
\hbox{$\scriptscriptstyle\rm al$}\kern -0.57 em A}R}}

\def\otimesal{\mathop{\hbox{\raise 1.5 ex
  \hbox{$\scriptscriptstyle\rm al$}
\kern -0.92 em \hbox{$\otimes$}}}}
\def\oplusal{\mathop{\hbox{\raise 1.5 ex
  \hbox{$\scriptscriptstyle\rm al$}
\kern -0.92 em \hbox{$\oplus$}}}}
\def\Gammal{\hbox{\raise 1.68 ex 
\hbox{$\scriptscriptstyle\rm al$}\kern -0.50 em $\Gamma$}}
\def\Bal{\hbox{\raise 1.68 ex 
\hbox{$\scriptscriptstyle\rm  al$}\kern -0.50 em $B$}}
\def\CARal{{\rm C\hskip 0.25 em \hbox{\raise 1.72 ex 
\hbox{$\scriptscriptstyle\rm al$}\kern -0.57 em A}R}}

\newcommand{\traa}[1]{\mskip-6mu\upharpoonright_{#1}}

\def\cE{{\mathcal E}}

\def\WF{{\rm WF}}
\makeatletter
\newcommand*{\defeq}{\mathrel{\rlap{%
                     \raisebox{0.3ex}{$\m@th\cdot$}}%
                     \raisebox{-0.3ex}{$\m@th\cdot$}}%
                     =}
\makeatother
\makeatletter
\newcommand*{\eqdef}{=\mathrel{\rlap{%
                     \raisebox{0.3ex}{$\m@th\cdot$}}%
                     \raisebox{-0.3ex}{$\m@th\cdot$}}%
                     }
\makeatother

\newcommand{\bea}{\begin{aligned}}
\newcommand{\beal}{\begin{array}{l}}
\newcommand{\eeal}{\end{array}}
\newcommand{\eea}{\end{aligned}}
\newcommand{\bec}{\begin{cases}}
\newcommand{\eec}{\end{cases}}
\def\Diff{{\rm Diff}}
\def\BT{{\rm BT}}\def\pe{{\overline \p}}
\def\cN{\mathcal{N}}
\def\wf{{\rm WF}}
\def\cinfb{C^{\infty}_{\rm b}}

\def\bS{\mathbb{S}^{d-1}}
\newcommand{\coo}[1]{T^{*}#1\backslash \{0\}}
\def\tig{\tilde{g}}
 \def\tih{{\tilde h}}

\def\vol{{\rm vol}}

\DeclareMathOperator{\Ker}{Ker}
\DeclareMathOperator{\Dom}{Dom}
\DeclareMathOperator{\supp}{supp}
\def\vara{{\rm\textsl{a}}}

\def\varq{{\rm\textsl{q}}}
\def\E{{\rm E}}

\begin{document}
\title[Hadamard states on Lorentzian manifolds of bounded geometry]{\Large Hadamard states for the Klein-Gordon equation\\ on Lorentzian manifolds of bounded geometry}
\author{\fontsize{10}{12}\selectfont Christian \textsc{G\'erard}}
\address{D\'epartement de Math\'ematiques, Universit\'e Paris-Sud XI, 91405 Orsay Cedex, France}
\email{christian.gerard@math.u-psud.fr}
\author{Omar \textsc{Oulghazi}}
\address{D\'epartement de Math\'ematiques, Universit\'e Paris-Sud XI, 91405 Orsay Cedex, France}
\email{omar.oulghazi@math.u-psud.fr}
\author{Micha{\l} \textsc{Wrochna}}
\address{Universit\'e Grenoble Alpes, Institut Fourier, UMR 5582 CNRS, CS 40700, 38058 Grenoble \textsc{Cedex} 09, France}
\email{michal.wrochna@univ-grenoble-alpes.fr}
\begin{abstract}
We consider the Klein-Gordon equation on a class of \emph{Lorentzian manifolds with Cauchy surface of bounded geometry}, which is shown to include examples such as exterior Kerr, Kerr-de Sitter spacetime {and the maximal globally hyperbolic extension of the Kerr outer region}. In this setup, we give an approximate diagonalization  and a microlocal decomposition of the Cauchy evolution using a time-dependent version of the pseudodifferential calculus on Riemannian manifolds of bounded geometry. We apply this result to construct all pure regular Hadamard states (and associated Feynman inverses), where regular refers to the state's two-point function having Cauchy data given by pseudodifferential operators. This allows us to conclude that there is a one-parameter family of elliptic pseudodifferential operators that encodes both the choice of (pure, regular) Hadamard state and the underlying spacetime metric. 
\end{abstract}

\maketitle

\section{Introduction \& summary of results}\label{sec0}\init

\subsection{Introduction} Modern formulations of quantum field theory on curved spacetimes allow for a precise distinction between local, model-independent features, and global aspects specific to the concrete physical setup. In the case of non-interacting scalar fields, the study of the latter is directly related to the \emph{propagation of singularities} for the Klein-Gordon equation, as well as to specific {global} properties of its solutions, such as {two-point function \emph{positivity}}. Thus, a careful implementation of methods from microlocal analysis that takes into account asymptotic properties of the spacetime is essential in the rigorous construction of quantum fields. The present paper is aimed at generalizing known methods, in particular \cite{GW1}, by providing the necessary tools to work on a much wider class of backgrounds that includes examples such as Kerr and Kerr-de Sitter spacetimes.

Before formulating the problem in more detail, let us first recall how various notions from quantum field theory are related to inverses of the Klein-Gordon operator and to special classes of bi-solutions.

\subsubsection{Klein-Gordon equation} Consider a Klein-Gordon operator 
\[
P = - \nabla^{a}\nabla_{a}+ V(x)
\]
 on a Lorentzian manifold $(M, g)$, where $V: M\to \rr$ is a smooth function. Assuming {\em global hyperbolicity}\footnote{See Subsect. \ref{secp3.2}.} of $(M, g)$, the operator $P$ has two essential properties, the proofs of which date back to Leray \cite{L, C,BGP}.
 
The first one is  the existence of {\em retarded/advanced inverses}  of $P$, i.e.   operators $G_{\rm ret/adv}$, mapping $\coinf(M)$ into $\cinf(M)$ such that
 \[
 P\circ G_{\rm adv/ret}= G_{\rm ret/adv}\circ P = \one, \ \ \supp G_{\rm ret/adv}u\subset J^{\pm}(\supp u),
\]
 where $J^{\pm}(K)$ is the future/past causal shadow\footnote{The future/past causal shadow of $K\subset M$ is the set of points reached from $K$ by future/past directed causal (i.e. non-spacelike) curves.} of a set $K\subset M$. The second is the unique solvability of the Cauchy problem: if $\{\Sigma_t\}_{t\in\rr}$ is a foliation of $M$ by space-like Cauchy hypersurfaces, and   $\rho(t): \cinf(M)\ni \phi\mapsto (\phi\traa{\Sigma_t}, \i^{-1}\p_{n}\phi\traa{\Sigma_t})\in \cinf(\Sigma; \cc^{2})$ is the Cauchy data operator on $\Sigma_t$, {then, for any fixed $s$, there exists a unique solution of the Cauchy problem
 \beq\label{cauchy}
\begin{cases}
P\phi=0,\\
\rho(s)\phi=f
\end{cases}
\eeq
for any given $f\in\cinf_{0}(\Sigma; \cc^{2})$ (and moreover, $\rho(t)\phi\in \cinf_{0}(\Sigma; \cc^{2})$ for any $t\in\rr$). }
In the present setup the two properties are actually essentially equivalent.

These two facts are basic to the theory  of {\em quantum Klein-Gordon fields} on the curved space\-time $(M,g)$, see  e.g. \cite{D}, which we now briefly recall (see Subsect. \ref{secp4.1} for more details).

\subsubsection{Quantum Klein-Gordon fields} By a {\em phase space} we will mean a complex vector space equipped with a non-degenerate hermitian form. The operator $G= G_{\rm ret}- G_{\rm adv}$ is anti-hermitian for the natural scalar product $(u| v)_{M}= \int_{M}\bar{u}v\,d\vol_{g}$, which allows to equip $\coinf(M)$ with the hermitian form  $\bar{u}\cdot Q v= \i (u| Gv)_{M}$. One can show that the  kernel of $G$ equals to $P\coinf(M)$, hence  if  $\cV= \frac{\coinf(M)}{P\coinf(M)}$, $(\cV, Q)$ is a phase space --- it is in fact the fundamental structure that defines the classical content of the theory.

% Moreover  $G\coinf(M)$ equals to the space $\Sol(P)$ of smooth  complex {\em space-compact} solutions of  $P\phi=0$, i.e. the solutions of \eqref{cauchy} for smooth compactly supported Cauchy data. This allows to  identify $(\cV, Q)$ with $(\Sol(P), q)$ by $G$. In terms of Cauchy data one has 
% \[
% \bar{\phi}\cdot q \phi= \bar{f}\cdot q_{\Sigma}f= \int_{\Sigma}\bar{f}_{1} f_{0}+ \bar{f}_{0}f_{1}d\sigma_{\Sigma},
% \]
% where $d\sigma_{\Sigma}$ is the induced measure on $\Sigma$.

 This allows one to introduce the polynomial {\rm CCR} $^{*}$-algebra ${\rm CCR}(\cV, Q)$, by definition generated by the identity $\one$ and elements called the {\em (abstract) charged fields}, which are of the form $\psi([u]), \psi^{*}([u])$ for $[u]\in  \frac{\coinf(M)}{P\coinf(M)}$ and are subject to the relations:
 \[
 \begin{array}{rl}
 i)& [u]\mapsto \psi([u])\hbox{ is }\cc-\hbox{anti-linear},\\[2mm]
 ii)& [u]\mapsto \psi^{*}([u])\hbox{ is }\cc-\hbox{linear},\\[2mm]
 iii)&\begin{array}{l}
 \big[\psi([u]), \psi([v])\big]= \big[\psi^{*}([u]), \psi^{*}([v])\big]=0, \\[2mm]
  \big[\psi([u]), \psi^{*}([v])\big]= \i (u|Gv)_{M}\one, 
  \end{array}\\[5mm]
  iv)&\psi([u])^{*}= \psi^{*}([u]).
 \end{array}
 \]
 The algebraic approach to quantum field theory provides a way to represent the above {\em canonical commutation relations} in terms of closed operators on some concrete Hilbert space. The standard way to obtain such a representation  is to specify a \emph{state}. 

 \subsubsection{Hadamard states}
  A state $\omega$ on ${\rm CCR}(\cV, Q)$ is a positive linear functional $\omega$ on ${\rm CCR}(\cV, Q)$ such that $\omega(\one)=1$. A particularly natural class of states for linear Klein-Gordon fields is the class of {\em quasi-free states} {(see e.g. \cite[Sec. 17.1]{DG} and references therein)}, which are entirely determined by the expectation values:
 \beq\label{stato}
\omega\big(\psi([u])\psi^{*}([v])\big)\eqdef  (u|\Lambda^{+}v)_{M}, \ \ \omega\big(\psi^{*}([v])\psi([u])\big)\eqdef  (u| \Lambda^{-}v)_{M}.
 \eeq
 This definition implies in particular that $P\circ \Lambda^{\pm}= \Lambda^{\pm}\circ P=0$.   
 It is also natural to require that $\Lambda^{\pm}: \coinf(M)\to \cinf(M)$, in which case $\Lambda^{\pm}$ have distributional kernels $\Lambda^{\pm}(x, x')\in \cD'(M\times M)$, called the {\em two-point functions} 
 of the state $\omega$.

 Among all quasi-free states, {\em Hadamard states} are considered as  the physically acceptable ones, because their short distance behavior {resembles that of} the vacuum state on Minkowski space\-time \cite{KW}.  Since the work of Radzikowski \cite{R},
Hadamard states are characterized by a condition on  the {\em wave front set} of their two-point 
functions $\Lambda^{\pm}$, see Def. \ref{def:hadamard} for the precise statement.
The use of wave front sets had a deep impact on quantum field theory on curved space\-times, for example on the perturbative construction of interacting models; see e.g. \cite{BF,dang2,HW,KM} and also \cite{BDH,dang1,dang3} for some recent related mathematical developments.

The microlocal formulation of the Hadamard condition in \cite{R} is intimately linked to the notion of {\em distinguished parametrices} introduced by Duistermaat and H\"{o}rmander in their influential paper \cite{DH}.  Distinguished parametrices are parametrices of $P$ (inverses modulo smoothing errors), which are determined uniquely (modulo smoothing errors) by the wave front set of their Schwartz kernel.  Duistermaat and H\"{o}rmander demonstrated that there are exactly four classes of distinguished parametrices, the {\em advanced}/{\em retarded} and {\em Feynman}/{\em anti-Feynman} ones, see Subsect. \ref{feyninv} for details.
The (uniquely defined) retarded/advanced inverses $G_{\rm ret/adv}$ are examples of retarded/advanced parametrices. 

In contrast, there is no canonical choice of a Feynman/anti-Feynman inverse on a generic spacetime. This is actually very closely related to the problem of specifying a distinguished Hadamard state, see e.g. \cite{FV2}. More specifically, the link between Hadamard states and Feynman inverses discovered by Radzikowski is that if $\Lambda^{\pm}$ are the two-point functions of a Hadamard state then the operator 
\[
G_{\rm F}= \i^{-1}\Lambda^{+}+ G_{\rm adv}= \i^{-1}\Lambda^{-}+ G_{\rm ret}
\]
is a Feynman inverse\footnote{One could call $G_{\rm F}$ a `time-ordered Feynman inverse' to make the distinction with the Feynman propagator of Gell-Redman, Haber and Vasy \cite{GHV,positive}, which is a generalized inverse of $P$ considered as a Fredholm operator on suitably chosen spaces; here we just stick to the shorthand terminology.} of $P$.

There already exist a large number of existence results for Hadamard states. First of all, the deformation argument of Fulling, Narcowich and Wald \cite{FNW} shows that Hadamard states exist on any globally hyperbolic space\-time{. T}his construction has however  the disadvantage of being very indirect, which poses problems in applications. An alternative existence proof on arbitrary globally hyperbolic spacetimes was given in \cite{GW1}. It has however another severe drawback which is that it fails to produce \emph{pure states} (see Subsect. \ref{secp4.1.1}) in general. 

Specific examples of Hadamard states on spacetimes with special (asymptotic) symmetries include passive states for stationary space\-times \cite{SV}, states constructed from data at null infinity on various classes of asymptotically flat or asymptotically de Sitter space\-times \cite{Mo,DMP1,BJ,VW} and on cosmological space\-times\footnote{Let us also mention that on static and cosmological spacetimes with compact Cauchy surface, a different construction of Hadamard states was recently proposed by Brum and Fredenhagen \cite{BM}.} \cite{DMP2,JS,BT}. Furthermore, a remarkable recent result by Sanders \cite{sanders} proves the existence and Hadamard property of the so-called Hartle-Hawking-Israel state on spacetimes with a static bifurcate Killing horizon.

{Finally, Junker \cite{J1, J2} and Junker and Schrohe \cite{JS} used the {\em pseudodifferential calculus} on a Cauchy hypersurface $\Sigma$ to construct Hadamard states in the case of $\Sigma$ compact. The construction was then reworked in \cite{GW1} to yield classes of Hadamard states for $P$ in the non-compact case.}  Let us emphasize that outside of the case of $\Sigma$ compact, the calculus of properly supported pseudodifferential operators, which exists on any smooth manifold, is not sufficient to address  the {\em positivity condition} $\Lambda^{\pm}\geq 0$ and the {\em CCR condition} $\Lambda^{+}- \Lambda^{-}= \i G$ which have to be satisfied by the two-point functions $\Lambda^{\pm}$ in order to be consistent with \eqref{stato} (see Subsect. \ref{secp4.1.4}). This was tackled in \cite{GW1}  by assuming that the Cauchy surface $\Sigma$ is diffeomorphic to $\rr^d$ (so that the spacetime $M$ is diffeomorphic to $\rr^{1+d}$), with some uniformity conditions on $g$ at spatial infinity, which allowed to use the {\em uniform pseudodifferential calculus} on $\rr^{d}$. 

In the case of spacetimes with Cauchy surfaces that are either compact or diffeomorphic to $\rr^d$, the constructions in  \cite{J1,J2,GW1} have the advantage that they yield examples of Hadamard states which (in contrast to the general existence argument from \cite{GW1}) are pure, and it turns out that with some additional effort it is possible to obtain large classes of those. Furthermore, the two-point functions of these states are given by rather explicit formulae (in contrast to \cite{FNW}), from which one can recover the space\-time metric, see Subsect. \ref{ss:qfsg} for a discussion in the present, more general context. Unfortunately, many space\-times of interest, like for instance {\em blackhole space\-times} fall outside the hypotheses in \cite{J1,J2,GW1}.

%The states constructed in \cite{GW1} have covariances  given by (matrices of) pseudodifferential operators on $\Sigma$. In particular we characterized all {\em pure} Hadamard states with pseudodifferential covariances. 

\subsection{Content of the paper} In this paper we rework and extend the results of {\cite{J1,J2,GW1}} in two essential directions. First of all, we greatly generalize the framework of \cite{GW1} by basing our analysis on the pseudodifferential calculus on manifolds of {\em bounded geometry}, due to {Kordyukov \cite{Ko} and Shubin \cite{Sh2}}. This allows us to work on a much larger class of spacetimes, including examples such as Kerr and Kerr-de Sitter spacetimes. Secondly, the construction of Hadamard states is now obtained as a consequence of a {\em microlocal decomposition} of the Cauchy evolution operator $\cU_{A}(t,s)$ associated to $P$. Beside simplifying the proofs, this allows us to derive many formulas of independent interest, including for instance expressions for the Feynman inverses canonically associated to the Hadamard states we construct.

%This allows  to associate to any such microlocal decomposition a unique {\em Feynman inverse}.
 Let us now describe in more detail the content of the paper.

The background on Riemannian manifolds of bounded geometry is presented in Sect. \ref{secp0}. We use an equivalent definition of bounded geometry which is much more convenient in practice. In rough terms, it amounts to the existence of chart diffeomorphisms $\{\psi_x\}_{x\in M}$ such that the pull-back metric (on $\rr^{n}$) $(\psi_x^{-1})^* g$ is {(together with all derivatives) bounded above and below and equivalent to the flat metric (together with all derivatives), uniformly w.r.t. $x\in M$.}

This leads naturally to the notion of  {\em Lorentzian manifolds of bounded geometry} and of Cauchy hypersurfaces of bounded geometry, developed in Sect.  \ref{secp3}, which is an interesting topic in its own right. The main ingredient is the choice of a reference Riemannian metric $\hat g$ used to define bounded tensors. We  then introduce  in Subsect. \ref{secp3.3} a class of space\-times and associated Klein-Gordon operators for which parametrices for the Cauchy problem can be constructed by pseudodifferential calculus:

\begin{hypothesis}\label{hypothesis} We assume that there exists a neighborhood $U$ of a Cauchy surface $\Sigma$ in $(M,g)$, such that:

\hangindent=1.15cm $(\rm H)$ \ $(U,g)$ is conformally embedded in a Lorentzian manifold of bounded geometry $(M,\tilde g)$ and the conformal factor $\tilde c^2$ is such that $\nabla_{\hat g}\ln \tilde c$ is a bounded $(1,0)$-tensor, moreover $\Sigma$ is a so-called \emph{Cauchy hypersurface of bounded geometry} in $(M,\tilde g)$;

\hangindent=1.15cm $(\rm M)$ \ $\tilde c^2 V$ is a bounded $(0,0)$-tensor.
\end{hypothesis}

We refer to Subsect. \ref{secp3.3} for the detailed definitions. It turns out, see Sect. \ref{sec-examples}, that most standard examples of space\-times, like cosmological space\-times, Kerr, Kerr-de Sitter, {the maximal globally hyperbolic extension of Kerr}, or cones, double cones and wedges in Minkowski space belong to this class of space\-times.
 
 The pseudodifferential calculus on a manifold of bounded geometry is recalled in Sect. \ref{secp1}. A new result of  importance for the analysis in the later sections of the paper is a version of {\em Egorov's theorem}, see Thm. \ref{pth1}.

Sect. \ref{secp2} contains the main analytical results of the paper. The condition that $\Sigma$ is a Cauchy hypersurface of bounded geometry allows to identify the neighborhood $U$ with $I\times \Sigma$ (with $I$ an open interval), and  the Klein-Gordon equation on $U$ can be reduced to the standard form
\beq\label{equio}
\p_{t}^{2}\phi+ r(t, x)\p_{t}\phi+ a(t, x, \p_{x})\phi=0,
\eeq
where $a(t, x,\p_{x})$ is a second order, elliptic differential operator on $\Sigma$.  Denoting by $\cU_{A}(t,s)$ the Cauchy evolution operator for \eqref{equio},  mapping $\rho(s)\phi$ to $\rho(t)\phi$, we construct what we call a {\em microlocal decomposition} of $\cU_{A}$, i.e. a decomposition
\beq\label{decompo}
\cU_{A}(t,s)= \cU_{A}^{+}(t,s)+ \cU_{A}^{-}(t,s),
\eeq
where $\cU_{A}^{\pm}$ have the following properties, see Thm. \ref{newprop.2}:
\ben
\item $\{\cU_{A}^{\pm}(t,s)\}_{t,s\in I}$ are two-parameter groups {(i.e. $\cU_{A}^{\pm}(t,t')\cU_{A}^{\pm}(t',s)=\cU_{A}^{\pm}(t,s)$ for all $t,t',s\in I$)}  and $\cU_{A}^{\pm}(t,t)\eqdef  c^{\pm}(t)$ are projections,
\item $\cU_{A}^{\pm}(t,s)$ propagate the wave front set in the upper/lower energy shells $\cN^{\pm}$, i.e. the two respective connected components of the characteristic set of \eqref{equio},
\item the kernels of $\cU_{A}^{\pm}(t,s)$ are symplectically orthogonal for the canonical symplectic form preserved by the evolution.
\een
We demonstrate in Thm. \ref{feyninvth} that to such a decomposition one can  associate a unique {\em Feynman inverse} for $P$.

Sect. \ref{secp4} is devoted to the construction of Hadamard states from a microlocal decomposition, which can be summarized as follows. We use the `time-kernel' notation for two-point functions $\Lambda^\pm$, that is we write $\Lambda^\pm(t,s)$ to mean the associated operator-valued Schwartz kernel in the time variable. We say that a state is \emph{regular} if $\Lambda^\pm(t,t)$ is a matrix of pseudo\-differential operators on $\Sigma$ for some $t$.  

\begin{theoreme}\label{thm:main} Let $(M,g)$ be a spacetime satisfying Hypothesis \ref{hypothesis} and consider the reduced Klein-Gordon equation \eqref{equio}. Let $t_0\in I$. {Then} there exists a pure regular Hadamard state with two-point functions given by
\beq\label{eq:thisform}
\Lambda^\pm(t,s)=\mp\pi_0\cU_A^\pm(t,s)\pi_1^*,
\eeq 
where  $\pi_0,\pi_1$ are the projections to the respective two {components} of Cauchy data and $\{\cU_{A}^{\pm}(t,s)\}_{t,s\in I}$ is a microlocal decomposition, such that
\beq\label{eq:fort0}
\cU_{A}^{\pm}(t_0,t_0)= \mat{\mp(b^{+}- b^{-})^{-1}b^{\mp}}{\pm(b^{+}- b^{-})^{-1}}{\mp b^{+}(b^{+}- b^{-})^{-1}b^{-}}{\pm b^{\pm}(b^{+}- b^{-})^{-1}}(t_0)
\eeq
for some pair $b^\pm(t_0)$ of elliptic first order pseudodifferential operators. Moreover, the two-point functions of any pure regular Hadamard state are of this form.\end{theoreme}

The detailed results are stated in Thm. \ref{thp4.1} and \ref{michal's-identities}, see also Prop. \ref{reduc-prop} for the arguments that allow to get two-point functions for the original Klein-Gordon equation on the full spacetime $(M,g)$ rather than for the reduced equation \eqref{equio} on $I\times \Sigma$. 

Since one can get many regular states out of a given one by applying suitable \emph{Bogoliubov transformations} as in \cite{GW1}, Thm. \ref{thm:main} yields in fact a large class of Hadamard states.
%Some auxiliary computations are collected in Appendix \ref{secapp1}. 

%Conversely we show that any pure Hadamard state whose covariances are given on $\Sigma$ by pseudodifferential operators is associated to a microlocal decomposition as in \eqref{decompo}. 

%ici
\subsection{From quantum fields to spacetime geometry} \label{ss:qfsg}

In our approach, microlocal splittings are obtained by setting 
\[
\cU_A^\pm(t,s)\defeq\cU_A(t,t_0)\cU_A^\pm(t_0,t_0)\cU_A(t_0,s)
\]
where $\cU_A^\pm(t_0,t_0)$ is defined by formula \eqref{eq:fort0} with $b^\pm(t)$ constructed for $t\in I$ as approximate solutions (i.e. modulo smoothing terms) of the operatorial equation
\begin{equation}\label{eq-enew}
\big(\p_{t}+ \i b^{\pm}+ r\big)\circ \big(\p_{t}- \i b^{\pm}\big)= \p_{t}^{2}+ r\p_{t}+a,
\end{equation}
and satisfying some additional conditions, see Sect. \ref{secp2} (in particular Thm. \ref{th2.1}) for details. {We note that the approximate factorization \eqref{eq-enew}  was already used by Junker in his construction of Hadamard states \cite{J1,J2}.}

In summary, there is a pair of time-dependent elliptic pseudodifferential operators $b^\pm(t)$ that uniquely determines the choice of a pure regular Hadamard state. It is interesting to remark that $b^\pm(t)$ also determines the spacetime metric. First, by subtracting the members of  \eqref{eq-enew} one gets $r(t)$ modulo smoothing errors. Then \eqref{eq-enew}  gives $a(t)$ modulo smoothing terms. But since $a(t)$ and $r(t)$ are differential operators (the latter is just a multiplication operator), they can be determined exactly. Furthermore, the reduced operator  on the r.h.s. of \eqref{eq-enew} is  just the Klein-Gordon operator in Gaussian normal coordinates near a Cauchy hyper\-surface $\Sigma_{t_0}$ (see Subsect. \ref{secp2.2}), so the metric can be read in these coordinates from the knowledge of $a$ and $r$.

This way, both quantum fields (derived from pure Hadamard states) and the underlying spacetime metric are encoded by a time-dependent elliptic pseudodifferential operator $b^+(t)\oplus b^-(t)$. As long as one considers only pure Hadamard states (and spacetimes for which Gaussian normal coordinates make sense globally), this provides in particular a solution to the problem discussed in \cite{ST} {which consists in finding a description of Hadamard states without having to specify the spacetime metric explicitly}. It would be thus interesting to try to build a theory where $b^+(t)\oplus b^-(t)$ is treated as a dynamical quantity that accounts for both quantum degrees of freedom and spacetime geometry.  

{We also note that the construction does not indicate directly how to select states with specific symmetries (in the case when the spacetime has any), which would be desirable for applications and therefore deserves further investigation (see e.g. \cite{DD} for some recent attempts).} 

 \subsection{Notation}\label{sec0.not}
- if $X,Y$ are sets and $f:X\to Y$ we write  $f: X \xrightarrow{\sim}Y$ if $f$ is
bijective.  If $X, Y$ are equipped with topologies, {we write $f:X\to Y$ if the map is continuous, and $f: X \xrightarrow{\sim}Y$ if it is a homeomorphism.}

- the domain of a closed, densely defined operator  $a$  will be denoted by $\Dom a$.

- if $a$ is a selfadjoint operator on a Hilbert space $\cH$, we write $a>0$ if $a\geq 0$ and $\Ker a=\{0\}$. We denote by $\langle a\rangle^{s}\cH$ the completion of $\Dom |a|^{s}$ for the norm $\| u\|_{s}=\| (1+ a^{2})^{s/2}u\|$.

-  if $a, b$ are selfadjoint operators on a Hilbert space $\cH$, we write $a\sim b$ if
\[
  a,b >0, \ \Dom a^{\12}= \Dom b^{\12}, \ c^{-1}b\leq a \leq cb,
\]
for some  constant $c>0$.

  - similarly if  $I\subset \rr$ is an open interval and $\{\cH_{t}\}_{t\in I}$ is a family of Hilbert spaces with $\cH_{t}= \cH$ as topological vector spaces, and $a(t), b(t)$ are two selfadjoint operators on   $\cH_{t}$, we write $a(t)\sim b(t)$ if for each $J\Subset I$ there exist constants $c_{1, J}, c_{2, J}>0$  such that
  \beq\label{equitd}
  a(t), b(t) \geq c_{1, J}>0, \ \  c_{2, J}b(t)\leq a(t) \leq c_{2, J}^{-1}b(t), \ t\in J.
\eeq

- from now on the operator of multiplication by a function $f$ will be denoted by $f$, while the operators of partial differentiation will be denoted by $\pe_{i}$, so that $[\pe_{i}, f]= \p_{i}f$.

- we set $\langle x\rangle= (1+ x^{2})^{\12}$ for $x\in \rr^{n}$.
\section{Riemannian manifolds of bounded geometry}\label{secp0}\init
\subsection{Definition}\label{secp0.1}
We recall the notion of a Riemannian manifold of {\em bounded geometry}, see \cite{CG, Ro}. An important property of Riemannian manifolds of bounded geometry is that they admit a nice `uniform' pseudodifferential calculus, introduced in \cite{Sh2, Ko}, which will be recalled in Sect. \ref{secp1}.
\subsubsection{Notation}
 We denote by $\delta$ the flat metric on $\rr^{n}$ and by $B_{n}(y, r)\subset \rr^{n}$ the open ball of center $y$ and radius $r$.
 If $(M,g)$ is a Riemannian manifold and $x\in M$ we denote  by $B^{g}_{M}(x, r)$ (or $B^{g}(x, r)$ if the underlying manifold $M$ is clear from the context) the geodesic ball of center $x$ and radius $r$. 
 
 We denote by $r_{x}>0$  the injectivity radius at $x$ and by $\exp_{x}^{g}: B^{g(x)}_{T_{x}M}(0, r_{x})\to M$ the exponential map at $x$. 
 
 If $0<r<r_{x}$ it is well known that  $ \exp_{x}^{g}(B^{g(x)}_{T_{x}M}(0, r))= B^{g}_{M}(x, r)$ is an open neighborhood of $x$ in $M$.   Choosing a linear isometry  $e_{x}: (\rr^{n}, \delta)\to (T_{x}M, g(x))$ we obtain Riemannian normal coordinates at $x$ using the map $\exp_{x}^{g}\circ e_{x}$.

  If $T$ is a  $(q,p)$ tensor  on $M$, we can define the canonical norm of $T(x)$, $x\in M$, denoted by $\|T\|_{x}$, using appropriate tensor powers of $g(x)$ and $g^{-1}(x)$.  $T$ is {\em bounded} if $\sup_{x\in M}\| T\|_{x}<\infty$.
  
  Let $U\Subset \rr^{n}$ be open,  relatively compact with smooth boundary. We denote by 
$\cinf_{\rm b}(U)= \cinf(\rr^{n})\traa{U}$ the space of smooth functions on $U$, bounded with all derivatives.  

If $V$ is another open set like $U$ and $\chi: U\to V$ is a diffeomorphism, we will abuse slightly notation and write  that $\chi\in \cinf_{\rm b}(U)$ if all components of $\chi$ belong to $\cinf_{\rm b}(U)$ and all components of $\chi^{-1}$ belong to $\cinf_{\rm b}(V)$.

One defines similarly  smooth $(q,p)$ tensors on $U$, bounded with all derivatives. 
For coherence with later notation, this space will be denoted by ${\rm BT}^{p}_{q}(U, \delta)$, where $\delta$ is the flat metric on $U$. We equip ${\rm BT}^{p}_{q}(U, \delta)$ with its Fr\'echet space topology.

\begin{definition}\label{def0.1}
 A Riemannian manifold $(M,g)$ is {\em of bounded geometry} if
 \ben
 \item  the injectivity radius $r_{g}\defeq  \inf_{x\in M} r_{x}$ is strictly positive,
 \item $\nabla_{g}^{k}R_{g}$ is a bounded tensor for all  $k\in \nn$, where $R_{g}$, $\nabla_{g}$ are  the Riemann curvature tensor and  covariant derivative associated to $g$.
 \een
\end{definition}
We give an alternative characterization, which is often more useful in applications. 

\begin{theoreme}\label{thp0.1}
A Riemannian manifold $(M,g)$ is of bounded geometry iff for each $x\in M$, there exists $U_{x}$ open neighborhood of $x$ and
\[
\psi_{x}: U_{x} \xrightarrow{\sim} B_{n}(0,1)
\]
a smooth diffeomorphism with $\psi_{x}(x)=0$ such that if $g_{x}\defeq  (\psi_{x}^{-1})^{*}g$ then:

\noindent
{\rm (C1)} the family $\{g_{x}\}_{x\in M}$ is  bounded in ${\rm BT}^{0}_{2}(B_{n}(0,1), \delta)$,

\noindent
 {\rm (C2)} there exists $c>0$ such that:
\[
c^{-1}\delta\leq g_{x}\leq c \delta, \ x\in M.
\]

 A family $\{U_{x}\}_{x\in M}$ resp. $\{\psi_{x}\}_{x\in M}$ as above will be called a family of {\em good chart neighborhoods}, resp. {\em good chart diffeomorphisms}.
\end{theoreme}

\proof Let us first prove the   $\Rightarrow$ implication. We choose 
\[
U_{x}= \exp^{g}_{x} \circ e_{x}(B_{n}(0, \frac{r}{2}))= B^{g}_{M}(x, \frac{r}{2}),
\]
for $e_{x}: \rr^{n}\to T_{x}M$ a linear isometry  and $\psi_{x}^{-1}(v)= \exp_{x}^{g}(\frac{r}{2}e_{x}v)$ for  $v\in B_{n}(0,1)$. 

It is known (see e.g. \cite[Sect. 3]{CGT}) that  
if $(M, g)$ is of bounded geometry, then  $\{g_{x}\}_{x\in M}$ is  bounded in ${\rm BT}^{0}_{2}(B_{n}(0,1), \delta)$.  In fact by \cite[Prop. 2.4]{Ro}  the Christoffel symbols expressed in normal coordinates at $x$ are uniformly bounded with all derivatives. Since 
$\nabla_{i}g_{jk}= \p_{i}g_{jk}- \Gamma^{l}_{ij}g_{lk}=0$,
this implies that all derivatives of $g_{x}$ in normal coordinates are bounded, hence (C1) holds. Moreover, by \cite[Lemma 2.2]{Ro} we know that 
\[
m_{x}\defeq  \left(\sup_{X}\|(\exp^{g}_{x})_{*}X\|_{g(x)}+ \|(\exp^{g}_{x})_{*}X\|^{-1}_{g(x)}\right),
\]
where $X$ ranges over all unit vector fields on $B_{n}(0,1)$, is uniformly bounded in $x\in M$. This is equivalent to property (C2).

Let us now prove $\Leftarrow$.
We first check  that $\nabla^{k}R$ is a  bounded tensor for $k\in \nn$.  Since $\psi_{x}: (U_{x}, g)\to (B_{n}(0,1), g_{x})$ is isometric,  it suffices to show that
\beq\label{e3.1}
\sup_{x\in M}\|\nabla^{k}_{g_{x}}R_{g_{x}}(0)\|<\infty.
\eeq
In \eqref{e3.1}, the norm is   associated to $g_{x}$, but by condition (C2)  we can replace it by the norm associated to the flat metric $\delta$. Then the l.h.s. of \eqref{e3.1} is a  fixed polynomial in the derivatives of $g_{x}$ and $g_{x}^{-1}$ computed at $0$, which  are uniformly bounded in $x\in M$, by condition (C1). Therefore \eqref{e3.1} holds.

It remains to prove that the injectivity radius $r_{g}$ is strictly positive. Let us denote for a moment by $r(x, N, h)$ the injectivity radius at $x\in N$ for $(N, h)$  a Riemannian manifold. % and by $r_{\rm inj}(N,h)= \inf_{x\in N}r(x, N, h)$ the global injectivity radius.  

Clearly
\begin{equation}
\label{e3.2}
r(x, U_{x}, g)\leq r(x, M,g), \ \  x\in M.
\end{equation}
By the isometry property of $\psi_{x}$ recalled above, we have:
\beq\label{e3.3}
r(x, U_{x}, g)= r(0, B_{n}(0,1), g_{x}).
\eeq
Conditions (C1), (C2) and standard estimates on differential equations imply that $\inf_{x\in M}r(0, B_{n}(0,1), g_{x})>0$ hence $r_{g}>0$ by \eqref{e3.2}, \eqref{e3.3}. \qed 
%By condition (C1), we know that if $\kappa_{x}$ is the sectional curvature of $g_{x}$, then there exists $\kappa>0$ such that $\sup_{x\in M}\kappa_{x}\leq \kappa$. 
% Since $g_{x}\geq c^{-1}\delta$, a  well known bound (see e.g. \cite{???}) asserts that
%\beq\label{e3.4}
%r(0, B_{n}(0,1), g_{x})\geq \min(\pi\sqrt{\kappa}^{-1}, r_{\rm inj}(B_{n}(0,1), c^{-1}\delta))= r>0.
%\eeq
%Using \eqref{e3.1}, \eqref{e3.2} this implies that $r_{g}>0$.
 \begin{lemma}\label{lomar.1}
 Let $\{U_{x}\}_{x\in M}$ be the good chart neighborhoods in Thm. \ref{thp0.1}. Then there exists $r>0$ such that $B^{g}_{M}(x, r)\subset U_{x}$ for all $x\in M$.
\end{lemma}
\proof From condition (C2) in Thm. \ref{thp0.1} we obtain the existence of some $r_{1}>0$ such that 
for any $x\in M$ $B_{g_{x}}(0, r_{1})\subset B_{n}(0, \12)$.  Since $\psi_{x}: (U_{x}, g)\to (B_{n}(0,1), g_{x})$ is isometric, this implies that $B_{g}(x, r_{1})\subset U_{x}$ as claimed. \qed

\begin{theoreme}\label{thomar.2} Let $(M,g)$ be a manifold of bounded geometry and  $\epsilon<\inf(1, r, r_{g})$, where $r$ is given in Lemma \ref{lomar.1}.  Set
 \[
\chi_{x}\defeq  \psi_{x}\circ \exp^{g}_{x}\circ e_{x}: B_{n}(0, \epsilon)\to \psi_{x}(B^{g}_{M}(x,\epsilon)).
\]
Then  for any multi-index $\alpha$ one has:
\beq\label{e3.5}
\sup_{x\in M, y\in B_{n}(0,\epsilon)}\|D_{y}^{\alpha}\chi_{x}(y)\|+ \sup_{x\in M, y\in \psi_{x}(B^{g}_{M}(x,\epsilon))}\|D_{y}^{\alpha}\chi_{x}^{-1}(y)\|<\infty.
\eeq
\end{theoreme}
\proof  Set $V_{x}=\psi_{x}(B^{g}_{M}(x, \epsilon))$. Since  $B^{g}_{M}(x, \epsilon)\subset U_{x}$ by Lemma \ref{lomar.1}, we see that $V_{x}\subset B_{n}(0,1)$. This implies \eqref{e3.5} for $\alpha=0$.

Let us now consider the case $|\alpha|=1$. Since $g_{x}= (\psi_{x}^{-1})^{*}g$, we have $\chi_{x}^{*}g_{x}=( \exp_{x}^{g}\circ e_{x})^{*}g$. Since $(M, g)$ is of bounded geometry, there exists $c>0$ such that 
\begin{equation}
\label{e3.6}
c^{-1}\delta\leq ( \exp_{x}^{g}\circ e_{x})^{*}g\leq c \delta.
\end{equation}
Using also condition (C2) of Thm. \ref{thp0.1}, we obtain $c_{1}>0$ such that
\[
c_{1}^{-1}\delta\leq g_{x}\leq c_{1}\delta,
\]
hence
\[
c_{1}^{-1}\chi_{x}^{*}\delta\leq \chi_{x}^{*}g_{x}\leq c_{1}\chi_{x}^{*}\delta.
\]
Since $\chi_{x}^{*}g_{x}= ( \exp_{x}^{g}\circ e_{x})^{*}g$, we obtain:
\beq\label{e3.7}
c_{1}^{-1}( \exp_{x}^{g}\circ e_{x})^{*}g\leq \chi_{x}^{*}\delta\leq c_{1}( \exp_{x}^{g}\circ e_{x})^{*}g.
\eeq
Combining \eqref{e3.6} and \eqref{e3.7} we obtain $c_{2}>0$ such that
\[
c_{2}^{-1}\delta\leq \chi_{x}^{*}\delta\leq c_{2}\delta.
\]
This is equivalent to \eqref{e3.5} for $|\alpha|=1$. 

To bound higher derivatives we use that $\chi_{x}$ is the exponential map transported by the chart diffeomorphism $\psi_{x}$. Denoting by $\Gamma^{k}_{ij,x}$ the Christoffel symbols for $g_{x}$, we obtain that if $v\in B_{n}(0, \epsilon)\subset T_{x}M$ and $|t|\leq 1$ and $(x^{1}(t), \dots, x^{n}(t))\defeq  \chi_{x}(tv)$ we have:
\[
\bec
\ddot{x}^{k}(t)= \Gamma^{k}_{ij, x}(x(t))\dot{x}^{i}(t)\dot{x}^{j}(t), \\
x(0)=0, \\
 \dot{x}(0)=v.
\eec
\]
Since $\{\Gamma_{ij,x}^{k}\}_{x\in M}$ is a bounded family in $\cinf_{\rm b}(B_{n}(0,1))$, it follows from standard arguments on dependence on initial conditions for differential equations that $\chi_{x}$ is uniformly bounded in $\cinf_{\rm b}(B_{n}(0, \epsilon))$ for $x\in M$. Since we already know that $D \chi_{x}^{-1}$ is bounded in  $C^{0}(V_{x})$,  we also obtain that $\chi_{x}^{-1}$ is bounded in $\cinf_{\rm b}(V_{x})$ uniformly in $x\in M$  as claimed. This completes the proof of the theorem. \qed

\subsection{Chart coverings and partitions of unity}\label{secp0.1b}
It is known (see \cite[Lemma 1.2]{Sh2}) that  if $(M, g)$ is of bounded geometry, there exist coverings by good chart neighborhoods:
\[
M= \bigcup_{i\in \nn}U_{i}, \ \  U_{i}= U_{x_{i}}, \ x_{i}\in M
\]
which are in addition {\em uniformly finite}, i.e. there exists $N\in \nn$ such that $\bigcap_{i\in I}U_{i}= \emptyset$ if $\sharp I> N$. Setting  $\psi_{i}= \psi_{x_{i}}$, we will call $\{U_{i}, \psi_{i}\}_{i\in \nn}$ a {\em good chart covering} of $M$. 

One can associate (see \cite[Lemma 1.3]{Sh2}) to a good chart covering a partition of unity:
\[
1=\sum_{i\in \nn}\chi_{i}^{2},  \ \chi_{i}\in \coinf(U_{i})
\]
such that $\{(\psi_{i}^{-1})^{*}\chi_{i}\}_{i\in \nn}$ is a bounded sequence in $\cinf_{\rm b}(B_{n}(0,1))$. Such a partition of unity will be called a {\em good partition of unity}.

\subsection{Bounded tensors,  bounded differential operators, Sobolev spaces}\label{secp0.2}
We now recall the definition of bounded tensors, bounded differential operators and of Sobolev spaces on $(M,g)$ of bounded geometry, see \cite{Sh2}.
\subsubsection{Bounded tensors}\label{secp0.2.1}

\begin{definition}\label{defp0.2}
 Let $(M,g)$  of  bounded geometry. We denote by ${\rm BT}^{p}_{q}(M,g)$ the spaces of  smooth $(q,p)$ tensors $T$ on $M$ such that if $T_{x}= (\exp_{x}^{g}\circ e_{x})^{*}T$ then the family 
 $\{T_{x}\}_{x\in M}$ is  bounded in ${\rm BT}^{p}_{q}(B_{n}(0, \frac{r}{2}), \delta)$. We equip ${\rm BT}^{p}_{q}(M, g)$ with its natural Fr\'echet space topology.
 \end{definition}
By Thm. \ref{thomar.2}  we can replace in Def. \ref{defp0.2} the geodesic maps $\exp_{x}^{g}\circ e_{x}$ by $\psi_{x}^{-1}$, where $\{\psi_{x}\}_{x\in M}$ is any family of good chart diffeomorphisms as in Thm. \ref{thp0.1}.

The Fr\'echet space topology on ${\rm BT}^{p}_{q}(M, g)$ is independent on the choice of  the family of good chart diffeomorphisms $\{\psi_{x}\}_{x\in M}$.
\subsubsection{Bounded differential operators}\label{defp0.22}
If $m\in \nn$ we denote by $\Diff^{m}(B_{n}(0,1), \delta)$ the space of {\em differential operators} of order $m$ on $B_{n}(0,1)$ with $\cinfb(B_{n}(0,1))$ coefficients, equipped with its Fréchet space topology.  
\begin{definition}
 Let $(M,g)$ of bounded geometry. We denote by $\Diff^{m}(M, g)$ the space of differential operators $P$ of order $m$ on $M$  such that  if $P_{x}= (\psi_{x}^{-1})^{*}P$ then the family $\{P_{x}\}_{x\in M}$ is bounded in $\Diff^{m}(B_{n}(0,1), \delta)$.

\end{definition}

\subsubsection{Sobolev spaces}\label{secp0.2.2}
Let $-\Delta_{g}$ be the Laplace-Beltrami operator on $(M, g)$, defined as the closure of its restriction to $\coinf(M)$. 
\begin{definition}
 For $s\in \rr$ we define the {\em Sobolev space} $H^{s}(M, g)$ as:
 \[
H^{s}(M, g)\defeq  \langle -\Delta_{g}\rangle^{-s/2}L^{2}(M, dg),
\]
with its natural Hilbert space topology.  
\end{definition}
It is known (see e.g. \cite[Sect. 3.3]{Ko}) that if $\{U_{i}, \psi_{i}\}_{i\in \nn}$ is a good chart covering and $1= \sum_{i}\chi_{i}^{2}$ is a subordinate good partition of unity, then an equivalent norm on $H^{s}(M, g)$ is given by:
\beq\label{e3.8bis}
\|u\|_{s}^{2}= \sum_{i\in \nn}\|(\psi_{i}^{-1})^{*}\chi_{i}u\|^{2}_{H^{s}(B_{n}(0,1))}.
\eeq
\subsection{Embeddings   of bounded geometry}\label{bemb}
We now recall the definition of embeddings of bounded geometry, see \cite{eldering}.
\begin{definition}\label{defp3.2b}
 Let $(M,g)$ an $n-$dimensional Riemannian manifold of bounded geometry, $\Sigma$ an $n-1$ dimensional manifold. An embedding $i: \Sigma\to M$ is called {\em of bounded geometry} if there exists a family  $\{U_{x}, \psi_{x}\}_{x\in M}$ of good chart diffeomorphisms for $g$ such that if $\Sigma_{x}\defeq  \psi_{x}(i(\Sigma)\cap U_{x})$ for we have
 \[
\Sigma_{x}= \{(v', v_{n})\in B_{n}(0,1) : \ v_{n}= F_{x}(v')\},
\]
where $\{F_{x}\}_{x\in M}$ is a bounded family in $\cinf_{\rm b}(B_{n-1}(0,1))$.
\end{definition}
The following fact  is shown in \cite[Lemma 2.27]{eldering}. 
\begin{lemma}\label{bemb-lemma}
 Assume $i: \Sigma\to M$ is an embedding of bounded geometry of $\Sigma$ in $(M, g)$. Then $(\Sigma, i^{*}g)$ is of bounded geometry.
\end{lemma}
\begin{lemma}\label{bemb-lemma-bis}
 Let $i: \Sigma\to M$ an embedding of bounded geometry. Then there exists a family  $\{U_{x}, \psi_{x}\}_{x\in M}$ of good chart diffeomorphisms as in Def. \ref{defp3.2b} such that if $x\in i(\Sigma)$ one has
 \[
\Sigma_{x}=  \psi_{x}(i(\Sigma)\cap U_{x})=\{(v', v_{n})\in B_{n}(0,1) : \ v_{n}= 0\}.
\]
\end{lemma}
\proof  Since  the family $\{F_{x}\}_{x\in \Sigma}$ in Def. \ref{defp3.2b} is uniformly bounded  in $\coinf(B_{n-1}(0,1))$ we can find $\alpha, \beta>0$ such that  if $\phi_{x}(v', v_{n})= (v', \alpha(v_{n}- F_{x}(v')))$ we have $B_{n}(0, 1)\subset \phi_{x}(B_{n}(0,1))\subset B_{n}(0, \beta)$. Clearly $\{\phi_{x}\}_{x\in \Sigma}$ is a bounded family of diffeomorphisms in $\cinfb(B_{n}(0,1))$.
For $x\in \Sigma$ we replace $U_{x}$ by  $(\phi_{x}\circ \psi_{x})^{-1}B_{n}(0,1)$ and $\psi_{x}$ by $ \phi_{x}\circ \psi_{x}$.  For $x\not\in \Sigma$, $U_{x}$ and $\psi_{x}$ are left unchanged. \qed
\subsection{Equivalence classes of Riemannian metrics}\label{secp0.3}
The results of this subsection are due to \cite{Ou}.
\begin{proposition}\label{propp0.0}
 Let $(M,g)$ be of bounded geometry. Let $k$ be another  Riemannian metric on $M$ such that $k\in {\rm BT}^{0}_{2}(M,g)$ and $k^{-1}\in {\rm BT}^{2}_{0}(M, g)$. Then
 \ben
 \item $(M,k)$ is of bounded geometry;
 \item ${\rm BT}^{p}_{q}(M, g)= {\rm BT}^{p}_{q}(M,k)$, $H^{s}(M, g)= H^{s}(M, k)$ as topological vector spaces.
 \een 
 Let us write $k\sim g$  if the above conditions are satisfied. Then $\sim$ is an equivalence relation on the class of bounded geometry Riemannian metrics on $M$.
 \end{proposition}
 \proof 
 Let us first prove (1).  We equip $M$ with a good chart covering $\{U_{x}, \psi_{x}\}_{x\in M}$ for $g$. Then conditions (C1),  (C2) of Thm. \ref{thp0.1} are  satisfied by $k$, hence $(M,k)$ is of bounded geometry and $\{U_{x}, \psi_{x}\}_{x\in M}$ is a good chart covering for $k$. Using that $k\in {\rm BT}^{0}_{2}(M,g)$ and $k^{-1}\in {\rm BT}^{2}_{0}(M, g)$ this implies that ${\rm BT}^{p}_{q}(M, g)= {\rm BT}^{p}_{q}(M,k)$.  The statement about Sobolev spaces follows from  the equivalent norm given in \eqref{e3.8bis}.
 
 Let us show that $\sim$ is symmetric. If $g_{1}\sim g_{2}$, then we have seen that $\BT^{p}_{q}(M, g_{1})= \BT^{p}_{q}(M, g_{2})$. Since $(M, g_{2})$ is of bounded geometry, we have $g_{2}\in \BT^{0}_{2}(M, g_{2})= \BT^{0}_{2}(M, g_{1})$, $g_{2}^{-1}\in \BT^{2}_{0}(M, g_{2})=\BT^{2}_{0}(M, g_{1})$, hence $g_{2}\sim g_{1}$. The same argument shows that $\sim$  is transitive.  \qed

  We conclude this subsection with an easy fact.
  
\begin{proposition}\label{propp0.1}
 Let $g_{i}$, $i=1,2$ be two Riemannian metrics of bounded geometry having a common family of good chart diffeomorphisms $\{U_{x}, \psi_{x}\}_{x\in M}$. Then $g_{1}\sim g_{2}$.
\end{proposition}
 \proof This follows directly from the remark below Def. \ref{defp0.2} and the definition of the equivalence relation $g_{1}\sim g_{2}$.\qed

\subsection{Examples}\label{secp0.4}
We now recall  some well-known examples of manifolds of bounded geometry, which will be useful later on. 
\subsubsection{Compact manifolds and compact perturbations}
Clearly  any compact Riemannian manifold is of bounded geometry. Similarly if $(M,g_{1})$ is of bounded geometry and if $g_{2}=g_{1}$ outside some compact set, then $(M, g_{2})$ is of bounded geometry and $g_{1}\sim g_{2}$.
\subsubsection{Gluing of Riemannian manifolds}

Let $(M_{i}, g_{i})$, $i=1,2$ be two Riemannian manifolds of bounded geometry,  $K_{i}\subset M_{i}$  be open and relatively compact   and $j: K_{1}\to K_{2}$   an isometry. Then the Riemannian manifold $(M, g)$ obtained by gluing $M_{1}$ and $M_{2}$ along $K_{1}\overset{j}{\sim} K_{2}$  is of bounded geometry. 

\subsubsection{Cartesian products}
If $(M_{i}, g_{i})$ $i=1,2$ are Riemannian manifolds of bounded geometry then $(M_{1}\times M_{2},g_{1}\oplus g_{2})$ is of bounded geometry.
\subsubsection{Warped products} We provide a further useful argument that gives manifolds of bounded geometry in the form of warped products.
\begin{proposition}\label{warped}
 Let $(K, h)$ be a  Riemannian manifold of bounded geometry, and  $M= \rr_{s}\times K$, $g= ds^{2}+ F^{2}(s)h$, where:
 \ben
 \item $F(s)\geq c_{0}>0, \ \forall s\in \rr$,  for some $c_{0}>0$;
 \item $|F^{(k)}(s)|\leq c_{k}F(s), \ \forall s\in \rr, \ k\geq 1$.
 \een 
 Then $(M, g)$ is of bounded geometry.
\end{proposition}
\proof 
Let $r$ the injectivity radius of $(K, h)$, and $e_{y}: (\rr^{n-1}, \delta)\to (T_{y}K, h(y))$ for $y\in K$  be linear isometries. We set for $x=(\sigma, y)\in M$ and $c_{0}$ the constant in (1):
\[
\psi_{x}^{-1}:\ \begin{array}{l}
]-1, 1[\times B_{n-1}(0, \frac{r}{2}c_{0})\to ]-1, 1[\times B^{h}_{K}(y, \frac{rc_{0}}{2 F(\sigma)})\\[2mm]
(s, v)\mapsto \left(s+\sigma, \exp^{h}_{y}(F(\sigma)^{-1}e_{y}v)\right).
\end{array}
\]
We have:
\[
g_{x}=(\psi_{x}^{-1})^{*}g= ds^{2}+ \frac{F^{2}(s+ \sigma)}{F^{2}(\sigma)}h_{y}(e_{y}F(\sigma)^{-1}v)dv^{2},
\]
where $h_{y}= (\exp^{h}_{y}\circ e_{y})^{*}h$.  By (2) we have
$|\ln(\frac{F(s+ \sigma)}{F(\sigma)})|\leq \int_{\sigma}^{s+\sigma}| \frac{F'(u)}{F(u)}|du\leq c_{1}|s|$,  hence:
\[
\e^{-c_{1}}\leq \frac{F(s+ \sigma)}{F(\sigma)}\leq\e^{c_{1}}, \ \sigma\in \rr, s\in\,]-1, 1[.
\]
This implies that $g_{x}$ is uniformly equivalent to the flat metric $\delta$.  Moreover from conditions (1), (2) we obtain that $\{g_{x}\}_{x\in M}$ is bounded in $\BT^{0}_{2}(]-1, 1[\times B_{n-1}(0,  \frac{r}{2}c_{0}))$.

We now choose $\rho>0$ such that  $B_{n}(0, \rho)\subset \,]-1, 1[\times B_{n-1}(0,  \frac{r}{2}c_{0})$, 
and  compose $\psi_{x}^{-1}$ with a fixed diffeomorphism between $B_{n}(0, \rho)$ and $B_{n}(0, 1)$.   
Conditions  (C1), (C2) of Thm. \ref{thp0.1} are then satisfied. 
\qed  

\section{Lorentzian manifolds of bounded geometry}\label{secp3}\init
In this section we consider Lorentzian manifolds $(M, g)$. Reference Riemannian metrics on $M$ with be denoted by $\hat{g}$.
We still denote by $\exp^{g}_{x}$ the exponential map at $x\in M$ for $g$. The results of this section are due to \cite{Ou}.
\subsection{Definitions}\label{secp3.1}
\begin{definition}\label{defp3.1}
 A smooth Lorentzian manifold $(M, g)$ is of {\em bounded geometry} if there exists a Riemannian metric $\hat{g}$ on $M$ such that:
 \ben
 \item $(M, \hat{g})$ is of bounded geometry;
 \item $g\in {\rm BT}^{0}_{2}(M, \hat{g})$ and $g^{-1}\in {\rm BT}^{2}_{0}(M, \hat{g})$.
 \een
\end{definition}
Clearly the above conditions only depend on the equivalence class of $\hat{g}$ for the equivalence relation $\sim$ introduced in  Subsect. \ref{secp0.3}. The following theorem is a partial converse to this property.  
\begin{theoreme}\label{th3.1}
 Let $(M,g)$ a Lorentzian manifold and $\hat{g}_{i}$, $i= 1,2$ two Riemannian metrics on $M$ such that:
 \begin{romanenumerate}
 \item $(M, \hat{g_{i}})$ is of bounded geometry;
 \item $g\in {\rm BT}^{0}_{2}(M, \hat{g}_{i})$ and $g^{-1}\in {\rm BT}^{2}_{0}(M, \hat{g}_{i})$.
 \end{romanenumerate}
 Then  the following are equivalent:
 \ben
\item $\hat{g}_{1}\sim \hat{g}_{2}$,
\item there exists $c>0$ such that $c^{-1}\hat{g}_{2}(x)\leq \hat{g}_{1}(x)\leq c \hat{g}_{2}(x)$, $\forall x\in M$,
\item there exists $c>0$ such that $ \hat{g}_{2}(x)\leq c \hat{g}_{1}(x)$, $\forall x\in M$.
\een
\end{theoreme}
\proof 
We start by some preparations.
Let $(M, g)$ be a smooth Lorentzian manifold and $\hat{g}$ a  Riemannian metric on $M$ such that $(M, \hat{g})$ is of bounded geometry and $g\in \BT^{0}_{2}(M, \hat{g})$, $g^{-1}\in \BT^{2}_{0}(M, \hat{g})$. Let $\{U_{x}, \psi_{x}\}_{x\in M}$  be a family of good chart diffeomorphisms for $\hat{g}$ and  let $g_{x}= (\psi_{x}^{-1})^{*}g$.  

By the above property of $g$ and $g^{-1}$, we obtain that there exists $0<r, r'<1$ 
such that  $\exp_{0}^{g_{x}}$ is well defined on $B_{n}(0, r)$,  is a smooth diffeomorphism on its image, and moreover $B_{n}(0, r')\subset \exp_{0}^{g_{x}}B_{n}(0, r)$, and the family $\{\exp_{0}^{g_{x}}\}_{x\in M}$ is bounded in $\cinf_{\rm b}(B_{n}(0,r))$.

Let us identify $B^{\hat{g}(x)}(0, 1)\subset T_{x}M$ with $B_{n}(0,1)\subset \rr^{n}$ with isometries $e_{x}: (T_{x}M, \hat{g}(x))\to (\rr^{n}, \delta)$ and set
\[
\phi_{x}: B_{n}(0,1)\ni v\mapsto \exp^{g}_{x}\circ e_{x}(rv)\in M,
\]
$V_{x}\defeq  \phi_{x}(B_{n}(0,1))$, $\chi_{x}= \phi_{x}^{-1}$. Since $\exp^{g_{x}}_{0}$ equals $\exp^{g}_{x}$ transported by $\psi_{x}$, it follows from the 
properties of $\{\exp_{0}^{g_{x}}\}_{x\in M}$ shown above  that $\{V_{x}, \chi_{x}\}_{x\in M}$ is a family of good chart diffeomorphisms for $\hat{g}$.

Let now $\hat{g}_{i}$, $i= 1, 2$ as in the theorem and let $r= \inf(r_{1},r_{2})$, where $r_{i}$ is the radius $r$ above for $\hat{g}_{i}$.  
We choose isometries $e_{i,x}: (\rr^{n}, \delta)\to (T_{x}M, \hat{g}_{i}(x))$ and denote by $\{V_{i,x}, \chi_{i,x}\}_{x\in M}$ the families of good chart diffeomorphisms for $\hat{g}_{i}$ constructed above. 

Let us compute the map $T_{x}\defeq  \chi_{1,x}\circ \chi_{2, x}^{-1}$, which is defined on some neighborhood of $0$ in $B_{n}(0,1)$. Denoting by $\lambda_{r}: \rr^{n}\to \rr^{n}$ the multiplication by $r$, we have:
\beq\label{tita}
\bea
\chi_{1, x}\circ \chi_{2, x}^{-1}&= \lambda_{r}\circ (\exp^{g}_{x}\circ e_{1, x})^{-1}\circ( \lambda_{r}\circ (\exp^{g}_{x}\circ e_{2, x})^{-1} )^{-1}\\[2mm]
&= \lambda_{r} \circ e_{1, x}^{-1}\circ e_{2, x}\circ \lambda_{r}^{-1}= e_{1, x}^{-1}\circ e_{2,x}.
\eea
\eeq
We claim that property (1) is equivalent to
\begin{equation}
\label{titu}
\sup_{x\in M}\|T_{x}\| + \| T_{x}^{-1}\|<\infty,
\end{equation}
where $\|\cdot \|$ is the norm on  $L(\rr^{n})$ inherited from $\delta$. 
To  prove the claim we set:
\[
\hat{g}_{i, x}= (\chi_{1, x}^{-1})^{*}\hat{g}_{i}, \ \ \tilde{g}_{i, x}= (\chi_{i, x}^{-1})^{*}\hat{g}_{i},
\]
so that
\beq\label{lucette}
\tilde{g}_{2,x}= T_{x}^{*}\hat{g}_{2, x}.
\eeq
 We have seen above that $\{\chi_{i,x}\}_{x\in M}$ is a family of good chart diffeomorphisms for $\hat{g}_{i}$.
Therefore $\{\tilde{g}_{2, x}\}_{x\in M}$ and $\{\tilde{g}_{2, x}^{-1}\}_{x\in M}$ are bounded in $\BT^{0}_{2}(B_{n}(0, 1), \delta)$ and $\BT^{2}_{0}(B_{n}(0, 1), \delta)$.
Moreover by Prop. \ref{propp0.0} we know that  $\hat{g}_{2}\sim \hat{g}_{1}$ iff   $\{\hat{g}_{2, x}\}_{x\in M}$ and $\{\hat{g}_{2, x}^{-1}\}_{x\in M}$ are bounded in $\BT^{0}_{2}(B_{n}(0, 1), \delta)$ and $\BT^{2}_{0}(B_{n}(0, 1), \delta)$. 
By \eqref{lucette}, this is equivalent to the fact that
$\{(T_{x}^{-1})^{*}\delta\}_{x\in M}$ and $\{(T_{x}^{-1})^{*}\delta^{-1}\}$ are bounded in  $\BT^{0}_{2}(B_{n}(0, 1), \delta)$ and $\BT^{2}_{0}(B_{n}(0, 1), \delta)$.  Since $T_{x}$ are linear maps this is clearly equivalent to \eqref{titu}, which completes the proof of the claim.

Let now  $g_{i, x}\defeq  (\chi_{i, x}^{-1})^{*}g= (\lambda_{r}^{-1})^{*}(e_{i,x})^{*}(\exp_{x}^{g})^{*}g$. The same argument as in \eqref{tita} shows that:
\[
g_{2, x}= T_{x}^{*}g_{1, x}=  {}^{\rm t}T_{x}\circ g_{1,x}\circ T_{x}.
\]
Computing the determinant of  the quadratic forms $g_{i,x}(0)$ using $\delta$, this implies that
\[
(\det T_{x})^{2}=\det g_{2,x}(0)\det g_{1, x}^{-1}(0). 
\]
Since $g_{i}$ and $g_{i}^{-1}$ are bounded tensors, we obtain that  there exists $c>0$  such that $c^{-1}\leq |\det T_{x}|\leq c$.  This implies that \eqref{titu} is equivalent to
\begin{equation}
\label{titop}
\sup_{x\in M}\|T_{x}^{-1}\| <\infty.
\end{equation}
Finally the discussion above shows that property (2) is equivalent to $c^{-1}\tilde{g}_{2, x}\leq \hat{g}_{2, x}\leq c \tilde{g}_{2,x}$ $\forall x\in M$, which is  equivalent to \eqref{titu}. Property (3) is equivalent to $\hat{g}_{2, x}\leq c \tilde{g}_{2,x}$ $\forall x\in M$ which is equivalent to \eqref{titop}.  Since we have seen that (1), \eqref{titu} and \eqref{titop} are equivalent, the proof is complete. 
\qed

\subsection{Cauchy hypersurfaces of bounded geometry}\label{secp3.2}

%\textcolor{Red}{Clarify immersion or embedding. If too complicated, consider simply a submanifold}.
%\begin{definition}\label{defp3.2}
% Let $(M, \hat{g})$ an $n-$dimensional Riemannian manifold of bounded geometry, $\Sigma$ an $n-1$ dimensional manifold. An immersion $i: \Sigma\to M$ is called {\em of bounded geometry} if there exists a family  $\{U_{x}, \psi_{x}\}_{x\in M}$ of good chart diffeomorphisms for $\hat{g}$ such that for $\Sigma_{x}\defeq  \psi_{x}(i(\Sigma)\cap U_{x})$ we have either $\Sigma_{x}=\emptyset$ or
% \[
%\Sigma_{x}= \{(v',v_{n})\in B_{n}(0,1)\ :  \ v_{n}= F_{x}(v')\},
%\]
%where $\{F_{x}\}_{x\in M}$ is a bounded family in $\cinf_{\rm b}(B_{n-1}(0,1))$.
%\end{definition}

We adopt the convention that  a spacetime $(M,g)$ is a  Hausdorff, paracompact, connected time orientable Lorentzian manifold equipped with a time orientation.  Lorentzian manifolds are naturally endowed with a {\em causal structure}; we refer the reader to \cite[Chap. 8]{W} or \cite[Sect. 1.3]{BGP} for details. 

In the sequel we denote by $I^{\pm, g}_{M}(U)$, (resp. $J^{\pm, g}_{M}(U)$) for $U\subset M$ the  future/past time-like  (resp. causal) shadow of $U$. If $(M, g)$ is clear from the context we use instead the notation $I^{\pm}(U)$ (resp. $J^{\pm}(U)$). 
 We denote by $\cinf_{\rm sc}(M)$ the space of smooth {\em space-compact} functions, i.e. with support included in $J^{+}(K)\cup J^{-}(K)$ for some compact set $K\Subset M$.

A smooth hypersurface $\Sigma$ is a {\em Cauchy hypersurface} if any inextensible piecewise smooth time-like curve intersects $\Sigma$ at one and only one point.  

 A space\-time having a Cauchy hypersurface is called {\em globally hyperbolic} {(see \cite{bernalsanchez} for the equivalence with the alternative definition where $\Sigma$ is not required to be smooth)}. Global hyperbolic space\-times are natural Lorentzian manifolds on  which to study  Klein-Gordon operators.

\begin{definition}\label{defp3.3}
 Let $(M, g)$ be an $n-$dimensional   Lorentzian manifold of bounded geometry and $\hat{g}$ a Riemannian metric as in Def. \ref{defp3.1}. Assume also that  $(M, g)$ is  globally hyperbolic. Let $\Sigma\subset M$ a {\em spacelike} Cauchy hypersurface.   Then $\Sigma$ is called a {\em bounded geometry Cauchy hypersurface} if:
 \ben
 \item  the injection $i: \Sigma\to M$ is of bounded geometry for $\hat{g}$,
 \item if $n(y)$ for  $y\in \Sigma$ is the future directed unit normal  for $g$ to $\Sigma$ one has:
 \[
\sup_{y\in \Sigma}n(y)\cdot \hat{g}(y)n(y)<\infty.
\]
 \een
\end{definition}
We recall now a well-known result about geodesic normal coordinates to a Cauchy hypersurface $\Sigma$.
\begin{proposition}\label{propp0.2}
 Let $\Sigma$ be a space-like Cauchy hypersurface in a globally hyperbolic space\-time $(M, g)$. Then there exists a neighborhood $U$ of $\{0\}\times \Sigma$ in $\rr\times \Sigma$ and a neighborhood $V$ of $\Sigma$ in $M$ such that  the map:
 \[
\chi:\ \begin{array}{l}
U\to V\\
(s, y)\mapsto \exp_{y}^{g}(sn(y))
\end{array}
\] 
is  a diffeomorphism.
 Moreover $\chi^{*}g\traa{V}= -ds^{2}+ h_{s}$  where $h_{s}$ is a smooth, $s-$dependent family of Riemannian metrics on $\Sigma$. 
\end{proposition}
\proof The first statement is shown in \cite[Prop. 26]{O1}.  To prove the second statement, we can work  near a point in $\Sigma$ and introduce local coordinates $y$ on $\Sigma$. In \cite[Sect. 3.3]{W} it is shown that the normal geodesics are orthogonal to the hypersurfaces $\Sigma_{t}=\{s=t\}$. 
Since in  the normal coordinates, $\p_{s}$ is a  tangent vector to the normal geodesics, and $\p_{y^{i}}$ span $T \Sigma_{t}$ this implies 
 that the metric does not contain $ds dy^{i}$ terms. If $n$ is the future directed normal vector field to the family $\Sigma_{t}$, then $n \cdot g n=-1$ first on  $\Sigma_{0}$ and then on all $\Sigma_{t}$ by the geodesic equation. This completes the proof. \qed

 In the next theorem  we study properties of the normal coordinates for Cauchy hypersurfaces of bounded geometry.
\begin{theoreme}\label{th-omar}
 Let $(M, g)$ a Lorentzian manifold of bounded geometry and $\Sigma$ a bounded geometry Cauchy hypersurface. Then the following holds:
\ben
\item there exists $\delta>0$ such that the normal geodesic flow to $\Sigma$:
\[
\chi:\ \begin{array}{l}
]-\delta, \delta[\times \Sigma\to M\\
(s, y)\mapsto \exp_{y}^{g}(sn(y))
\end{array}
\] 
is well defined and is a smooth diffeomorphism on its image;
\item $\chi^{*}g= - ds^{2}+h_{s}$, where $\{h_{s}\}_{s\in\,]-\delta, \delta[}$ is a smooth family of Riemannian metrics on $\Sigma$ with
\[
\begin{array}{rl}
i)&(\Sigma, h_{0})\hbox{ is of bounded geometry},\\[2mm]
ii)& s\mapsto h_{s}\in \cinf_{\rm b}(]-\delta, \delta[, \BT^{0}_{2}(\Sigma, h_{0})), \\[2mm]
iii)& s\mapsto h^{-1}_{s}\in \cinf_{\rm b}(]-\delta, \delta[, \BT^{2}_{0}(\Sigma, h_{0})).
\end{array}
\]
 \een
\end{theoreme}
\proof 
Let us first prove (1).  The proof consists of several steps.

{\it Step 1}: since $g$ is of bounded geometry for the reference metric $\hat{g}$, we first see by standard arguments that there exists $\rho_{2}, c_{2}>0$ such that  for all $x\in M$, 
\[
\exp_{x}^{g}:B_{T_{x}M}^{\hat{g}(x)}(0, \rho_{2})\to M
\]
 is well defined  and $c_{2}$-Lipschitz
if we equip $B_{T_{x}M}^{\hat{g}(x)}(0, \rho_{2})$ with the distance associated to $\hat{g}(x)$ and $M$ with the distance associated with $\hat{g}$.

{\it Step 2}:
Recall that  $i: \Sigma\to M$  is the natural injection. For $y\in \Sigma$, we set $A_{y}= D_{(0, y)}\chi\in L(\rr\times T_{y}\Sigma, T_{y}M)$. We  have:
\[
\begin{array}{l}
A_{y}(\alpha, v)= \alpha n(y)+ D_{y}i v, \ \alpha\in \rr, v\in T_{y}\Sigma,\\[2mm]
A_{y}^{-1}w= (- n(y)\cdot g(y)w, (D_{y}i)^{-1}(w+ n(y)\cdot g(y)w n(y))).
\end{array}
\]
If we equip $T_{y}\Sigma$ with the metric $i^{*}\hat{g}(y)$ and $T_{y}M$ with $\hat{g}(y)$, we deduce from conditions (1) and (2) in Def. \ref{defp3.3} that the norms of $A_{y}$ and $A_{y}^{-1}$ are  uniformly bounded in $y$.

 By the local inversion theorem, there exists $\delta_{1}>0$ such that  for any $y\in \Sigma$
$\chi$ is well defined on $]-\delta_{1}, \delta_{1}[\times B^{\hat{g}}(y, \delta_{1})\cap \Sigma$ and is a diffeomorphism on its image.  

{\it Step 3}: let now $c_{1}= \sup_{y\in \Sigma}n(y)\cdot \hat{g}(y)n(y)<\infty$ and 
\[
\delta= \min(\delta_{1}, \rho_{2})(2 + 2c_{1}+ 4c_{1}c_{2})^{-1},
\]
where $\rho_{2}, c_{2}$ are introduced at the beginning of the proof. 
We claim that
\[
\chi: \,]-\delta, \delta[\times \Sigma\to M
\]
is a smooth diffeomorphism on its image, which will complete the proof of (1). By the above discussion, $\chi$ is a local diffeomorphism, so it remains to prove that $\chi$ is injective. Let $(s_{i}, y_{i})\in\,]-\delta, \delta[\times \Sigma$ such that $\chi(s_{1}, y_{1})= \chi(s_{2}, y_{2})=x$.

If $y\in \Sigma$ and $|s|<\delta$, we have $\| s n(y)\|_{\hat{g}}\leq c_{1}\delta<\rho_{2}$, hence by  the Lipschitz property of $\exp_{x}^{g}$ in Step 1 we have:
\[
d_{\hat{g}}(y, \exp^{g}_{y}(sn(y)))\leq c_{2}|s|\|n(y)\|_{\hat{g}}.
\]
This yields
\[
\bea
d_{\hat{g}}(y_{1}, y_{2})&\leq d_{\hat{g}}(y_{1}, x)+ d_{\hat{g}}(y_{2}, x)\\[2mm]
&=d_{\hat{g}}(y_{1}, \exp_{y_{1}}^{g}(s_{1}n_{y_{1}}))+ d_{\hat{g}}(y_{2}, \exp_{y_{2}}^{g}(s_{2}n_{y_{2}}))\\[2mm]
&\leq c_{2}|s_{1}|\|n_{y_{1}}\|_{\hat{g}}+ c_{2}|s_{2}|\|n_{y_{2}}\|_{\hat{g}}\\[2mm]
&\leq 2 c_{2}c_{1}\delta\leq \frac{\delta_{1}}{2}.
\eea
\]
It follows that $(s_{i}, y_{i})\in\,]-\delta_{1}, \delta_{1}[\times B^{\hat{g}}(y_{1}, \delta_{1})\cap \Sigma$. Since by Step 2 $\chi$ is injective on this set, we have $(s_{1}, y_{1})= (s_{2}, y_{2})$, which completes the proof of (1).

Let us now prove (2).
%Note  that $g\in \BT^{0}_{2}(M, \hat{g})$, $g^{-1}\in \BT^{2}_{0}(M, \hat{g})$ and  $i: \Sigma\to M$ is of bounded geometry for $\hat{g}$. Using Def. \ref{defp3.2}, this implies that 
%$i^{*}g\in \BT^{0}_{2}(\Sigma, i^{*}\hat{g})$, $i^{*}g^{-1}\in \BT^{2}_{0}(\Sigma, \hat{g})$. But $i^{*}g= h_{0}$, which proves {\it i)}, by Prop. \ref{propp0.0}.
For $x\in \Sigma$, we choose  $U_{x}, \psi_{x}$ as in Lemma \ref{bemb-lemma-bis}. We recall that $\Sigma_{x}= \psi_{x}(\Sigma\cap U_{x})$,  $g_{x}=(\psi_{x}^{-1})^{*}g$ and denote by $n_{x}$ the future-directed unit  normal vector field to $\Sigma_{x}$ for $g_{x}$. We have  $\Sigma_{x}= \{v\in B_{n}(0,1):  v_{n}=0\}\sim B_{n-1}(0, 1)$ and we can decompose  $n_{x}$ as $n_{x}= n'_{x}+ \lambda_{x}e_{n}$, where $n'_{x}$ is tangent to $\Sigma_{x}$.  Then  $\{g_{x}\}_{x\in \Sigma}$, $\{g^{-1}_{x}\}_{x\in \Sigma}$, $\{n'_{x}\}_{x\in \Sigma}$, $\lambda_{x}$ are bounded in $\BT^{0}_{2}(B_{n}(0,1), \delta)$, $\BT^{2}_{0}(B_{n}(0,1), \delta)$, $\BT^{1}_{0}(B_{n-1}(0,1), \delta)$  and $\BT^{0}_{0}(B_{n-1}(0,1), \delta)$ respectively.  

By standard estimates on differential equations, this implies that there exists $\delta'>0$ such that the normal geodesic flow
\beq\label{local-normal}
\chi_{x}:\ \begin{array}{l}
]-\delta', \delta'[\times B_{n-1}(0,\12)\to B_{n}(0,1)\\
(s, v')\mapsto \exp_{(v',0)}^{g_{x}}(sn_{x}(v', 0))
\end{array}
\eeq
is a  diffeomorphism on its image, with $\{\chi_{x}\}_{x\in \Sigma}$ bounded in $\cinf_{\rm b}(]-\delta', \delta'[\times B_{n-1}(0,\12))$. Moreover if $V_{x}\defeq  \chi_{x}(]-\delta', \delta'[\times B_{n-1}(0,\12))$, then $\chi_{x}^{-1}$ is the restriction to $V_{x}$ of a map $\phi_{x}: B_{n}(0, 1)\to \rr^{n}$ such that 
$\{\phi_{x}\}_{x\in \Sigma}$ is bounded in $\cinf_{\rm b}(B_{n}(0,1))$.

We have $\chi_{x}^{*}g_{x}= - ds^{2}+ h_{x}(s, v')dv'^{2}$, where $h_{x}(s, v')dv'^{2}$ is an $s-$dependent Riemannian metric on $B_{n-1}(0,1)$. 

To prove statement (2) it remains to check that  $\{h_{x}\}_{x\in \Sigma}$ and $\{h_{x}^{-1}\}_{x\in \Sigma}$ are bounded in $\BT^{0}_{2}(]-\delta', \delta'[\times B_{n-1}(0,\12))$ and $\BT^{2}_{0}(]-\delta', \delta'[\times B_{n-1}(0,\12))$ respectively.  This follows from the same properties of $g_{x}$, $g_{x}^{-1}$ and $\chi_{x}$ recalled above.   The proof is complete. \qed

\begin{remark}\label{rem-th-omar}
 Since the diffeomorphisms $\chi_{x}$ in  \eqref{local-normal} are bounded with all derivatives (in good coordinates for the reference Riemannian metric $\hat{g}$), we see that $\chi^{*}\hat{g}$ is equivalent to $ds^{2}+ h_{0}dy^{2}$on $I\times \Sigma$, or more precisely that one can extend $\chi^{*}\hat{g}$ to $\rr\times \Sigma$ such that the extension is equivalent to $ds^{2}+ h_{0}dy^{2}$ on $\rr\times \Sigma$.
\end{remark}

\subsection{A framework for Klein-Gordon operators}\label{secp3.3}
In Sects. \ref{secp2}, \ref{secp4} we will consider {\em Klein-Gordon} operators  on a globally hyperbolic space\-time $(M, g)$:
\beq\label{kgkg}
P= - \nabla^{a}\nabla_{a}+ V, \ V\in\cinf(M; \rr),
\eeq
and in particular  the {\em Cauchy problem} on  a Cauchy hypersurface $\Sigma$. In this subsection we formulate a rather general framework which will allow us later on to apply tools from the {\em pseudodifferential calculus} on manifolds of bounded geometry, see Sect. \ref{secp1} for the construction of {\em parametrices} for the Cauchy problem for $P$.

%We denote by $\Sol(P)$ the space of smooth (complex valued) space-compact solutions of $P\phi=0$. We recall  (see e.g. \cite{BGP}) that $(\Sol(P), \sigma)$ is a complex symplectic space for
%\[
%\bar{\phi}_{1}\cdot\sigma \phi_{2}= \int_{\Sigma}(\p_{n}\bar{\phi}_{1}\phi_{2}- \bar{\phi}_{1}\p_{n}\phi_{2})d\nu_{\Sigma},
%\]
%where $\p_{n}$ is the normal vector field to $\Sigma$, $d\nu_{\Sigma}$ the induced measure on $\Sigma$, the above integral being independent on the choice of the space-like Cauchy hypersurface $\Sigma$.

If $(M_{i}, g_{i})$ are two space\-times, a \emph{space\-time embedding} $i: (M_{1}, g_{1})\to (M_{2}, g_{2})$ is by definition an embedding that is isometric and  preserves the time-orientation. In addition, if $(M_{i}, g_{i})$ globally hyperbolic, one says that $i$  is {\em causally compatible} if:
\[
I^{\pm, g_{1}}_{M_{1}}(U)= i^{-1}(I^{\pm, g_{2}}_{M_{2}}(U)), \ \forall \, U\subset M_{1}.
\]

We fix a globally hyperbolic space\-time $(M,g)$, a Cauchy hypersurface $\Sigma$ and a function $V\in \cinf(M; \rr)$.   We assume that there exist:
\ben
\item  a neighborhood $U$ of $\Sigma$ in $M$, 
\item a Lorentzian metric $\tilde{g}$ on $M$,
\item a function $\tilde{c}\in \cinf(M; \rr)$, $\tilde{c}>0$,
\een
such that:

(H1)   $(M, \tilde{c}^{2}\tilde{g})$ is globally hyperbolic, $i: (U, g)\to (M, \tilde{c}^{2}\tilde{g})$ is causally compatible,

(H2)  $\tilde{g}$ is of bounded geometry for some reference Riemannian metric $\hat{g}$, $\Sigma$ is a Cauchy hypersurface of  bounded geometry in $(M, \tilde{g})$,

(H3) $d\ln \tilde{c}$ belongs to $\BT^{0}_{1}(M, \hat{g})$,

(M) $\tilde{c}^{2}V$ belongs to $\BT^{0}_{0}(M, \hat{g})$.

\begin{proposition}\label{propostup}
 Assume hypotheses {\upshape (H)}. Then  there exist:
\ben
  \item an open interval $I$ with $0\in I$, a diffeomorphism $\chi:  I\times \Sigma\to U$, 
  \item  a smooth family $\{h_{t}\}_{t\in I}$ of Riemannian metrics on $\Sigma$ with
 \[
 \begin{array}{l}
 (\Sigma, h_{0})\hbox{ is of bounded geometry},\\[2mm]
 I\ni t \mapsto h_{t}\in \cinfb(I; \BT^{0}_{2}(\Sigma, h_{0})), \ I\ni t \mapsto h^{-1}_{t}\in \cinfb(I; \BT^{2}_{0}(\Sigma, h_{0})),
 \end{array}
  \]
\item a function $c\in \cinf(I\times \Sigma)$, $c>0$ with 
\[
\nabla_{h_{0}}\ln c\in \cinfb(I; \BT^{1}_{0}(\Sigma, h_{0})), \ \p_{t}\ln c\in \cinfb(I; \BT^{0}_{0}(\Sigma, h_{0})),
\]
or equivalently
\[
dc\in \BT^{0}_{1}(I\times \Sigma, dt^{2}+ h_{0}),
\]
\een
such that
 \beq\label{omito}
 \chi^{*}g= c^{2}(t, y)(- dt^{2}+ h_{t}(y)dy^{2}) \hbox{ on }U. 
 \eeq
 If moreover hypothesis {\upshape (M)} holds then:
 \beq\label{omiti}
 c^{2}V\circ \chi^{-1}\in \cinfb(I; \BT^{0}_{0}(\Sigma, h_{0})).
\eeq
 \end{proposition}
\proof  We apply Thm. \ref{th-omar} to $\tilde{g}$ to obtain $I, U, \chi$. We set $c= \tilde{c}\circ \chi$, so that \eqref{omito} follows from 
$g= \tilde{c}^{2}\tilde{g}$ on $U$. Property  (2)  of $t\mapsto h_{t}$ follow from Thm. \ref{th-omar},  property  (3)  of $c$ from  hypothesis (H3) and the fact that $\chi^{*}\hat{g}$ is equivalent to $dt^{2}+ h_{0}dy^{2}$ on $I\times \Sigma$, by Remark \ref{rem-th-omar}.  Finally \eqref{omiti} follows from hypothesis (M). \qed

The following proposition is a converse to Prop. \ref{propostup}.
\begin{proposition}\label{propostip}
 Let $(M, \tilde{g})$ be a globally hyperbolic space\-time with 
 \[
 M= \rr_{t}\times \Sigma_{y}, \ \tilde{g}= -dt^{2}+ h_{t}(y)dy^{2}, 
 \]
  such that $\Sigma$  is a Cauchy hypersurface in $(M, \tilde{g})$. Let $c\in \cinf(M)$, $c>0$ and $W\in \cinf(M; \rr)$.
    Assume that conditions (2), (3)  and identity \eqref{omito} in Prop. \ref{propostup} are satisfied   by   $\{h_{t}\}_{t\in I}, c$  for  some  bounded open interval $I$ and $\chi={\rm Id}$.  
  
  Then   for any  $J\Subset I$ conditions (H1), (H2), (H3) are satisfied for $g= c^{2}\tilde{g}$, $\tilde{c}= c$ and $U= J\times \Sigma$ . 
  If moreover $\tilde{V}\in \cinf(M; \rr)$ is such that \eqref{omiti} is satisfied for $\chi= {\rm Id}$, then there exist $V\in \cinf(M; \rr)$ such that $V= \tilde{V}$ on $J\times \Sigma$ and  $V$ satisfies condition (M).
\end{proposition}
\proof 
We extend  the maps $t\mapsto h_{t}$ and  $t\mapsto c(t, \cdot)$ from $I$ to $\rr$, in such a way that conditions (2) and (3)are satisfied with $I$ replaced by $\rr$, taking $h_{t}= h_{0}$, $c(t, \cdot)= c(0, \cdot)$ for $|t|$ large.  
As reference Riemannian metric on $M$ we take $\hat{g}= dt^{2}+ h_{t}(y)dy^{2}$. The fact that $(M, \hat{g})$ is of bounded geometry is easy. The remaining conditions in (H2), (H3) follow immediately from (2) and (3).  If \eqref{omiti} holds, we  can similarly  construct $V$ with $V= \tilde{V}$ on $I\times \Sigma$, $V=0$ for $|t|$ large such that $V$ satisfies (M). \qed
%We consider now  a more general situation,  essentially equivalent to Subsect. \ref{secp3.2} after a conformal transformation.  
%
%Let $(M, g)$ be a globally hyperbolic space\-time with a Cauchy hypersurface $\Sigma$. We assume that there exists  an open interval $I$ and an open neighborhood $U\subset M$ of $\Sigma$ such that $(U, g)$ is isometric to $(I\times \Sigma, -c^{2}(t,x)dt^{2}+  h_{t}dx^{2})$.
%
%

% We  also fix a real valued function $m\in \cinf(M, \rr)$ will play the role of a mass in the Klein-Gordon equation on $(M, g)$.
%
% 
%Setting $\tilde{h}_{t}\defeq  c^{-2}h_{t}$, 
%%so that $|\tilde{h}|^{\12}= c^{-d}|h|^{\12}$. 
%\def\tih{{\tilde h}}
%\def\cinfb{\cinf_{\rm b}}
%we assume  that:
%\ben
%\item $(\Sigma, \tih_{0})$ is of bounded geometry,
%\item $\tih\in \cinfb(]-T, T[, \BT^{0}_{2}(\Sigma, \tih_{0}))$, $\tih^{-1}\in \cinfb(]-T, T[, \BT^{2}_{0}(\Sigma, \tih_{0}))$,
%
%\item $\nabla_{i}\ln c\in \cinfb(]-T, T[, \BT^{0}_{1}(\Sigma, \tih_{0}))$, $\p_{t}\ln c\in \cinfb(]-T, T[, \BT^{0}_{0}(\Sigma, \tih_{0}))$.
%\item $mc^{2}\in \cinfb(]-T, T[, \BT^{0}_{0}(\Sigma, \tih_{0}))$.
%\een
\section{Examples}\init\label{sec-examples}\init
In this section we give several examples of space\-times to which the framework of Subsect. \ref{secp3.3} applies.

\subsection{Cosmological space\-times}
Let $(\Sigma, h)$ a Riemannian manifold, $\vara\in \cinf(\rr; \rr)$ and consider $M= \rr_{t}\times \Sigma_{y}$ with metric 
\[
g=-dt^{2}+ \vara^{2}(t)h_{ij}(y)dy^{i}dy^{j}.
\]
If $(\Sigma, h)$ is of bounded geometry, $(M, g)$ satisfies conditions  (H) for $\Sigma= \{t=0\}$, $\tilde{c}=1$, $U= I\times \Sigma$, $I\Subset \rr$ an interval. Condition (M) is satisfied in particular for $V= m^{2}$, $m\in\rr$. 
\begin{remark}
 The construction of  propagators  and Hadamard states for Klein-Gordon equations on cosmological space\-times can be  done without  the pseudodifferential calculus used  in Sects. \ref{secp2}, \ref{secp4} in the general case.  Instead one can rely on the functional calculus for $\epsilon= (-\Delta_{h})^{\12}$. All objects constructed in Sects. \ref{secp2}, \ref{secp4}, like the propagators $\cU_{A}^{\pm}(t,s)$ (see Subsect. \ref{secp2.4}) or the covariances $\lambda^{\pm}(t)$ (see Thm. \ref{thp4.1}) can be written as functions of $(t, s)$ and of the selfadjoint operator $\epsilon$. This amounts to what is known in the physics literature as the mode decomposition, see e.g. \cite{JS,Ol,BT,Av} for related results.  
% This is known in the physics litterature as the {\em mode decomposition}.
\end{remark}

\subsection{Kerr and Kerr-de Sitter exterior space\-times}\label{kdstoto}
\subsubsection{The Kerr-de Sitter family}
Let us recall  the family of Kerr-de Sitter metrics.  One sets  $M = \rr_{t}\times I_{r}\times\mathbb{S}^{2}_{\theta, \varphi}$, where $I$ is some open interval and $\theta\in [0, \pi]$, $\varphi\in \rr/2\pi \zz$ are the  spherical coordinates on $\mathbb{S}^{2}$.  The metric is given  in the coordinates $(t, r, \theta, \varphi)$ (Boyer-Lindquist coordinates) by:
\[
\bea
g&=\rho^{2}\left(\frac{dr^{2}}{\Delta_{r}}+ \frac{d\theta^{2}}{\Delta_{\theta}}\right)+ \frac{\Delta_{\theta}\sin^{2} \theta}{(1+ \alpha)^{2}\rho^{2}}\left(\vara dt^{2}- (r^{2}+\vara^{2})d\varphi\right)^{2}\\[2mm]
&\phantom{=}- \frac{\Delta_{r}}{(1+ \alpha)^{2}\rho^{2}}(dt- \vara\sin^{2}\theta d\varphi)^{2}\\[2mm]
&\eqdef g_{tt}dt^{2}+ g_{\varphi\varphi}d\varphi^{2}+ 2 g_{t\varphi}dtd\varphi+ g_{rr}dr^{2}+ g_{\theta\theta}d\theta^{2},
\eea
\]
for
\[
\bea
\Delta_{r}&= \left(1- \frac{\alpha}{\vara^{2}}r^{2}\right)(r^{2}+ \vara^{2})- 2Mr,\\[2mm]
\Delta_{\theta}&= 1 + \alpha \cos^{2}\theta,\
\rho^{2}= r^{2}+ \vara^{2}\cos^{2}\theta,\\[2mm]
\sigma^{2}&= (r^{2}+\vara^{2})^{2}\Delta_{\theta}- \vara^{2}\Delta_{r}\sin^{2}\theta.
\eea
\]
Here  $\alpha= \frac{\Lambda \vara^{2}}{3}$, $M, \vara, \Lambda>0$ are respectively the mass of the blackhole, its angular momentum and the cosmological constant. The Kerr metric corresponds to $\Lambda=0$.

If $\Lambda=0$ (Kerr) one assumes that $|\vara|<M$ (slow Kerr)  which implies that   for $r_{h}= M + \sqrt{M^{2}-\vara^{2}}$  one has: 
\[
r_{h}>0, \ \Delta_{r}(r_{h})=0, \ \Delta_{r}>0 \hbox{ on } ]r_{h},+\infty[
\]
and one takes $I= ]r_{h}, +\infty[$.
If $\Lambda\neq 0$ (Kerr-de Sitter)  one assumes that there exists $r_{h}<r_{c}$ such that 
\[
\begin{array}{rl}
i)&r_{h}>0, \, \Delta_{r}>0 \hbox{ on } ]r_{h}, r_{c}[,\,  \Delta_{r}(r_{h})=\Delta_{r}(r_{c})=0, \\[2mm]
ii)&  \p_{r}\Delta_{r}(r_{h})>0, \,\p_{r}\Delta_{r}(r_{c})<0,\\[2mm]
 iii)& \sup_{]r_{h}, r_{c}[}\Delta_{r}>\sup_{[0, \pi]}\Delta_{\theta},
\end{array}
\]
and one takes $I= ]r_{h}, r_{c}[$.
The set  $S$ of parameters $(\vara, M, \Lambda)$ such that such $r_{h}, r_{c}$ exist  is open and contains 
the set $\{|\vara|<M, \, \Lambda=0\}$ (slow Kerr) and $\{\vara=0, \, 9\Lambda M^{2}<1\}$ (Schwarzschild-de Sitter).

It is easy to check that if $(\vara, M,  \Lambda)\in S$ then there exists $c>0$ such that $\sigma^{2}(r, \theta)\geq c$ for all $\theta\in [0, \pi]$.

The  part of the boundary $r=r_{h}$ of $M$ is  the (outer) {\em black hole horizon}, the part $r= r_{c}$ in the Kerr-de Sitter case is the {\em cosmological horizon}.  Condition {\it iii)} means that  the region $\Delta_{r}>\Delta_{\theta}$ where $\frac{\p}{\p t}$ is time-like is not empty; one chooses the time orientation so that $\frac{\p}{\p t}$  is future oriented in this region. 
The space\-time $M$ is usually called the {\em outer  region} of the  Kerr or Kerr-de Sitter space\-time.
\begin{center}
 \includegraphics{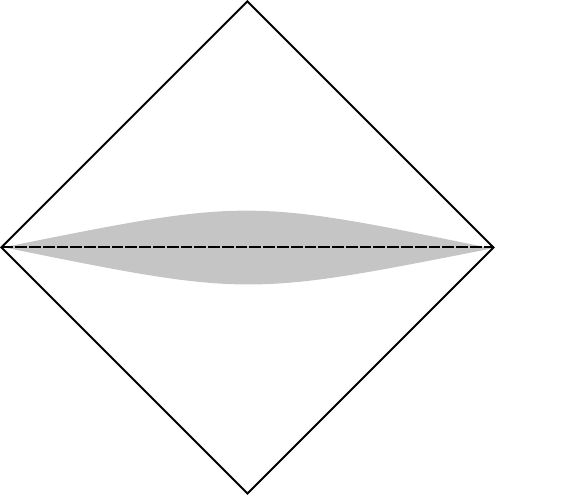}
\put(-95, 120){M}
\put(-90, 80){$U$}
\put(-40,75){$\Sigma$}
\put(-52,120){$+\infty$ or $r_{c}$}
\put(-132,120){$r_{h}$}
\put(-52,25){$+\infty$ or $r_{c}$}
\put(-130,25){$r_{h}$}
%\put(-170, 120){$+\infty$}
%\put(-170, 55){$+\infty$}
\\
Fig. 1 Kerr-de Sitter exterior region
\end{center}

% \begin{figure}[H]
%%\input{kerr-desitter.pdf_tex}
%\end{figure}
\subsubsection{Verification of conditions (H)}\label{sssecchange}
The first step consists in expressing  the metric in rotating coordinates.
We have: 
\[
 \bea
g&= (g_{tt}- g_{t\varphi}^{2}g_{\varphi\varphi}^{-1})dt^{2}+ g_{\varphi\varphi}(d\varphi+ g_{t\varphi}g_{\varphi\varphi}^{-1}dt)^{2}+ g_{rr}dr^{2}+ g_{\theta\theta}d\theta^{2}.
\eea
\]
We set $R= g_{t\varphi}g_{\varphi\varphi}^{-1}$, $\tilde{\varphi}= \varphi+ t R(r, \theta)$. Denoting again 
$\tilde{\varphi}$ by $\varphi$ we obtain:
\[
\bea
g&= (g_{tt}- g_{t\varphi}^{2}g_{\varphi\varphi}^{-1})dt^{2}+ g_{\varphi\varphi}(d\varphi -t \p_{r}R dr- t\p_{\theta }Rd\theta)^{2}+ g_{rr}dr^{2}+ g_{\theta\theta}d\theta^{2}.
\eea
\]
Then one introduces {\em Regge-Wheeler coordinates} on $I$, defining $s= s(r)$ by
 \[
 \frac{ds}{dr}= (1+ \alpha)\frac{r^{2}+ \vara^{2}}{\Delta_{r}}.
 \]
 (The integration constant is irrelevant).
The space\-time $M$ becomes $ \rr_{t}\times \rr_{s}\times\mathbb{S}^{2}_{\omega}$ and we choose the Cauchy hypersurface:
\[
\Sigma= M\cap \{t=0\}\sim \rr_{s}\times\mathbb{S}^{2}_{\omega}.
\]
 We set now:
\beq\label{defdetc}
\tilde{c}^{2}\defeq -g_{tt}+ g_{t\varphi}^{2}g_{\varphi\varphi}^{-1},
\eeq
and write 
\[
g= \tilde{c}^{2}\tilde{g}\hbox{ for }\tilde{g}= - dt^{2}+ h_{t}, \ h_{t}\hbox{ Riemannian metric on }\Sigma,
\]
with $ h_{t}\eqdef  h_{0}-2 t \hat{h}_{1}+ t^{2} \hat{h}_{2}$.
   \begin{proposition}\label{theokds}
 \ben
 \item $(\Sigma, h_{0})$ is of bounded geometry;
 
 \item for $J= [-\epsilon, \epsilon]$ and $\epsilon>0$ small enough  one has
 \[ J\ni t \mapsto h_{t}\in \cinfb(J; \BT^{0}_{2}(\Sigma, h_{0})), \ J\ni t \mapsto h^{-1}_{t}\in \cinfb(J; \BT^{2}_{0}(\Sigma, h_{0})),
   \]
\item One has
\[
\begin{array}{l}
\nabla_{h_{0}}\ln \tilde{c}\in \BT^{1}_{0}(\Sigma, h_{0}), \ \tilde{c}\in \BT^{0}_{0}(\Sigma, h_{0}). \end{array}
\]
 \een
\end{proposition}

\begin{remark}
 By Prop. \ref{propostip} we see that   conditions (H) are satisfied. Moreover  $V=m^{2}$ satisfies condition (M).
\end{remark}

Some  technical computations used in the proof of Prop. \ref{theokds} are collected in Subsect. \ref{subsecapp1}, where the reader can also find the definitions of the function classes $S^{p}_{\rm KdS}$ and $S^{m, p}_{\rm K}$, see Def. \ref{defkds2}.\medskip

\noindent \textbf{Proof of Prop. \ref{theokds}.}
A routine  computation gives:
 \beq\label{kds.e1}
 \tilde{c}^{2}=\frac{\Delta_{r}\Delta_{\theta}\rho^{2}}{(1+ \alpha)^{2}\sigma^{2}},\  g_{\varphi\varphi}= \frac{\sin^{2}\theta \sigma^{2}}{(1+ \alpha)^{2}\rho^{2}},\ 
 g_{rr}= \frac{\rho^{2}}{\Delta_{r}}, \ g_{\theta\theta}=\frac{\rho^{2}}{\Delta_{\theta}}.
 \eeq
 We set also:
\[
F(s)\defeq (1+ \alpha)^{2}\frac{(r^{2}+ \vara^{2})^{2}}{\Delta_{r}}, \ \ G(s, \theta)\defeq  \frac{\sigma^{2}}{(r^{2}+\vara^{2})^{2}\Delta_{\theta}},
\]  
and 
\[
d\omega^{2}= d \theta^{2}+ \frac{1+ \alpha \cos^{2}\theta}{1+\alpha}\sin^{2}\theta d\varphi^{2}.
\]
By Lemma \ref{limo} $d\omega^{2}$ is a smooth Riemannian metric on $\mathbb{S}^{2}$.
From the identity in Lemma \ref{limo} we  have
\[
\bea
h_{0}&=  \frac{\sigma^{2}}{(r^{2}+\vara^{2})^{2}\Delta_{\theta}}ds^{2}+ \frac{(1+ \alpha)^{2}\sigma^{2}}{\Delta_{r}\Delta_{\theta}\rho^{2}}(g_{\theta\theta}d\theta^{2}+ g_{\varphi\varphi}d\varphi^{2})\\[2mm]
&=G(s, \theta)\left(ds^{2}+ \frac{F(s)}{\Delta_{\theta}}d\omega^{2}+F(s)w\right),
\eea
\]
for
\[w=\left(\frac{\vara^{2}}{(1+ \alpha)\rho^{2}}+ \frac{2m\vara^{2}r}{(1+ \alpha)^{2}\rho^{4}}\right)(\sin^{2}\theta d\varphi)^{2}\in {\rm T}_{2}^{0}(\Sigma).
\]
 From Lemma \ref{kds.lem2} {\it v)} we obtain that $\inf F(s)>0$ and $|\p_{s}^{\alpha}F(s)|\leq C_{\alpha}F(s)$, hence  if $k_{0}= ds^{2}+ F(s)d\omega^{2}$,  $(\Sigma,k_{0})$ is of bounded geometry by Prop. \ref{warped}.  

Next we see  from Lemma \ref{kds.lem2} {\it vi)} that $G, G^{-1}\in \BT^{0}_{0}(\Sigma, k_{0})$ since $\inf G>0$ and $\p^{\alpha}_{s}(F(s)^{-\12}\p_{\omega})^{\beta}G$ is bounded on $\Sigma$ for any $(\alpha, \beta)\in \nn^{3}$.

The factor  in front of $(\sin^{2}\theta d\varphi)^{2}$ in $w$ belongs to $S^{0}_{\rm KdS}$ resp. to $S^{-2, 0}_{\rm K}$. 
The same argument as the one used for $G$, using  the estimates in Lemma \ref{kds.lem2} shows that $F(s)w\in \BT^{0}_{2}(\Sigma, k_{0})$.
  This implies that $h_{0}\in \BT^{0}_{2}(\Sigma, k_{0})$. Since $w\geq 0$ we
   immediately have that $h_{0}^{-1}\in \BT^{2}_{0}(\Sigma, k_{0})$, i.e. $h_{0}\sim k_{0}$, which proves (1).

To prove (2) we need to compute $ \hat{h}_{1}$ and $ \hat{h}_{2}$. We have:
\[
\bea
 \hat{h}_{1}&=  \tilde{c}^{-2}g_{\varphi\varphi}R_{r}drd\varphi+ \tilde{c}^{-2}g_{\varphi\varphi}R_{\theta}(\sin 2\theta d\theta )d\varphi\\[2mm]
&=\frac{\sigma^{4}}{\Delta_{r}\Delta_{\theta}\rho^{4}} R_{r}dr (\sin^{2}\theta d\varphi)+ \frac{\sigma^{4}}{\Delta_{r}\Delta_{\theta}\rho^{4}}R_{\theta}(\sin 2\theta d\theta)(\sin^{2}\theta d\varphi)\\[2mm]
&=\frac{\sigma^{4}}{\Delta_{\theta}\rho^{4}(1+ \alpha)(r^{2}+\vara^{2})} R_{r}ds (\sin^{2}\theta d\varphi)+ \frac{\sigma^{4}}{\Delta_{r}\Delta_{\theta}\rho^{4}}R_{\theta}(\sin 2\theta d\theta)(\sin^{2}\theta d\varphi)\\[2mm]
&\eqdef h_{1, s\varphi}ds (\sin^{2}\theta d\varphi)+ h_{1, \theta\varphi}(\sin 2\theta d\theta)(\sin^{2}\theta d\varphi).
\eea
\]
Similarly:
\[
\bea
 \hat{h}_{2}&= \tilde{c}^{-2}g_{\varphi\varphi}(R_{r})^{2}dr^{2}+ \tilde{c}^{-2}g_{\varphi\varphi}(R_{\theta})^{2}(\sin2 \theta d\theta)^{2} + 2  \tilde{c}^{-2}g_{\varphi\varphi}R_{r}R_{\theta}dr(\sin2\theta d\theta)\\[2mm]
&=\frac{\sigma^{4}}{\Delta_{r}\Delta_{\theta}\rho^{4}}\sin^{2}\theta (R_{r})^{2}dr^{2} +\frac{\sigma^{4}}{\Delta_{r}\Delta_{\theta}\rho^{4}}\sin^{2}\theta(R_{\theta})^{2}(\sin 2\theta d\theta)^{2}\\[2mm]
&\phantom{=}+ 2\frac{\sigma^{4}}{\Delta_{r}\Delta_{\theta}\rho^{4}}\sin^{2}\theta R_{r}R_{\theta}dr (\sin 2 \theta d\theta)
\\[2mm]
&=\frac{\sigma^{4}}{\Delta_{\theta}\rho^{4}(r^{2}+ \vara^{2})^{2}(1+ \alpha)^{2}}\sin^{2}\theta \Delta_{r}(R_{r})^{2}ds^{2}+ \frac{\sigma^{4}}{\Delta_{r}\Delta_{\theta}\rho^{4}}\sin^{2}\theta (R_{\theta})^{2}(\sin 2\theta d \theta)^{2}\\[2mm]
&\phantom{=}+ 2\frac{\sigma^{4}}{(1+ \alpha)(r^{2}+ \vara^{2})\Delta_{\theta}\rho^{4}}\sin^{2}\theta R_{r}R_{\theta}ds(\sin 2\theta d \theta)\\[2mm]
&\eqdef h_{2, ss}ds^{2}+ h_{2, \theta\theta}(\sin 2\theta d\theta)^{2}+ 2h_{2, s\theta}ds(\sin 2\theta d\theta).
\eea
\]
We now collect the properties of the coefficients of $ \hat{h}_{1}$, $ \hat{h}_{2}$. From \eqref{kds.e8} and estimates similar to those in Lemma \ref{kds.lem2} we obtain:
\[
\begin{array}{l}
h_{1, s\varphi}\in S^{0}_{\rm KdS},\hbox{ resp. }\in S^{-1, 0}_{\rm K}, \ h_{1, \theta\varphi}\in S^{0}_{\rm KdS},\hbox{ resp. }\in S^{0, 0}_{\rm K},\\[2mm]
h_{2, ss}\in S^{-1}_{\rm KdS},\hbox{ resp. }\in S^{-4, -1}_{\rm K},\ h_{2, \theta\theta}\in S^{-1}_{\rm KdS},\hbox{ resp. }\in S^{-2, -1}_{\rm K},\\[2mm]
 h_{2, s\theta}\in S^{-1}_{\rm KdS},\hbox{ resp. }\in S^{-3, -1}_{\rm K}.
\end{array}
\]
Since  $\sin 2\theta d\theta$ and $\sin^{2}\theta d\varphi$ are smooth forms on $\mathbb{S}^{2}$, this implies 
that $ \hat{h}_{i}\in \BT^{0}_{2}(\Sigma, h_{0})$, $i=1,2$.
If $J= [-\epsilon, \epsilon]$ for $\epsilon$ small enough we have hence
\[
J\ni t\mapsto h_{t}\in \cinf_{\rm b}(J, \BT^{0}_{2}(\Sigma, h_{0})), \ J\ni t\mapsto h^{-1}_{t}\in \cinf_{\rm b}(J, \BT^{2}_{0}(\Sigma, h_{0})),
\]
which proves (2).  

From \eqref{kds.e1} we obtain that $\tilde{c}^{2}\in S^{-1}_{\rm KdS}$, resp. $\in S^{0, -1}_{\rm K}$. 
This implies (3). \qed

\subsection{Kerr-Kruskal space\-time}\label{secp3.4}
In this subsection we consider {the maximal globally hyperbolic extension of the outer Kerr region considered in Subsect. \ref{kdstoto}. For the sake of brevity we call it the {\em Kerr-Kruskal extension}.}  In the slow Kerr case ($|\vara|<M, \Lambda=0$), $\Delta_{r}$ has  two  roots $0<r_{-}<r_{+}$, ($r_{+}$ was  previously denoted by $r_{h}$). The region $r>r_{+}$ of $\rr_{t}\times \rr_{r}\times \mathbb{S}^{2}_{\omega}$ considered earlier is called the (Boyer-Lindquist) block I, the region $r_{-}<r<r_{+}$ is called the block II. 

The construction of the Kerr-Kruskal extension of block I is as follows (see \cite[Chap. 2]{O2} for details):  a block II is glued  to the future of block I along $r= r_{+}, t>0$ using Kerr-star coordinates, and a block II', i.e. a block II with reversed time orientation, is glued to the past of block I along $r= r_{+}, t<0$ using star-Kerr coordinates.  Then a block I',  i.e. a block I with reversed time orientation, is glued to the past of block II and the future of block II'. The four blocks can be smoothly glued together at $r= t=0$ (the so-called {\em crossing sphere}), see \cite[Sect. 3.4]{O2}.  The time orientation of block I can be extended to a global time orientation,  and it can be shown that the resulting space\-time $(M^{\rm ext}, g)$ is globally hyperbolic, with $\Sigma^{\rm ext}=\{t=0\}$ as a Cauchy hypersurface.
\begin{center}
 \includegraphics{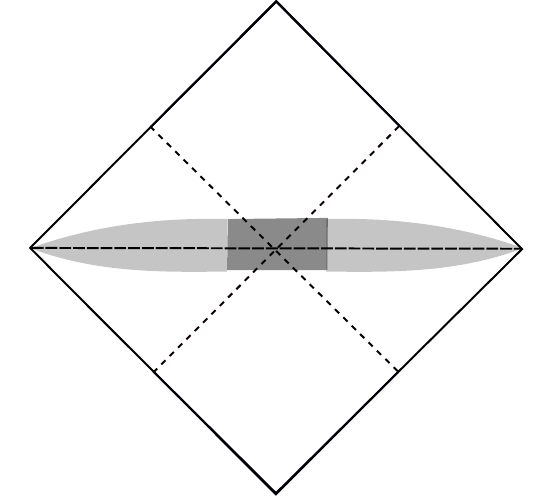}
\put(-50, 85){I}
\put(-87, 105){II}
\put(-125, 85){I'}
\put(-85, 75){$U_{0}$}
\put(-50, 75){$U$}
\put(-130, 75){$U'$}
\put(-87, 40){II'}
\put(-34,75){$\Sigma^{\rm ext}$}
\put(-30,100){$+\infty$}
\put(-30,45){$+\infty$}
\put(-70,95){$r_{+}$}
\put(-70,50){$r_{+}$}
\put(-97,50){$r_{+}$}
\put(-100,95){$r_{+}$}
\put(-52,120){$r_{-}$}
\put(-120,120){$r_{-}$}
\put(-50,20){$r_{-}$}
\put(-120,20){$r_{-}$}
\put(-140,100){$+\infty$}
\put(-145,45){$+\infty$}
%\put(-170, 120){$+\infty$}
%\put(-170, 55){$+\infty$}
\\
Fig. 2 Kerr-Kruskal extension
\end{center}

% \begin{figure}[H]
%\input{kerr-kruskal.pdf_tex}
%\caption{ Kerr-Kruskal extension}\label{fig-kerr-kruskal}
%%\caption{toto}
%%\label{fig2}

%\end{figure}

We claim that the Kerr-Kruskal extension $M^{\rm ext}$ satisfies the conditions (H).  In fact let $U$ be a neighborhood of $\Sigma$  in block I of the form $\{|t|<\epsilon, \ r>R\}$,  such that Prop. \ref{theokds} holds on $[-\epsilon, \epsilon]$, and let $U'$ be its copy in block I'. We also fix a relatively compact neighborhood  $U_{0}$  of the crossing sphere such that $V=U'\cup U_{0}\cup U$ is a neighborhood of $\Sigma^{\rm ext}$ in $M^{\rm ext}$.  It is clear that the hypotheses of Prop. \ref{propostip} are satisfied, since  they are satisfied over $U$ and $U'$, and  $U_{0}$ is relatively compact. 

\subsection{Double cones, wedges and lightcones  in Minkowski}
In this subsection we consider the Klein-Gordon operator $P= -\nabla^{a}\nabla_{a}+ m^{2}$ on  double cones, wedges and lightcones  in Minkowski space\-time.
\subsubsection{Double cones}
\begin{center}
 \includegraphics{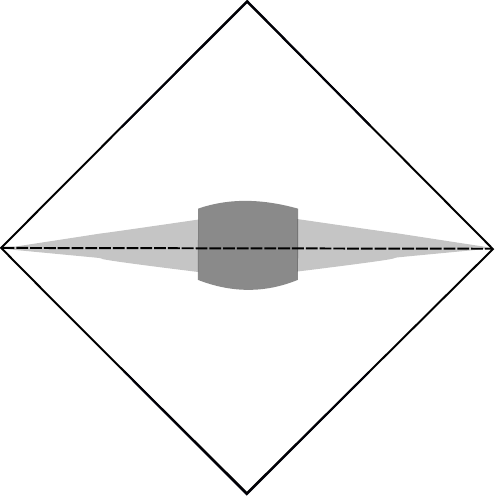}
 %%\put(-90, 90){$U_{0}$}
\put(-50, 75){$U$}
\put(-80, 75){$U_{0}$}
\put(-30,80){$\Sigma$}
 \\Fig. 3 The double cone
\end{center}
%\begin{figure}[h!]
%\input{doublecone.pdf_tex}
%\caption{The double cone}\label{fig-doublecone}

%\end{figure}

The standard double cone is
\[
M=\{(t, x)\in \rr^{1+d} : \ |t|<1-|x|\}, \ ds^{2}= -dt^{2}+ dx^{2}.
\]
We follow the framework of Subsect. \ref{secp3.3} with $\Sigma= M\cap\{t=0\}$, $V= m^{2}$. 
We set
\[
U= \{|t|< \delta(1-|x|), \ t^{2}+ (1- |x|)^{2}<\delta\}\hbox{ for }0<\delta\ll 1,
\]
and fix a relatively compact open set $U_{0}$ such that $U\cup U_{0}$ is a neighborhood of $\Sigma$, see Fig. 3.  It suffices to check conditions (H) over $ U$, since $U_{0}$ is relatively compact in $M$.  We  introduce polar coordinates $x= r \omega$ and set
 \[
r=1- \e^{-X}\cos T, \  \ t= \e^{-X}\sin T.
\]
 We are reduced to
 \[
\bea
U&=\,]-\alpha, \alpha[_{T}\times \,]C, +\infty[_{X}\times \bS_{\omega},\ \Sigma= \{T=0\}, \\[2mm]
ds^{2}&=\e^{-2X}\cos(2T)\left(-dT^{2}+  dX^{2}+ 2 \tan(2T)dTdX+ (\e^{X}- \cos T)^{2}d\omega^{2}\right)
 \eea
\]
 We take $\tilde{c}(T, X)= \e^{-X}\cos(2T)$ and choose the reference Riemannian metric 
 \[
 \hat{g}= dT^{2}+ dX^{2}+ \e^{2X}d\omega^{2},
 \]
 which is of  bounded geometry by Prop. \ref{warped}.  The Lorentzian metric
 \[
 \tilde{g}= -dT^{2}+  dX^{2}+ 2 \tan(2T)dTdX+ (\e^{X}- \cos T)^{2}d\omega^{2}
 \]
  is of bounded geometry for $\hat{g}$.
 Clearly $\Sigma=\{T=0\}$ is a bounded hypersurface of $(U, \hat{g})$. Its normal vector field for $\tilde{g}$ is $\p_{T}$, from  which it follows that $\Sigma$ is a Cauchy surface of bounded geometry, hence (H2) is satisfied.
One easily checks that  $\tilde{c}$  satisfies (H3) and that (M) is satisfied for $V= m^{2}$. 
  \subsubsection{Wedges}
  
\begin{center}
 \includegraphics{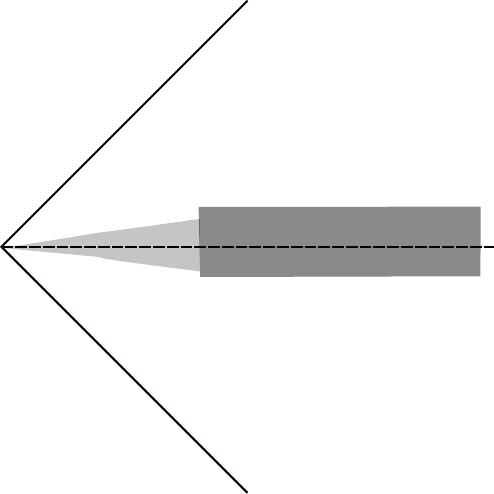}
  \put(-80, 80){$U_{\infty}$}
\put(-110, 80){$U_{0}$}
\put(-30,75){$\Sigma$}
 \\ Fig. 4 The standard wedge
\end{center}
 
%\begin{figure}[h!]
%\input{wedge.pdf_tex}
%
%\caption{The standard wedge}\label{fig-wedge}
%\end{figure}

The standard wedge is
 \[
M= \{(t, x_{1}, x')\in \rr^{1+d} : \ |t|<x_{1}\}, \ \  ds^{2}= -dt^{2}+ dx_{1}^{2}+ dx'^{2}.
\]
We take again $\Sigma= M\cap \{t=0\}$. We take: 
\[
U_{0}= \{|t|<\delta x_{1}, \  \ t^{2}+x^{2}_{1}<1\},\ U_{\infty}= \{|t|<\delta, \ 2<x_{1} \}. 
\]
We check hypotheses (H) over $U_{0}$ as above, replacing $1-r$ by $x_{1}$ and $\omega$ by $x'$. Hypotheses (H) over $U_{\infty}$ are immediate since $g$ is the Minkowski metric. Thus, (H) is satisfied over $U_{0}\cup U_{\infty}$.
Hypothesis (M) is again satisfied for $V= m^{2}$.

\subsubsection{Lightcones in Minkowski}
\begin{center}
 \includegraphics{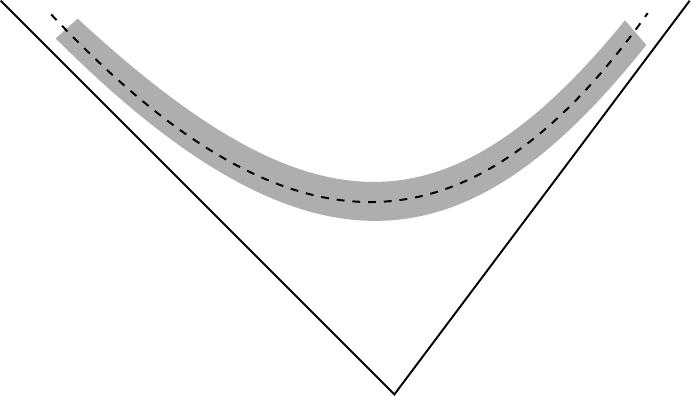}
 \put(-110, 50){$U$}
 \put(-80,60){$\Sigma$}\\
 Fig.5 The future lightcone
\end{center}
%\begin{figure}[h!]
%\input{lightcone.pdf_tex}
%\caption{The future lightcone}\label{fig-lightcone}
%
%\end{figure}

The future lightcone is
\[
M=\{(t, x)\in \rr^{1+d} : \ t>|x|\}, \ ds^{2}= -dt^{2}+ dx^{2}.
\]
We choose $\Sigma= M\cap \{t^{2}-x^{2}= 1\}$, use polar coordinates $x= r\omega$ and set
\[
r= \e^{-T}\sh X, \ t= \e^{-T}\ch X,
\]
so that 
\[
\bea
M &= \rr_{T}\times \rr_{X}\times \bS_{\omega},\ \Sigma= \{T= 0\}, \\[2mm]
ds^{2}&=\e^{- 2T}(-dT^{2}+ dX^{2}+ \sh^{2}X d\omega^{2}).
\eea
\]
We take $U= ]-\delta, \delta[_{T}\times \rr_{X}\times \bS_{\omega}$ as neighborhood of $\Sigma$. As before it suffices to check hypotheses (H) over $U\cap \{|X|>1\}$.  We take $\tilde{c}(T, X)= \e^{-T}$ and choose the reference Riemannian metric
\[
\hat{g}= dT^{2}+ dX^{2}+ \e^{2|X|}d\omega^{2}
\]
which is of bounded geometry by Prop. \ref{warped}.   Then $\tilde{g}= - dT^{2}+ dX^{2}+  \sh^{2}X d\omega^{2}$  and  hypotheses (H) are clearly satisfied, as is hypothesis (M) for $V= m^{2}$.

\section{Pseudodifferential  calculus on manifolds of bounded geometry}\label{secp1}\init
\subsection{Introduction}\label{secp1.1}

In this section we  recall the uniform pseudodifferential calculus on a manifold of bounded geometry, due to Kordyukov \cite{Ko} and Shubin \cite{Sh}.
This calculus generalizes for example the pseudodifferential calculus on a compact manifold and the uniform pseudodifferential calculus on $\rr^{n}$.
An important result for us  is the  generalization of {\em Seeley's theorem} \cite{Se}, originally proved on a  compact manifold. 

More precisely, if $A\in \Psi^{m}(M)$ is an elliptic pseudodifferential operator of order $m\geq 0$ on $M$,  symmetric and strictly positive on $\coinf(M)$, then $A$ has a unique selfadjoint extension, still denoted by $A$, with domain $H^{m}(M)$. Then  Seeley's theorem asserts that $A^{z}$ is a holomorphic family of pseudodifferential operators of order $m\Re z$. 

The extension of Seeley's theorem to  pseudodifferential operators on manifolds of bounded geometry is due to \cite{alnv1}, which we will closely follow. 

Another important result proved in this section is {\em Egorov's theorem}.  It is usually formulated as the fact that if $A$ is a pseudodifferential operator  and $\cU$ a {\em Fourier integral operator}  then $B= \cU^{-1}A\cU$ is again a pseudodifferential operator.
In our case we will take $\cU= \cU_{\epsilon}(t, s)$ equal to the evolution group generated by a smooth  time-dependent family $\epsilon(t)$ of elliptic first order $\Psi$DOs, with real principal symbol.

It will be convenient to consider also {\em time-dependent} pseudodifferential operators $A= A(t)\in \cinfb(I; \Psi^{m}(M))$ for $I\subset \rr$ an open interval.   It turns out that the  framework of \cite{alnv1}  is general enough to accommodate this extension without much additional work.

\subsection{Symbol classes}\label{secp1.2}
In this subsection we recall well-known definitions about symbol classes.
\subsubsection{Symbol classes on $\rr^{n}$}\label{secp1.2.1}
Let $U\subset \rr^{n}$ be  an open set, equipped with the flat metric $\delta$ on $\rr^{n}$.

 we denote by $S^{m}(T^{*}U)$, $m\in \rr$,  the space of $a\in \cinf(U\times \rr^{n})$ such that
\[
 \langle \xi\rangle^{-m+|\beta|}\p_{x}^{\alpha}\p_{\xi}^{\beta}a(x, \xi)\hbox{is  bounded on }U\times \rr^{n}, \ \forall \alpha, \beta\in \nn^{n}, 
\]
equipped with its canonical seminorms $\| \cdot\|_{m,\alpha, \beta}$. 

We set 
\[
S^{-\infty}(T^{*}U)\defeq  \bigcap_{m\in \rr}S^{m}(T^{*}U),\ \ S^{\infty}(T^{*}U)\defeq  \bigcup_{m\in \rr}S^{m}(T^{*}U),
\]
 with their canonical Fr\'echet space topologies.

If $m\in \rr$ and $a_{m-i}\in S^{m-i}(T^{*}U)$ we write
\[
a\simeq \sum_{i\in \nn}a_{m-i}
\]
if for each $p\in \nn$ 
\begin{equation}
\label{ep1.1}
r_{p}(a)\defeq  a- \sum_{i=0}^{p}a_{m-i}\in S^{m-p-1}(T^{*}U).
\end{equation}
It is well-known (see e.g. \cite[Sect. 3.3]{Sh}) that if $a_{m-i}\in S^{m-i}(T^{*}U)$, there exists $a\in S^{m}(T^{*}U)$, unique modulo $S^{-\infty}(T^{*}U)$ such that $a\simeq \sum_{i\in \nn}a_{m-i}$.

We denote by $S^{m}_{\rm h}(T^{*}U)\subset S^{m}(T^{*}U)$ the  space of $a$ such that $a(x, \lambda\xi)= \lambda^{m}a(x, \xi)$, for $x\in U$, $|\xi|\geq C$, $C>0$.

We denote by $S^{m}_{\rm ph}(T^{*}U)\subset S^{m}(T^{*}U)$ the space of $a$ such that $a\simeq \sum_{i\in \nn}a_{m-i}$ for a sequence $a_{m-i}\in S^{m-i}_{\rm h}(T^{*}U)$. 

Following \cite{alnv1}, we equip $S^{m}_{\rm ph}(T^{*}U)$ with  the topology defined by  the seminorms of $a_{m-i}$ in $S^{m-i}(T^{*}U)$ and $r_{p}(a)$ in $S^{m-p-1}(T^{*}U)$, (see \eqref{ep1.1}). This topology is strictly stronger than the topology induced by $S^{m}(T^{*}U)$.

The space  $S^{m}_{\rm ph}(T^{*}U)/S^{m-1}_{\rm ph}(T^{*}U)$ is isomorphic to $S^{m}_{\rm h}(T^{*}U)$, and the image of $a$ under the quotient map is called the {\em principal part} of $a$ and denoted by $a_{\rm pr}$. 

Finally we note that  if $U= B_{n}(0,1)$ (more generally if $U$ is relatively compact with smooth boundary), there exists a  continuous extension map $E: S^{m}(T^{*}U)\to S^{m}(T^{*}\rr^{n})$ such that $E a\traa{T^{*}U}=a$. Moreover $E$ maps $S^{m}_{\rm ph}(T^{*}U)$  into $S^{m}_{\rm ph}(T^{*}\rr^{n})$  and is continuous for the topologies of $S^{m}_{\rm ph}(T^{*}U)$ and $S^{m}_{\rm ph}(T^{*}\rr^{n})$, which means that all the maps
\[
a\mapsto (Ea)_{m-i}, \ a\mapsto r_{p}(Ea),
\]
are continuous.
\subsubsection{Time-dependent symbol classes on $\rr^{n}$}\label{timdep}
 let $I\subset \rr$ an open interval. We will also need to consider {\em time-dependent} symbols $a(t, x, \xi)\in \cinf(I\times T^{*}U)$. 
 
 The space $\cinf_{\rm b}(I; S^{m}(T^{*}U))$ is naturally defined as the space of $a\in  \cinf(I\times T^{*}U)$ such that
 \[
 \langle \xi\rangle^{-m+|\beta|}\p_{t}^{\gamma}\p_{x}^{\alpha}\p_{\xi}^{\beta}a(x, \xi)\hbox{is bounded on }I\times U\times \rr^{n}, \ \forall \alpha, \beta\in \nn^{n}, \ \gamma\in \nn,
\]
equipped with its canonical seminorms $\| \cdot\|_{m,\alpha, \beta, \gamma}$.  The  notation $a\sim \sum_i {a_{m-i}}$ and the subspaces $\cinfb(I; S^{m}_{\rm ph}(T^{*}U))$ are defined accordingly, by requiring uniform estimates on $I$ of all time derivatives.
\subsubsection{Symbol classes on $M$}\label{secp1.2.2}

\begin{definition}\label{defp.1}
 We denote by $S^{m}(T^{*}M)$ the space of $a\in \cinf(T^{*}M)$ such that for each $x\in M$,  $a_{x}\defeq (\psi_{x}^{-1})^{*}a\in S^{m}(T^{*}B_{n}(0,1))$ and the family $\{a_{x}\}_{x\in M}$ is bounded
 in $S^{m}(T^{*}B_{n}(0,1))$.  We equip  $S^{m}(T^{*}M)$ with the seminorms
 \[
\| a\|_{m, \alpha, \beta}= \sup_{x\in M}\| a_{x}\|_{m, \alpha, \beta}.
\]
Similarly we denote by $S^{m}_{\rm ph}(T^{*}M)$ the space of $a\in S^{m}(T^{*}M)$ such that for each $x\in M$, $a_{x}\in S^{m}_{\rm ph}(T^{*}B_{n}(0,1))$ and the family $\{a_{x}\}_{x\in M}$ is bounded
 in $S^{m}_{\rm ph}(T^{*}B_{n}(0,1))$. We equip $S^{m}_{\rm ph}(T^{*}M)$ with the seminorms
 \[
\| a\|_{m, i, p, \alpha, \beta}= \sup_{x\in M}\| a_{x}\|_{m,i, p,\alpha, \beta}.
\]
where $\| \cdot \|_{m,i, p,\alpha, \beta}$ are the seminorms defining the topology of $S^{m}_{\rm ph}(T^{*}B_{n}(0,1))$.

\end{definition}
It is easy to see that the definition of $S^{m}(T^{*}M)$, $S^{m}_{\rm ph}(T^{*}M)$ and their  Fr\'echet space topologies are  independent on the choice of the $\{U_{x}, \psi_{x}\}_{x\in M}$, with the above properties.

The notation $a\simeq \sum_{i\in \nn}a_{m-i}$ for $a_{m-i}\in S^{m-i}_{\rm ph}(T^{*}M)$ is defined as before.  If $a\in S^{m}_{\rm ph}(T^{*}M)$, we denote again by $a_{\rm pr}$ the image of $a$ in $S^{m}_{\rm ph}(T^{*}M)/S^{m-1}_{\rm ph}(T^{*}M)$.

If $I\subset\rr$ is an open interval, the spaces $\cinfb(I; S^{m}(T^{*}M))$ and $\cinfb(I; S^{m}_{\rm ph}(T^{*}M))$ are defined as in \ref{timdep}.

\subsection{Pseudodifferential operators}\label{secp1.3}
We now recall standard facts about the associated pseudodifferential operators, see \cite{Ko, Sh, alnv1}.

\subsubsection{Pseudodifferential operators on $\rr^{n}$}\label{secp1.3.2}
If $a\in S^{m}(T^{*}\rr^{n})$, we denote by $\Opw(a)$ its Weyl quantization, defined by
\[
 \Opw(a)u(x)= (2\pi)^{-n}\int \e^{\i (x-y)\cdot \xi}a\big(\textstyle\frac{x+y}{2}, \xi\big)u(y)dy d\xi.
\]
We recall the following well-known properties:

\ben
\item $\Opw(a): \coinf(\rr^{n})\to \cE'(\rr^{n})$ is continuous,
\item  $\Op: S^{m}(T^{*}\rr^{n})\to \bigcap_{s\in \rr}B(H^{s}(\rr^{n}), H^{s-m}(\rr^{n}))$ is continuous, where $H^{s}(\rr^{n})$ is the Sobolev space of order $s$ on $\rr^{n}$.
\item there exists a bilinear continuous map 
\[
 S^{\infty}(T^{*}\rr^{n})\times S^{\infty}(T^{*}\rr^{n})\ni (a,b)\mapsto a\sharp b\in S^{\infty}(T^{*}\rr^{n})
 \]
 such that $\Opw(a)\Opw(b)= \Opw(a\sharp b)$.
\een
\subsubsection{Time-dependent pseudodifferential operators on $\rr^{n}$}
If $I\subset \rr$ is an open interval and $a= a(t)\in\cinfb(I; S^{m}(T^{*}M))$ we can consider the time-dependent pseudodifferential operator $\Opw(a(t))$.   We have
\ben
\item $\Opw(a(t)): \cinfb(I; \coinf(\rr^{n}))\to \cinfb(I;\cE'(\rr^{n}))$ is continuous,
\item  $\Op: \cinfb(I; S^{m}(T^{*}\rr^{n}))\to \bigcap_{r,s\in \rr}B(H^{r}(I;H^{s}(\rr^{n})), H^{r}(I; H^{s-m}(\rr^{n})))$ is continuous, where $H^{r}(I;H^{s}(\rr^{n}))$ is the Sobolev space of bi-order $r,s$ on $I\times\rr^{n}$.
\een
\subsubsection{Quantization maps}\label{secp1.3.3}
We now recall the quantization procedure on a manifold of bounded geometry. 
Let $\{U_{i}, \psi_{i}\}_{i\in \nn}$ be a good chart covering of $M$   and  
\[
\sum_{i\in \nn}\chi_{i}^{2}= \one
\]
a subordinate good partition of unity, see 
 Subsect. \ref{secp0.1b}. Let
\[
(\psi_{i}^{-1})^{*}dg\eqdef  m_{i}dx,
\]
so that $\{m_{i}\}_{i\in \nn}$ is bounded in $\cinfb(B_{n}(0,1))$. 
 We set also: 
\[
\begin{array}{rl}
T_{i}:&L^{2}(U_{i}, dg)\to L^{2}(B_{n}(0,1), dx),\\[2mm]
&u\mapsto m_{i}^{\12}(\psi_{i}^{-1})^{*}u,
\end{array}
\]
so that $T_{i}:L^{2}(U_{i}, dg)\to L^{2}(B_{n}(0,1), dx)$ is unitary. 
\begin{definition}\label{def-de-op}
 Let $a= a(t)\in \cinfb(I; S^{m}(T^{*}M))$. We set
 \[
\Op(a)\defeq  \sum_{i\in \nn}\chi_{i}T_{i}^{*}\circ \Opw(Ea_{i})\circ T_{i}\chi_{i},
\]
where $a_{i}= a_{x_{i}}$ (see Def. \ref{defp.1}), and $E$ is the extension map (see Subsect. \ref{secp1.2}).
Clearly $\Op(a):  \cinfb(I;\coinf(M))\to  \cinfb(I;\cE'(M))$ is continuous.
\end{definition}
Such a map $\Op$ obtained from a good chart covering and partition of unity will be called a {\em good quantization map}.

Note that $\Op (1)= \one$, and that $\Op(a)^{*}(t)= \Op(\bar{a})(t)$ on $\coinf(M)$, where $A^{*}$ is the adjoint of $A$ for the scalar product
\[
(u|v)_M= \int_{M}\bar{u}v\,d\vol_{g}.
\]
Note that if $A(t)\in\cinfb(I;\Op(S^{\infty}(T^{*}M)))$, then its distributional kernel  $A(t, x, y)$ is supported in 
\[
\{(x, y)\in M\times M : \ d(x, y)\leq C\},
\]
for some $C>0$, where $d$ is the geodesic distance on $M$. It follows  that $\Op(a): \coinf(M)\to \coinf(M)$, hence $\Op(a)\circ \Op(b)$ is well defined. 
However because of the above support property $\Op(S^{\infty}(T^{*}M))$ is not stable under composition. 
To obtain an algebra of operators, it is necessary to add to $\Op(S^{\infty}(T^{*}M))$ an ideal of smoothing operators, which we introduce below.

The Sobolev spaces $H^{s}(M, g)$ defined in \ref{secp0.2.2}  will be simply denoted by $H^{s}(M)$. We will set:
\beq\label{defhinfty}
H^{\infty}(M)= \bigcap_{m\in \zz}H^{m}(M), \ H^{-\infty}(M)= \bigcup_{m\in \zz}H^{m}(M),
\eeq
equipped with their natural topologies. 
\begin{definition}
 We set:
 \[
\cW^{-\infty}(M)\defeq  \bigcap_{m\in \nn}B(H^{-m}(M), H^{m}(M)),
\]
equipped with its natural topology given by the seminorms 
\[
\|A\|_{m}=\|(- \Delta_{g}+1)^{m/2} A(- \Delta_{g}+1)^{m/2}\|_{B(L^{2}(M))}.
\]
Similarly we equip
\[
\cinfb(I; \cW^{-\infty}(M))
\]
with the topology given by the seminorms
\[
\|A\|_{m,p}= \sup_{t\in I, k\leq p}\|\p_{t}^{k}A(t)\|_{m}.
\]

\end{definition}
The following result, showing the independence modulo $\cinfb(I; W^{-\infty}(M))$ of $\Op(\cinfb(I; S^{\infty}(T^{*}M)))$  of the above choices of $\{U_{i}, \psi_{i}, \chi_{i}\}$, is easy to prove.
\begin{proposition}\label{pp1}
 Let $\Op'$ another  good quantization map. Then
 \[
\Op - \Op': \cinfb(I; S^{\infty}(T^{*}M))\to \cinfb(I; \cW^{-\infty}(M)).
\]
is continuous.
\end{proposition}
\subsubsection{The axioms of a Weyl algebra}\label{secp1.3.4}
In \cite{alnv1},  a  set of abstract axioms was introduced, with the aim of defining pseudodifferential operators on a manifold in a very general framework. The main result of \cite{alnv1}  is the extension of  Seeley's theorem  \cite{Se}. 
We will now check the abstract axioms of \cite[Sect. 1]{alnv1} in our situation. Namely, we need to specify a tuple $(\cup_{k\geq 1}\cW^{-\infty}_k, \cH, \varq, \sharp)$ that satisfies the following properties (we refer the reader to \cite[1.2]{alnv1} for the precise formulation in the general case):\medskip

{\em Axiom {\it (i)}: the LF-algebra and the Hilbert space:} One requires that $\cH$ is a Hilbert space and $\cW^{-\infty}=\cup_{k\geq 1}\cW^{-\infty}_k$ is a LF-algebra\footnote{This means that $\cW^{-\infty}$ is a strict inductive limit of Fr\'echet spaces and is endowed with an algebra structure with some additional grading and continuity properties, see \cite[1.2]{alnv1}.}, continuously embedded in $B(\cH)$ and such that the adjoint operation $^*$ maps $\cW^{-\infty}\to\cW^{-\infty}$ continuously. We choose $\cH= L^{2}(I; L^{2}(M, dg))$ and $\cW^{-\infty}= \cW^{-\infty}_{k}=  \cinfb(I; \cW^{-\infty}(M))$.  The LF-algebra properties are immediate. Furthermore, we have indeed $\cinfb(I; \cW^{-\infty}(M))\subset B(\cH)$ and 
 \[ \big(\cinfb(I; \cW^{-\infty}(M))\big)^{*}=  \cinfb(I; \cW^{-\infty}(M)).
 \]

{\em Axiom {\it (ii)}: existence of an injective, self-adjoint operator in $\cW^{-\infty}$:} We choose the (time-independent) operator $R= \e^{-(\Delta_{g}+1)}$. Clearly $R= R^{*}\in \cinfb(I;\cW^{-\infty}(M))$. 

%We have also
%\[
%\cinfb(I;\cW^{-\infty}(M))\cH=L^{2}(I; H^{-\infty}(M))= \bigcap_{m\in \nn}L^{2}(I; H^{m}(M)).
%\]
%

{\em Axiom {\it (iii)}: quantization map $\varq$:} We choose $\varq(a)\defeq  \Op(a)$. One needs to check in our case that 
\[
\begin{array}{l}
\Op(a): L^{2}(I; H^{-\infty}(M))\to L^{2}(I; H^{-\infty}(M)),\\[2mm]
\Op: \cinfb(I;S^{-\infty}(T^{*}M))\to \cinfb(I;\cW^{-\infty}(M)),
\end{array}
\]
which is straightforward from the properties of $\Op$ already listed.

{\em Axiom {\it (iv)}}: It is easy to check using for example  the norm given in \eqref{e3.8bis} that $\Op(a)\in \cinfb(I; B(H^{s}(M), H^{s-m}(M)))$ for $a\in \cinfb(I;S^{m}(T^{*}M))$. This implies that 
\[
\Op(\cinfb(I;S^{\infty}(T^{*}M)))\cinfb(I;\cW^{-\infty}(M))\subset \cinfb(I;\cW^{-\infty}(M)),
\]
which is in our setting the required property of the quantization map $\varq$. 

{\em Axiom {\it (v)}: existence of a symbolic calculus:} from the symbolic calculus in $\Opw(S^{m}(T^{*}\rr^{n}))$ we obtain the existence of a bilinear map \[
(a, b)\mapsto a\sharp b\hbox{ defined on } \cinfb(I;S^{\infty}(T^{*}M))
\] such that 
\[
\Op(a)\Op(b)- \Op(a\sharp b)\in \cinfb(I;\cW^{-\infty}(M)), \hbox{ for }a, b\in \cinfb(I; S^{\infty}(T^{*}M)).
\]
Concretely we have
\[
a\sharp b= \sum_{i\in \nn}\chi_{i}^{2}\psi_{i}^{*}(a_{i}\sharp b_{i}),
\] 
where $a_{i}\sharp b_{i}$ is recalled at the beginning of Subsect. \ref{secp1.2}.  The fact that $a\sharp b$ as an asymptotic expansion in terms of homogeneous bi-differential operators follows from the analogous property of the symbolic calculus on $\rr^{n}$.

{\em Axiom {\it (vi)}: boundedness of  $\Psi$DOs:} from the analogous property on $\rr^{n}$ we easily obtain that 
\[
\Op: \cinfb(I;S^{0}(T^{*}M))\to B(L^{2}(I;L^{2}(M)))\hbox{is continuous}.
\]

{\em Axiom {\it (vii)}:} One requires that the map 
\[
\cinfb(I; S^{m}(T^{*}M))\times \cinfb(I;\cW^{-\infty}(M))\ni (a, T)\mapsto \Op(a)\circ T\in \cinfb(I; \cW^{-\infty}(M))
\]
is continuous. This follows from axiom {\it {(vi)}} in our situation. \medskip

Two further important conditions are introduced in \cite{alnv1}. 

The first condition, called condition $(\sigma)$ in \cite{alnv1} amounts to the property that if $a\in \cinf(I;S^{m}_{\rm ph}(T^{*}M))$ and $\Op(a)\in \cinfb(I;\cW^{-\infty}(M))$, then $a$ belongs to $\cinfb(I; S^{-\infty}(T^{*}M))$.  In our case we deduce from the properties of the $\Psi$DO calculus on $\rr^{n}$ that  the sequence $\{a_{i}\}_{i\in \nn}$ is uniformly bounded in $\cinfb(I; S^{-\infty}(T^{*}B_{n}(0,1)))$, which implies that $a\in\cinfb(I; S^{-\infty}(T^{*}M))$.

The second condition, called condition $(\psi)$ in \cite{alnv1}, is the {\em spectral invariance} of the algebra $\one +\cinfb(I;\cW^{-\infty}(M))$. This condition  is stated and proved in the lemma below.
\begin{lemma}\label{lp.1}
 Let $R_{-\infty}\in \cinfb(I;\cW^{-\infty}(M))$ such that $\one - R_{-\infty}$ is invertible in $B(L^{2}(I;L^{2}(M)))$. Then 
 \[
(\one - R_{-\infty})^{-1}= \one -R_{1, -\infty}\hbox{ for }R_{1, -\infty}\in \cinfb(I;\cW^{-\infty}(M)).
\] \end{lemma}
\proof 
On $L^{2}(I; L^{2}(M))\sim \int^{\oplus}_{I}L^{2}(M)dt$ we have:
\[
\one - R_{-\infty}= \int^{\oplus}_{I}\one - R_{-\infty}(t)dt,
\]
hence
\[
(\one - R_{-\infty})^{-1}= \int^{\oplus}_{I}(\one - R_{-\infty}(t))^{-1}dt,
\]
and
\[
\bea
\|(\one - R_{-\infty})^{-1}\|_{B(L^{2}(I; L^{2}(M)))}&= {\rm ess\:  sup}_{t\in I}\|(\one - R_{-\infty}(t))^{-1}\|_{B(L^{2}(M))}\\[2mm]
&=\sup_{t\in I}\|(\one - R_{-\infty}(t))^{-1}\|_{B(L^{2}(M))},
\eea
\]
since 
\[
I\ni t\mapsto (\one - R_{-\infty}(t))^{-1}\in B(L^{2}(M))
\]
is norm continuous. We have
\beq\label{derivo}
(\one- R_{-\infty}(t))^{-1}= \one + R_{-\infty}(t) + R_{-\infty}(t)(\one -R_{-\infty}(t))^{-1}R_{-\infty}(t).
\eeq
Since $(\one -R_{-\infty}(t))^{-1}\in B(L^{2}(M))$ and $R_{-\infty}(t)\in \cW^{-\infty}(M)$, we see that $R_{-\infty}(t)(\one -R_{-\infty}(t))^{-1}R_{-\infty}(t)\in \cW^{-\infty}(M)$. To prove that $R_{1, -\infty}\in \cinfb(I; \cW^{-\infty}(M))$ we differentiate \eqref{derivo} w.r.t. $t$ using the Leibniz rule and the identity 
\[
\p_{t}(\one- R_{-\infty}(t))^{-1}=  (\one- R_{-\infty}(t))^{-1}\p_{t}R_{-\infty}(t)(\one- R_{-\infty}(t))^{-1}. \ \Box
\]

\subsubsection{Time-dependent pseudodifferential operators on $M$}\label{secp1.3.5}
We can now define classes of  time-dependent pseudodifferential operators on $M$, by applying the abstract framework of  \cite[Sect. 1]{alnv1}. 
We will only consider {\em classical} pseudodifferential operators, i.e. operators obtained from poly-homogeneous symbols. 
\begin{definition}
 We set for $m\in \rr$:
 \[
\cinfb(I; \Psi^{m}(M))\defeq  \Op(\cinfb(I;S^{m}_{\rm ph}(T^{*}M)))+ \cinfb(I;\cW^{-\infty}(M)).
\]
\end{definition}
\begin{remark}\label{remnot}
 An element of $\cinfb(I; \Psi^{m}(M))$ will usually be denoted by $A$,  while $A(t)$ for $t\in I$ will be an element of $\Psi^{m}(M)$. Writing for example 
$L^{2}(I; L^{2}(M))$ as $\int^{\oplus}_{I}L^{2}(M)dt$, we have 
\[
A= \int_{I}^{\oplus}A(t)dt.
\]
 
\end{remark}

Note that $\cinfb(I;\Psi^{-\infty}(M))= \cinfb(I;\cW^{-\infty}(M))$. If necessary we equip the space $\cinfb(I;\Psi^{m}(M))$ with the quotient topology obtained from the map \[
\cinfb(I;S^{m}_{\rm ph}(T^{*}M))\times\cinfb(I;\cW^{-\infty}(M))\in (a, R)\mapsto \Op(a)+R\in \cinfb(I;\Psi^{m}(M)).
\]
 It follows that the injection:
\[
\cinfb(I;\Psi^{m}(M))\rightarrow \bigcap_{s\in \rr}\cinfb(I;B(H^{s}(M), H^{s-m}(M)))
\]
is continuous.
\def\glurg{\cinfb(I;}
\begin{definition}
 Let $A= \Op(a)+R_{-\infty}\in \cinfb(I;\Psi^{m}(M))$.  We denote by $\sigma_{\rm pr}(A)\in \cinfb(I;S^{m}_{\rm h}(T^{*}M))$ the {\em principal symbol} of $A$ defined as 
 \[
 \sigma_{\rm pr}(A)\defeq [a]\in\cinfb(I; S^{m}_{\rm ph}(T^{*}M))/\cinfb(I;S^{m-1}_{\rm ph}(T^{*}M)).
 \]  By property $(\sigma)$ and Prop. \ref{pp1} $\sigma_{\rm pr}(A)$ is independent on the decomposition of $A$ as $\Op(a)+R_{-\infty}$ and on the choice of the good quantization map $\Op$.
\end{definition}
\begin{definition}\label{defp1}
 $A\in \cinfb(I;\Psi^{m}(M))$ is {\em elliptic} if  there exists $C>0$ such that
 \[
|\sigma_{\rm pr}(A)(t, x, \xi)|\geq C(\xi \cdot g^{-1}(x)\xi)^{m/2}, \ t\in I, (x, \xi)\in T^{*}M.
\]
\end{definition}
The main property of elliptic operators is that they admit {\em parametrices}, i.e. inverses modulo $\glurg\cW^{-\infty}(M))$. 
\begin{proposition}\label{pp2b}
 Let $A\in \cinfb(I;\Psi^{m}(M))$  be elliptic. Then there exists $B\in \glurg\Psi^{-m}(M))$, unique modulo $\glurg\cW^{-\infty}(M))$ such that 
 \[
 AB-\one\in \glurg\cW^{-\infty}(M)), \ \ BA-\one\in \glurg(\cW^{-\infty}(M)).
 \]
 Such an operator $B$ is called a {\em parametrix} of $A$ and denoted by $A^{(-1)}$.
\end{proposition}
\proof  The proof given in \cite[Thm. 3.3]{Ko} or  \cite[Prop. 3.4]{Sh2} extends immediately to the time-dependent situation. \qed

We recall that  the notation $a\sim b$ for $a,b$ are two selfadjoint operators on  a Hilbert space $\cH$ is defined in Subsect. \ref{sec0.not}.

\begin{proposition}\label{pp2}
 Let $A\in \glurg\Psi^{m}(M))$, $m\geq 0$ be elliptic  such that $A(t)$  is symmetric on $H^{\infty}(M)$ for all $t\in I$. Then
 \ben
 \item $A(t)$ is essentially selfadjoint on $H^{\infty}(M)$ and 
 \[
\Dom A^{\rm cl}(t)= H^{m}(M).
\]
\item If in addition $\sigma_{\rm pr}(A)(t, x, \xi)\geq c(\xi \cdot g^{-1}(x)\xi)^{m/2}$ for some $c>0$, then $A^{\rm cl}(t)$ is bounded below, uniformly for $t\in I$. Moreover there exists $R\in \glurg(\cW^{-\infty}(M))$ such that 
\[
A(t)+ R_{-\infty}(t)\sim (-\Delta_{g}+1)^{m/2}, \hbox{ uniformly for }t\in I.
\]
\item $A$ (considered as a linear  operator on $L^{2}(I; L^{2}(M))$) is essentially selfadjoint on $L^{2}(I; H^{\infty}(M))$ and
\[
\Dom A^{\rm cl}=L^{2}(I; H^{m}(M)). 
\]
\een
\end{proposition}
\proof  statement (1) follows from \cite[Prop. 2.2]{alnv1} and the alternative characterization of Sobolev spaces given in \cite[Sect.3]{alnv1}. To prove (2) we may assume that $A= \Op(a)$ since $\cW^{-\infty}(M)\subset B(L^{2}(M))$. Then
\[
A(t)= \sum_{i\in \nn}\chi_{i}T_{i}^{*}A_{i}(t)T_{i}\chi_{i},
\]
where $\{A_{i}\}_{i\in \nn}$ is a bounded family in  $\Opw(\glurg S^{m}(T^{*}\rr^{n})))$ 
such that 
\[
\sigma_{\rm pr}(A_{i})(t, x, \xi)\geq c|\xi|^{m}, \hbox{ uniformly  for } i\in\nn, \ t\in I.
\] 
From the $\Psi$DO calculus on $\rr^{n}$ we deduce that $A_{i}(t)\geq c'\one$ uniformly in $i\in \nn$ and  which shows that $A(t)$ is bounded below uniformly in $t\in I$.
This also implies that  for $c\gg 1$  one has $A(t)+ c\sim (-\Delta_{g}+1)^{m/2}$. By functional calculus we can find $\chi\in \coinf(\rr)$ such that  $A(t)+ \chi(A(t))\sim A(t)+c$. By elliptic regularity we know that $\chi(A)\in\glurg \cW^{-\infty}(M))$, which completes the proof of (2).  (3) follows from (1). \qed

We  now state the main result of this subsection, which follows directly from   \cite{alnv1}, for the simpler case of real powers.

\begin{theoreme}\label{thp2.1}
Let $A\in \glurg \Psi^{m}(M))$ be elliptic, selfadjoint  with $A(t)\geq c\one$ for $c>0$, $t\in I$. Then $A^{s}\in \glurg\Psi^{ms}(M))$ for any $s\in \rr$ and
\[
\sigma_{\rm pr}(A^{s})(t)= \sigma_{\rm pr}(A(t))^{s}.
\]
\end{theoreme}
\proof We consider $A$ as a selfadjoint operator on $L^{2}(I; L^{2}(M))$ and apply  \cite[Thm. 8.9]{alnv1}, noting that $A^{s}(t)= A(t)^{s}$. \qed

The following lemma will be used in Subsect. \ref{secp4.3}.

\begin{lemma}\label{lemmatoche}
 Let $A\in \Psi^{\infty}(M)$ such that $A: \cE'(M)\to \cinf(M)$. Then $A\in \cW^{-\infty}(M)$.
\end{lemma}
\proof We can assume that $A= \Op(a)$ for $a\in S^{m}_{\rm ph}(T^{*}M)$, i.e (see Def. \ref{def-de-op}):
\[
A= \sum_{i\in \nn}\chi_{i}T_{i}^{*}\circ \Opw(Ea_{i})\circ T_{i}\chi_{i},
\]
where $\{Ea_{i}\}_{i\in\nn}$ is  bounded in $S^{m}_{\rm ph}(T^{*}\rr^{n})$.  We can fix  cutoff functions $\tilde{\chi}_{i}$ such that 
$T_{i}\chi_{i}= \tilde{\chi}_{i}T_{i}\chi_{i}$, $\{\tilde{\chi}_{i}\}_{i\in \nn}$ is bounded in $\coinf(B(0,1))$ and define $b_{i}$ by 
$\tilde{\chi}_{i}\circ \Opw(Ea_{i})\circ \tilde{\chi}_{i}= \Opw(b_{i})$.  The family $\{b_{i}\}_{i\in \nn}$ is bounded in $S^{m}_{\rm ph}(T^{*}\rr^{n})$ hence
for each $p\in \nn$ one has:
\[
b_{i}= \sum_{k=0}^{p}b_{i, m-k}+ r_{i, p}, 
\]
where $\{b_{i, m-k}\}_{i\in \nn}$, resp.  $\{r_{i, p}\}_{i\in \nn}$ is  bounded  in $S^{m-k}_{\rm h}(T^{*}\rr^{n})$ resp.  $S^{m-p-1}(T^{*}\rr^{n})$.
Since $A: \cE'(M)\to \cinf(M)$ it follows that $\Opw(b_{i}): L^{2}(\rr^{n})\to H^{-m+k}(\rr^{n})$ for any $k\in \nn$.  Taking $k=1$ we obtain that $\Opw(b_{i,m}): L^{2}(\rr^{n})\to H^{-m+1}(\rr^{n})$, hence $b_{i,m}=0$ since $b_{i,m}$ is homogeneous of degree $m$.  Iterating this argument we obtain that $b_{i}= r_{i,p}$ hence $\{\Opw(b_{i})\}_{i\in \nn}$ is bounded in $B(H^{s}(\rr^{n}), H^{s-m+p}(\rr^{n}))$.  But this implies that $A\in B(H^{s}(M), H^{s-m+p}(M))$, using the characterization of Sobolev spaces in \ref{secp0.2.2}. Since $p$ is arbitrary we have $A\in \cW^{-\infty}(M)$.\qed

\subsection{Egorov's theorem}\label{secp1.5}
Let us consider an operator $\epsilon(t)= \epsilon_{1}(t)+ \epsilon_{0}(t)$, such that:
\[
(\E) \ \begin{array}{l}
\epsilon_{i}(t)\in \cinfb(I; \Psi^{i}(M)), \ i=0,1,\\[2mm]
\epsilon_{1}(t) \hbox{ is elliptic, symmetric and  bounded from below on }H^{\infty}(M).
\end{array}
\]
(see Def. \ref{defp1}).
By Prop. \ref{pp2} we know that $\epsilon_{1}(t)$ with domain $\Dom \epsilon(t)= H^{1}(M)$ is selfadjoint, hence $\epsilon(t)$ with the same domain is closed, with non empty resolvent set. We denote by $\cU_{\epsilon}(t,s)$ the associated  propagator defined by:

\[
\bec
\frac{\p}{\p t}\cU_{\epsilon}(t,s)= \i  \epsilon(t)\cU_{\epsilon}(t,s),\ t,s\in I,\\[2mm]
\frac{\p}{\p s}\cU_{\epsilon}(t,s)= -\i \cU_{\epsilon}(t,s)\epsilon(s),\ t,s\in I,\\[2mm]
\cU_{\epsilon}(s,s)=\one, \ s\in I.
\eec
\]
Note that   the propagator $\cU_{\epsilon_{1}}(t,s)$ exists  and is unitary on $L^{2}(M)$, by e.g. \cite[Thm. X.70]{RS2}. Since $\epsilon(t)- \epsilon_{1}(t)$ is uniformly bounded in $B(L^{2}(M))$, one easily deduces  the existence of $\cU_{\epsilon}(t,s)$, which is strongly continuous in $(t,s)\in I^{2}$ with values in $B(L^{2}(M))$, uniformly bounded on $I^{2}$ in $B(L^{2}(M))$.

\begin{lemma}\label{ego.1}
Assume $(\E)$. Then
\ben
\item 
 $\cU_{\epsilon}(t,s)\in B(H^{m}(M))$ for $m\in \zz\cup\{\pm\infty\}$, $I^{2}\ni (t,s)\mapsto \cU_{\epsilon}(t,s)$ is strongly continuous on $H^{m}(M)$,  
 \item  if $r_{-\infty}\in \cW^{-\infty}(M)$ then
$\cU_{\epsilon}(t,s)r_{-\infty}, \ r_{-\infty}\cU_{\epsilon}(t, s)\in \cinfb(I^{2}_{t,s}, \cW^{-\infty}(M))$.

% \item if moreover $b(t)\in S^{0}(\rr, \Psi^{1}(M))$ and $b(t)- b^{*}(t)\in S^{-1- \delta}(\rr, \Psi^{0}(M))$ for $\delta>0$ then 
% $U_{b}(t,s)$ is uniformly bounded in $B(L^{2}(M))$.
 \een
\end{lemma}
\proof Note that  (2) follows from (1). If clearly suffices to prove (1) for $m$ finite. We set $a= (-\Delta_{g}+\one)^{\12}$ and compute 
\[
\bea
&\p_{t}\big(\cU_{\epsilon}(s,t)a^{m}\cU_{\epsilon}(t,s)a^{-m}\big)\\[2mm]
&= -\i \cU_{\epsilon}(s,t)[\epsilon(t), a^{m}]\cU_{\epsilon}(t,s)a^{-m}\\[2mm]
&= -\i \cU_{\epsilon}(s,t)\times [\epsilon(t), a^{m}]a^{-m}\times a^{m}\cU_{\epsilon}(t,s)a^{-m}.
\eea
\]
We know that $a^{m}\in \Psi^{m}(M)$, hence $[\epsilon(t), a^{m}]a^{-m}\in \cinf(I; \Psi^{0}(M))$. Moreover $\cU_{\epsilon}(t,s)$ is locally  bounded in 
$B(L^{2}(M))$ on $I^{2}$. Therefore
\[
\p_{t}\| \cU_{\epsilon}(s,t)a^{m}\cU_{\epsilon}(t,s)a^{-m}u\| \leq C\|\cU_{\epsilon}(s,t)a^{m}\cU_{\epsilon}(t,s)a^{-m}u\|,\ (t,s)\in I^{2}, \ u \in L^{2}(M).
\]
  For $m<0$, taking $u\in \Dom a^{-m}$ and using Gronwall's inequality yields (1). For $m>0$ we argue similarly, replacing the unbounded operator $a^{m}$ by $a_{\delta}^{m}= a^{m}(\one + \i \delta a)^{-m}$ for $\delta>0$. We obtain  from Gronwall's inequality that:
\[
\| \cU_{\epsilon}(s,t)a_{\delta}^{m}\cU_{\epsilon}(t,s)a^{-m}\| \leq C\| a_{\delta}^{m}a^{-m}\|, \ (t,s)\in I^{2}.
\]
 We conclude the proof by using that $\| a^{m}u\|= \sup_{0<\delta}\| a_{\delta}^{m}u\|$.\qed

The following theorem is a version of {\em Egorov's theorem}.
\begin{theoreme}\label{pth1}
Let $a\in \Psi^{m}(M)$ and $\epsilon(t)$ satisfying $(E)$. Then
\[
a(t,s)\defeq  \cU_{\epsilon}(t,s)a \cU_{\epsilon}(s,t)\in \cinfb(I^{2}, \Psi^{m}(M)).
\]
Moreover 
\[
\sigma_{\rm pr}(a)(t,s)= \sigma_{\rm pr}(a)\circ \Phi(s,t),
\]
where $\Phi(t,s): T^{*}M\to T^{*}M$ is the flow of the time-dependent Hamiltonian $\sigma_{\rm pr}(\epsilon)(t)$.
\end{theoreme}
\proof 
The proof  consists of several steps.

{\it Step 1}: we write $a=\Op(c)+ a_{-\infty}$, $c\in S_{\rm ph}^{m}(T^{*}M)$, $a_{-\infty}\in \cW^{-\infty}$. Then
\[
\bea
&\cU_{\epsilon}(t,s)a \cU_{\epsilon}(s,t)= \cU_{\epsilon}(t,s)\Op(c)\cU_{\epsilon}(s,t)+ \cU_{\epsilon}(t,s)a_{-\infty}\cU_{\epsilon}(s,t)\\[2mm]
&= \cU_{\epsilon}(t,s)\Op(c)\cU_{\epsilon}(s,t)+ \cinfb(I^{2}, \cW^{-\infty}(M)),
\eea
\]
by Lemma \ref{ego.1}. Therefore we can assume that $a= \Op(c)$.

{\it Step 2}:  we write $\epsilon(t)= \Op(b)(t)+ \epsilon_{-\infty}(t)$ for  $b(t)\in \cinfb(I; S^{1}_{\rm ph}(T^{*}M))$, $\epsilon_{-\infty}(t)\in \cinfb(I^{2},\cW^{-\infty}(M))$. 

 We write:
 \[
\cU_{\epsilon}(t,s)\eqdef  \cU_{\Op(b)}(t,s)\cV(t,s),
\]
where
\[
\bec
\p_{t}\cV(t,s)= -\i \cU_{\Op(b)}(s,t)\epsilon_{-\infty}(t)\cU_{\epsilon}(t,s)\eqdef  \tilde{\epsilon}_{-\infty}(t,s),\\[2mm]
\cV(s,s)=\one.
\eec
\]
By Lemma \ref{ego.1} we know that $ \tilde{\epsilon}_{-\infty}(t,s)\in \cinfb(I^{2}, \cW^{-\infty}(M))$, hence
\[
\cV(t,s)= \one +  \cinfb(I^{2}, \cW^{-\infty}(M)).
\]
It follows that:
\[
\bea
&\cU_{\epsilon}(t,s)\Op(c)\cU_{\epsilon}(s,t)= \cU_{\Op(b)}(t,s)\cV(t,s)\Op(c)\cV(s,t)\cU_{\Op(b)}(s,t)\\[2mm]
&=\cU_{\Op(b)}(t,s) \Op(c)\cU_{\Op(b)}(s,t)+ \cinfb(I^{2}, \cW^{-\infty}(M)),
\eea
\]
again by Lemma \ref{ego.1}. Therefore it suffices to consider  
\[
a_{1}(t,s)\defeq  \cU_{\Op(b)}(t,s) \Op(c)\cU_{\Op(b)}(s,t).
\]
{\it Step 3}: 
We  try to construct $d(t,s)\in \cinfb(I^{2}, S^{m}_{\rm ph}(T^{*}M))$ such that
\beq\label{ep1.6}
\bec
\p_{t}\Op(d)(t,s)= - [\Op(b)(t),\i \Op(d)(t,s)], \   t, s\in I,\\[2mm]
\Op(d)(s,s)= \Op(c), \ s\in I,
\eec
\eeq
modulo error terms in $\cW^{-\infty}(M)$.
As in \cite[Sec. 0.9]{taylor}, we  write 
\[
\begin{array}{l}
c\simeq \sum_{i\in \nn}c_{m-i}, \ c_{m-i}\in S^{m-i}_{\rm h}(T^{*}M),\\[2mm]
b(t)\simeq \sum_{i\in \nn}b_{1-i}(t), \ b_{1}(t)= \sigma_{\rm pr}(\epsilon)(t), \ b_{1-i}(t)\in \cinf(I; S^{1-i}_{\rm ph}(T^{*}M)),
\end{array}
\]
and solve \eqref{ep1.6} with the ansatz 
\[
d(t,s)\simeq\sum_{i\in \nn}d_{m-i}(t,s), \ d_{m-i}\in \cinf(I; S^{m-i}_{\rm h}(T^{*}M)).
\]
 We obtain the sequence of transport equations:
\[
\bea
(E0) \ &\begin{cases}
\p_{t}d_{m}(t,s)+\{\sigma_{\rm pr}(\epsilon(t)),
 d_{m}(t,s)\}=0, \\ d_{m}(s,s)= c_{m}, 
\end{cases} \\[2mm]
(Ei) \ &\begin{cases}
\p_{t}d_{m-i}(t,s)+ \{\sigma_{\rm pr}(\epsilon(t)), d_{m}(t,s)\}= \sum_{-j+ m-k+ 1-l= m-i}P_{j}(d_{m-k}, b_{1-l})(t,s), \\ d_{m-i}(s,s)=0,
\end{cases}
\eea
\]
where  $\{\cdot, \cdot\}$ is the Poisson bracket and 
\[
P_{j}: S^{p_{1}}_{\rm h}(T^{*}M)\times  S^{p_{2}}_{\rm h}(T^{*}M)\to S^{p_{1}+ p_{2}-j}_{\rm h}(T^{*}M)
\]
is a  bi-differential operator homogeneous of degree $j$ (see \cite[Sect. 1.1]{alnv1}).

This sequence of transport equations can be solved inductively in 
\[
\cinfb(I; S^{m-i}_{\rm h}(T^{*}M)),
\]
using that $\sigma_{\rm pr}(\epsilon(t))$ is real-valued and elliptic.  We have in particular:
\beq\label{ep1.7}
d_{m}(t,s)= c_{m}\circ \Phi(s,t).
\eeq
 Now we choose $d(t,s)\simeq \sum_{i\in \nn}d_{m-i}(t,s)$ and obtain:
\[
\bec
\p_{t}\Op(d)(t,s)= - [\Op(b)(t),\i \Op(d)(t,s))]+ \cinfb(I^{2}, \cW^{-\infty}(M))\\[2mm]
\Op(d)(s,s)= \Op(c)+ \cinfb(I; \cW^{-\infty}(M)).
\eec
\]
It follows that
\[
\begin{array}{l}
\p_{t}\left(\cU_{\Op(b)}(s,t)\Op(d)(t,s)\cU_{\Op(b)}(t,s)\right)\\[2mm]
= \cU_{\Op(b)}(s,t)\left(\p_{t}\Op(d)(t,s)- \i [\Op(b)(t),\Op(d)(t,s)]\right)\cU_{\Op(b)}(t,s)\\[2mm]
\in \cinfb(I^{2}, \cW^{-\infty}(M)),\\[2mm]
\cU_{\Op(b)}(s,s)\Op(d)(s,s)\cU_{\Op(b)}(s,s)= \Op(c)+\cW^{-\infty}(M).
\end{array}
\]
Hence by integrating from $s$ to $t$ and using again Lemma \ref{ego.1}:
\[
\Op(d)(t,s)= a_{1}(t,s)+ \cinfb(I^{2}, \cW^{-\infty}(M)).
\]
Hence $a_{1}(t,s)\in \cinf(I^{2}, \Psi^{m}(M))$ as claimed. By \eqref{ep1.7} we have:
\[
\sigma_{\rm pr}(a_{1}(t,s))= \sigma_{\rm pr}(a)\circ \Phi(s,t).
\]
The proof is complete.
\qed
\subsection{The wave front set}\label{secp1.6}
In this subsection we recall the characterization of the wave front set of a distribution $u\in \cE'(M)$ using pseudodifferential operators on $M$. One says that $A\in \Psi^{m}(M)$ is {\em elliptic} at $(x_{0}, \xi_{0})\in T^{*}M\backslash\{0\}$ if
\[
\sigma_{\rm pr}(A)(x_{0}, \xi_{0})\neq 0.
\]
\begin{proposition}\label{pp3}
 Let $u\in \cD'(M)$. Then $(x_{0}, \xi_{0})\in T^{*}M\backslash\{0\}$ does not belong to $\WF(u)$ iff there exists $A\in \Psi^{0}(M)$, elliptic at $(x_{0}, \xi_{0})$ and $\chi\in \coinf(M)$ with $\chi(x_{0})\neq 0$ such that  $A\chi u\in H^{\infty}(M)$, or equivalently $\chi A \chi u\in \coinf(M)$.
\end{proposition}
Let us also recall some more notation. If $M_{i}$, $i=1,2$ are two manifolds one identifies $T^{*}(M_{1}\times M_{2})$ and $T^{*}M_{1}\times T^{*}M_{2}$. If $K: \coinf(M_{2})\to \cD'(M_{1})$ is continuous, denoting again by $K\in \cD'(M_{1}\times M_{2})$ its distributional kernel, one sets:
\[
\WF(K)'\defeq  \{(X_{1}, X_{2})\in (T^{*}M_{1}\times T^{*}M_{2})\backslash\{0\} : \ (X_{1}, \bar{X}_{2})\in \WF(K)\},
\]
where $\bar{(x, \xi)}= (x, -\xi)$.
\begin{proposition}\label{pp4}
 Let $\cU_{\epsilon}(t,s)$ be as in Thm. \ref{pth1}. Then:
 \[
 \begin{array}{l}
 \WF (\cU_{\epsilon}(t,s)u)= \Phi(t,s)(\WF(u)), \ u\in H^{-\infty}(M),\\[2mm]
 \WF(\cU_{\epsilon}(t,s))'= \{(X, Y)\in T^{*}M\backslash\{0\}\times T^{*}M\backslash\{0\}  : \ X= \Phi(t,s)(Y)\}
\end{array}
 \]
\end{proposition}
\proof This follows immediately from Prop. \ref{pp3}, Thm. \ref{pth1} and the fact that $\cU_{\epsilon}(t,s)$ preserves $H^{\infty}(M)$. \qed
\section{Parametrices and propagators}\label{secp2}\init
\subsection{Introduction}
In this section we consider a class of model Klein-Gordon equations  of the form
\[
{\rm (KG)}\ \ \pe_{t}^{2}\phi+ r(t,x)\pe_{t}\phi+ a(t, x, \pe_{x})\phi=0
\] 
on $I_{t}\times \Sigma$, where $I\subset \rr$ is an open interval.  We will see that the Klein-Gordon equations introduced in Subsect. \ref{secp3.3} can be reduced to such model equations. 
We will consider the associated {\em Cauchy evolution operator} $\cU_{A}(t,s)$, mapping $\rho(s)\phi$ to $\rho(t)\phi$ for $\rho(t)\phi\defeq\begin{pmatrix}\phi(t) \\ \i^{-1}\partial_t \phi(t)\end{pmatrix}$.
It is well-known (see e.g. \cite{Ch}) that $\cU_{A}(t,s)$ can expressed microlocally as the sum of   two Fourier integral operators,  associated with the symplectic flow $\Phi^{\pm}(t,s)$ generated by $\pm(a_{2}(t, x, \xi))^{\12}$, where $a_{2}(t, x, \xi)$ is the principal symbol of $a(t, x, \pe_{x})$.

This fact is not sufficient for our purposes,  namely the construction of pure Hadamard states for Klein-Gordon fields.
We need a more precise decomposition of $\cU_{A}(t,s)$ as a  sum 
\[
\cU_{A}(t,s)= \cU_{A}^{+}(t,s)+ \cU_{A}^{-}(t,s)
\]
 which we call a {\em microlocal decomposition} (see Subsect. \ref{secp2.4}). The essential properties required of $\cU_{A}^{\pm}(t,s)$ is that they are evolutions groups, propagate the wave front set by the flows $\Phi^{\pm}(t,s)$ and  that their ranges are  {\em symplectically orthogonal} for the natural symplectic form preserved by $\cU_{A}(t,s)$.

On a technical level, we avoid the use of the Fourier integral operators machinery and rely instead on propagators $\cU_{b}(t,s)$ generated by time-dependent $\Psi$DOs, which were studied in Subsect. \ref{secp1.5}.  
As a by-product of the construction of $\cU_{A}^{\pm}(t,s)$, we also obtain a {\em Feynman inverse} for the operator $P$ in (KG), canonically associated with the corresponding state, see Subsect. \ref{feyninv}.
 
\subsection{The model Klein-Gordon equation}\label{secp2.1}
In this subsection we give the precise assumptions on our model Klein-Gordon operator ${\rm (KG)}$, to which the Klein-Gordon operators considered in Subsect. \ref{secp3.3} can be reduced.

We fix an open interval $I\subset \rr$ with $0\in I$ and a  smooth $d-$dimensional manifold $\Sigma$, equipped  with a Riemannian metric $k_{0}$, such that $(\Sigma, k_{0})$ is of bounded geometry.   

We fix  the following objects: 
\ben
\item a time-dependent Riemannian metric $h_{t}$ on $\Sigma$ such that $h_{t}\in C_{\rm b}^{\infty}(I; \BT^{0}_{2}(\Sigma, k_{0}))$ and $h_{t}^{-1}\in C_{\rm b}^{\infty}(I; \BT^{2}_{0}(\Sigma, k_{0}))$,
\item a  differential operator $a(t, x, \pe_{x})\in C_{\rm b}^{\infty}(I; \Diff^{2}(\Sigma, k_{0}))$ such that 
\[
\begin{array}{rl}
i)&\sigma_{\rm pr}(a)(t,x, \xi)= \xi\cdot h_{t}^{-1}(x)\xi, \\[2mm]
ii)& a(t, x, \pe_{x})= a^{*}(t, x, \pe_{x})
\end{array}
\]
where the adjoint is defined using the time-dependent scalar product
\beq\label{pscal}
(u|v)= \int_{\Sigma}\bar{u}v|h_{t}|^{\12}dx.
\eeq
\een
We define then the {\em model Klein-Gordon operator}:
\[
P= \pe_{t}^{2}+ r(t,x)\pe_{t}+ a(t, x, \pe_{x}),
\]
for $r(t,x)\defeq  |h_{t}|^{-\12}\p_{t}|h_{t}|^{\12}$. This way $P$ is formally selfadjoint  for the scalar product:
\[
(u|v)_{M}= \int_{I\times \Sigma}\bar{u}v |h_{t}|^{\12}dtdx.
\] 
% In the model case the symplectic form $\sigma$ on $\Sol(P)$ takes the form:
% \[
%\overline{\phi}_{1}\cdot \sigma \phi_{2}= \int_{\Sigma} \left(\p_{t}\overline{\phi}_{1}(t)\phi_{2}(t)- \overline{\phi}_{1}(t)\p_{t}\phi_{2}(t)\right)|h_{t}|^{\12}dx,
%\]
%the rhs being independent on $t$.

\subsection{Solutions to a Riccati equation}\label{secp2.3}
Let us abbreviate $a(t, x, \pe_{x})$, $r(t, x)$ simply by $a$, $r$.
The essential step in the construction of parametrices of the Cauchy problem for the model Klein-Gordon equation introduced in Subsect. \ref{secp2.1} is to find  time-dependent operators $b^{\pm}(t)\in \cinf(I; \Psi^{1}(\Sigma))$ such that the associated evolution operators $\cU_{b^{\pm}}(t,s)$ solve
\[
\left (\pe_{t}^{2}+ r\pe_{t}+ a\right)\cU_{b^{\pm}}(t,s)=0, \hbox{ modulo smoothing errors.}
\]
The above equation is equivalent to the following Riccati equation:
\beq\label{ep2.1}
\i \p_{t}b^{\pm}- b^{\pm 2}+ a + \i rb^{\pm}=0,
\eeq
again modulo smoothing errors. In \cite{GW1} \eqref{ep2.1} was solved in the special case $\Sigma= \rr^{d}$, $r=0$, using the uniform pseudodifferential calculus on $\rr^{d}$, {and an equivalent equation (see \eqref{eq-e1}) was solved before by Junker in the case of $\Sigma$ compact \cite{J1,J2}}. In this subsection we extend the construction to the case when $\Sigma$ is a manifold of bounded geometry, using the pseudodifferential calculus described in Sect. \ref{secp1}, allowing also for $r\neq 0$.

Applying Prop. \ref{pp2}  to $a$, we can find $c>0$ and $c_{-\infty}\in \cinfb(I; \cW^{-\infty}(\Sigma))$ such that  $a(t)+ c_{-\infty}(t)\geq c 1$ for $t\in I$. We set $\epsilon(t)= (a(t)+ c_{-\infty}(t))^{\12}$, so that $\epsilon^{2}(t)= a(t)+\cinfb(I; \cW^{-\infty}(\Sigma))$.
Since $a$ is elliptic, we know from Thm. \ref{thp2.1} that $\epsilon\in \cinfb(\rr, \Psi^{1}(\Sigma))$, with principal symbol $(\xi\cdot h_{t}^{-1}(x)\xi)^{\12}$.   
\begin{theoreme}\label{th2.1}
 There exists $b\in \cinfb(I; \Psi^{1}(\Sigma))$, unique modulo $\cinfb(I; \cW^{-\infty}(\Sigma))$ such that
 \[
 \begin{array}{rl}
 i)& \ b= \epsilon+ \cinfb(I; \Psi^{0}(\Sigma)),\\[2mm]
 ii)& \ (b+ b^{*})^{-1}= (2\epsilon)^{-\12}(\one + r_{-1})(2\epsilon)^{-\12},\ r_{-1}\in \cinfb(I; \Psi^{-1}(\Sigma)),\\[2mm]
 iii)& \ (b+ b^{*})^{-1}\geq c \epsilon^{-1},\hbox{ for some }c\in \cinfb(I;\rr), \ c>0,\\[2mm]
  iv)& \ \i \p_{t}b^{\pm}- b^{\pm 2}+ a + \i rb^{\pm}=r_{-\infty}^{\pm}\in \cinfb(I; \cW^{-\infty}(\Sigma)),\\[2mm]
  &\hbox{ for }b^{+}\defeq  b, \ b^{-}\defeq  -b^{*}.
 \end{array}
  \]
 \end{theoreme}
\proof We follow the proof in \cite[Appendix A3]{GW1}.  We can first replace in \eqref{ep2.1} $a$ by $\epsilon^{2}$, modulo an error term in $\cinfb(I; \cW^{-\infty}(\Sigma))$. Discarding   error terms in $\cinfb(I;\cW^{-\infty}(\Sigma))$, we can assume that $\epsilon= \Op(c)$, $c\in \cinfb(I; S^{1}_{\rm ph}(T^{*}\Sigma))$, with $c_{\rm pr}(t, x, \xi)= (\xi\cdot h_{t}^{-1}(x)\xi)^{\12}$. We  look for $b$ of the form $b= \Op(c)+ \Op(d)$ for $d\in \cinfb(I; S^{0}_{\rm ph}(T^{*}\Sigma))$. Since $\Op(c)$ is elliptic, it admits  parametrices, see Prop. \ref{pp2b}. We  fix  a symbol $\hat{c}\in  \cinfb(I; S^{-1}_{\rm ph}(T^{*}\Sigma))$ such that $\Op(\hat{c})$ is a parametrix of $\Op(c)$.

The equation \eqref{ep2.1} becomes, modulo  error terms in $\cinfb(I;\cW^{-\infty}(\Sigma))$:
\beq\label{enp.11c}
\Op(d)= \frac{\i }{2}(\Op(\hat{c})\Op(\p_{t}c)+ \Op(\hat{c})r \Op(c))+ F(\Op(d)), 
\eeq
for:
\[
F(\Op(d))= \12\Op(\hat{c}) \left(\i \Op(\p_{t}d)+ [\Op(c), \Op(d)]+ \i r \Op(d)- \Op(d)^{2}\right).
\]

From symbolic calculus, we obtain that:
\[
F(\Op(d))= \Op(\tilde{F}(d))+\cinfb(I; \cW^{-\infty}(\Sigma)), 
\]
for
\[
\tilde{F}(d)= \12 \hat{c}\sharp\left(\i\p_{t}d+ c\sharp d- d\sharp c+ \i r\sharp d-  d\sharp d\right),
\]
where the operation $\sharp$ (the Moyal product) is recalled in \ref{secp1.3.4}. The equation \eqref{enp.11c} becomes:
\begin{equation}
\label{enp.11d}
d= a_{0}+ \tilde{F}(d),
\end{equation}
for 
\[
a_{0}= \frac{\i}{2}(\hat{c}\sharp \p_{t}c+ \hat{c}\sharp r\sharp c)\in \cinfb(I; S^{0}_{\rm ph}(T^{*}\Sigma)).
\]
The map $\tilde{F}$ has the following property: 
\beq\label{enp11.e}
\bea
&d_{1}, d_{2}\in \cinfb(I;S^{0}_{\rm ph}(T^{*}\Sigma)), \ d_{1}- d_{2}\in \cinfb(I;S^{-j}_{\rm ph}(T^{*}\Sigma))\\[2mm]
  & \Rightarrow \ \tilde{F}(d_{1})- \tilde{F}(d_{2})\in \cinfb(I;S^{-j-1}_{\rm ph}(T^{*}\Sigma)).
\eea
\eeq
This allows to
solve symbolically \eqref{enp.11d} by setting
\[
d_{-1}= 0,  \ d_{n}\defeq  a_{0}+ \tilde{F}(d_{n-1}), 
\]
and
\[
d\simeq \sum_{n\in \nn}d_{n}- d_{n-1}, 
\]
which is an asymptotic series since by \eqref{enp11.e} we see that $d_{n}- d_{n-1}\in \cinfb(I; S^{-n}_{\rm ph}(T^{*}\Sigma))$. 
It follows that $\Op(c+ d)$ solves \eqref{ep2.1} modulo $\cinfb(I; \cW^{-\infty}(\Sigma))$.

We observe then that  if $b\in \cinfb(I; \Psi^{\infty}(\Sigma))$  we have:
 \[
( \p_{t}b)^{*}= \p_{t}(b^{*})+ r b^{*}- b^{*}r,
 \]
 (recall that the adjoint is computed w.r.t the time-dependent scalar product \eqref{pscal}). This implies that 
 $-\Op(d)^{*}$ is also a solution of \eqref{ep2.1} modulo $\cinfb(I; \cW^{-\infty}(\Sigma))$.
 
 To complete the construction of $b^{\pm}$, we consider  
 \[
 s=\Op(c+ d)+ \Op(c+ d)^{*},
 \] 
 which is  selfadjoint, with principal symbol equal to $2(\xi\cdot h_{t}^{-1}(x)\xi)^{\12}$. By Prop. \ref{pp2},  there exists $r_{-\infty}\in \cinfb(I; \cW^{-\infty}(\Sigma))$ such that 
 \beq\label{equilo}
s+ r_{-\infty}\sim \epsilon,
\eeq
where we recall that the notation $\sim$ is defined in \eqref{equitd}.  
We set now:
\[
b\defeq  \Op(c+d)+ \12 r_{-\infty}.
\]
Properties {\it i)} and {\it iv)} follow from the same properties of $\Op(c+d)$. Property {\it iii)} follows from \eqref{equilo} and the Kato-Heinz theorem. To prove property {\it ii)} we write
\[
b+ b^{*}= (2\epsilon)^{\12}(\one +  \tilde{r}_{-1}) (2\epsilon)^{\12},
\]
where $ \tilde{r}_{-1}\in \cinfb(I; \Psi^{-1}(\Sigma))$, by Thm. \ref{thp2.1}. Since $(\one + \tilde{r}_{-1})$ is boundedly invertible, we have again by Thm. \ref{thp2.1}
\[
(\one + \tilde{r}_{-1})^{-1}= \one + r_{-1}, \ r_{-1}\in \cinfb(I; \Psi^{-1}(\Sigma)),
\]
which implies {\it ii)}. The proof is complete. \qed

Note that by {\it iv)} one has (by subtracting the two identities)
\[
r=\i^{-1}(b^++b^-)-(b^+-b^-)^{-1}\p_t (b^+-b^-)
\]
modulo smoothing errors. Thus, the pair $b^\pm$ contains full information about $r$, and thus about $a$ (using {\it iv)} again).

\subsection{Approximate diagonalization}\label{sec2.3}
In this subsection we perform a diagonalization modulo smoothing errors of the Cauchy evolution operator $\cU_{A}(t,s)$, see \ref{sec2.3.2}. 

We extend the notation in Sect. \ref{secp1} to matrix-valued symbols, operators, etc., by introducing the 
sets $\cinfb(I; \Psi^{m}(\Sigma, \cc^{n}_{p}))$, $n,p\in \nn$ etc. We will frequently omit the extra symbol $\cc^{n}_{p}$ when the nature of the objects is clear from the context. 
We also extend to this situation the notation $\cU_{\epsilon}(t,s)$ when  $\epsilon\in \cinfb(I; \Psi^{m}(\Sigma, \cc^{n}_{n}))$.
\subsubsection{KG equation as a first order system}\label{sec2.3.2}

As usual we write  
\[
(\partial_t^2+ r(t)\p_t + a(t))\phi(t)=0
\] as a first order system:
\beq\label{eq:evA}
\i^{-1}\partial_t \psi(t) = A(t) \psi(t), \quad \mbox{ where \ } A(t)=\mat{0}{\one}{a(t)}{\i r(t)}, 
\eeq
by setting
\[
\psi(t)=\begin{pmatrix}\phi(t) \\ \i^{-1}\partial_t \phi(t)\end{pmatrix}\eqdef  \rho(t)\phi.
\]
We equip $L^{2}(\Sigma; \cc^{2})$ with the time-dependent scalar product obtained from \eqref{pscal}, by setting:
\[
(f| g)\defeq  \int_{\Sigma}(\bar{f}_{1}g_{1}+ \bar{f}_{0}g_{0})|h_{t}|^{\12}dx.
\]
We will use it to  define adjoints of linear operators and to identify sesquilinear forms on $L^{2}(\Sigma; \cc^{2})$ with linear operators. Note that if $\phi_{i}$ are $C^\infty$ solutions with $\phi_{i}\traa{\Sigma}$ compactly supported then 
\[
\i \bar{\phi}_{1}\cdot\sigma\phi_{2}= (\rho(t)\phi_{1}| q \rho(t)\phi_{2})
\] for:
\[
q\defeq  \mat{0}{1}{1}{0}
\]
is independent on $t$.
The evolution operator $\cU_{A}(t,s)$ is {\em symplectic}:
\begin{equation}
\label{ep3.1}
q=\cU_{A}^{*}(s,t)q\cU_{A}(s,t), \ s,t\in I.
\end{equation}

\subsubsection{First reduction}\label{sec2.3.3}
 The Riccati equation 
\begin{equation}
\label{eq-ric}
\i \p_{t}b^{\pm}- b^{\pm 2}+ a + \i rb^{\pm}=r_{-\infty}^{\pm}
\end{equation} 
implies that:
\begin{equation}
\label{eq-e1}
(\pe_{t}+ \i b^{\pm}+ r)\circ (\pe_{t}- \i b^{\pm})= \pe_{t}^{2}+ r\pe_{t}+a - r_{-\infty}^{\pm},
\end{equation}
which is a factorization of the Klein-Gordon operator $P$ modulo smoothing errors. One can also deduce from \eqref{eq-e1} a time-dependent diagonalization of the evolution operator for $P$, which we now define.
We set
\[
\tilde\psi(t)\defeq \begin{pmatrix}\pe_t-\i b^{-}(t) \\ \pe_t-\i b^{+}(t)\end{pmatrix}\phi(t),
\]
and obtain $\tilde{\psi}(t)= S^{-1}(t)\psi(t)$ with
 \beq\label{e4.01}
 S^{-1}(t)= \i \mat{- b^{-}(t)}{1}{-b^{+}(t)}{1}, \quad  S(t)= \i^{-1}\mat{1}{-1}{b^{+}(t)}{-b^{-}(t)}(b^{+}(t)- b^{-}(t))^{-1},
 \eeq
which makes sense thanks to $b^{+}(t)- b^{-}(t)$ being invertible by Thm. \ref{th2.1}.
  We obtain from \eqref{eq-e1} that
  \beq\label{e4.0b}
\bea
& \mat{\pe_{t}+ \i b^{-}+ r}{0}{0}{\pe_{t}+ \i b^{+}+r}\tilde{\psi}(t)
 =\begin{pmatrix}
 \pe_{t}^{2}+ a+r\pe_{t}- r_{-\infty}^{-}\\ \pe_{t}^{2}+ a+ r\pe_{t}- r_{-\infty}^{+}\end{pmatrix}\phi(t)\\[2mm]
 &= \col{P\phi}{P\phi}-\mat{r_{-\infty}^{-}}{0}{r_{-\infty}^{+}}{0}S(t)\tilde{\psi}(t),  
\eea
\eeq 
Let
\[
B(t)=\tilde{B}(t)+ S_{-\infty}(t),
\]
for
\[
\tilde{B}(t)= \mat{-b^{-}+ \i r}{0}{0}{-b^{+}+\i r},\ 
 S_{-\infty}(t)= \mat{r_{-\infty}^{-}}{-r_{-\infty}^{-}}{r_{-\infty}^{+}}{-r_{-\infty}^{+}}(b^{+}- b^{-})^{-1}.
\]
Then  since $P\phi=0$ we deduce from \eqref{e4.0b} that:
 \[
( \pe_{t}- \i B(t))\tilde{\psi}(t)=0,
 \]
hence:
\beq\label{e4.00}
\cU_{A}(t,s)= S(t)\cU_{B}(t,s)S(s)^{-1}.
\eeq
We have thus a formula that relates  $\cU_{A}(t,s)$ and the evolution generated by a time-dependent operator $B(t)$ that is diagonal up to $\cW^{-\infty}(\Sigma)$ remainders and whose on-diagonal terms have principal symbols $\pm(\xi\cdot h_{t}^{-1}(x)\xi)^{\12}$.
 
Let us now discuss the symplectic properties of $\cU_{B}(t,s)$. Since 
\[
S(s)^{*}qS(s)= (b^{+}- b^{-})^{-1}(s)\mat{1}{0}{0}{-1}\eqdef  q_{B}(s)
\]
we obtain from \eqref{e4.00}, \eqref{ep3.1} that:
\[
q_{B}(t)=\cU_{B}^{*}(s,t)q_{B}(s)\cU_{B}(s,t), \ s,t\in I.
\]
\subsubsection{Second reduction}\label{sec2.3.4}
To get rid of the $(b^{+}- b^{-})^{-1}(s)$ factor in $q_{B}(s)$ we set 
\[
\cU_{C}(t,s)\defeq  (b^{+}- b^{-})^{-\12}(t)\cU_{B}(t,s)(b^{+}- b^{-})^{\12}(s).
\]
It follows that:
\beq\label{taz}
\cU_{A}(t,s)= T(t)\cU_{C}(t,s)T(s)^{-1}, 
\eeq
for:
\beq\label{defdeT}
\begin{array}{l}
T(t)\defeq  S(t)(b^{+}- b^{-})^{\12}(t)= \i^{-1}\mat{1}{-1}{b^{+}}{-b^{-}}(b^{+}- b^{-})^{-\12}, \\[2mm]
 T^{-1}(t)= \i (b^{+}- b^{-})^{-\12}\mat{-b^{-}}{1}{-b^{+}}{1}.
\end{array}
\eeq
Note that:
\beq\label{symplac}
T^{*}(t) qT(t)=  \mat{1}{0}{0}{-1}\eqdef  \hat{q},
\eeq
so that $\cU_{C}(t,s)$ is symplectic for $\hat{q}$:
\beq\label{staro}
\cU_{C}(t,s)^{*}\hat{q} \cU_{C}(t,s)=\hat{q}.
\eeq
The generator of $\cU_{C}(t,s)$ is:
\beq\label{ep3.5}
C(t)=\tilde{C}(t)+ R_{-\infty}(t),
\eeq
for
\beq\label{ep3.6}
\bea
\tilde{C}(t)&\defeq(b^{+}- b^{-})^{-\12}\tilde{B}(t)(b^{+}- b^{-})^{\12}-\i \p_{t}(b^{+}- b^{-})^{-\12}(b^{+}- b^{-})^{\12}\\[2mm]
&= \mat{- b^{-}+ r_{0}^{-}}{0}{0}{-b^{+}+ r_{0}^{+}}
\eea
\eeq
where 
\[
r_{0}^{\pm}= \i r+ [(b^{+}- b^{-})^{-\12}, b^{\pm}]-\i \p_{t}(b^{+}- b^{-})^{-\12}(b^{+}- b^{-})^{\12}\in \cinfb(I; \Psi^{0}(\Sigma)).
\]
and
\beq\label{ep3.6b}
R_{-\infty}=- (b^{+}- b^{-})^{-\12}S_{-\infty}(b^{+}- b^{-})^{\12}\in\cinfb(I; \cW^{-\infty}(\Sigma)).
\eeq

\begin{remark}
 Let us explain another motivation for the introduction of the maps $T(t)$. There are two natural  topologies on the space of Cauchy data for \eqref{eq:evA}. The first is the {\em energy space topology} given by the topology of  $H^{1}(\Sigma)\oplus L^{2}(\Sigma)$, ubiquitous in the PDE literature. The second is the {\em charge space topology}, given by the topology of $H^{\12}(\Sigma)\oplus H^{-\12}(\Sigma)$, related to the quantization of the Klein-Gordon equation. It is  easy to see that 
 $S(t)$ is an isomorphism from  $L^{2}(\Sigma)\oplus L^{2}(\Sigma)$ to $H^{1}(\Sigma)\oplus L^{2}(\Sigma)$, while $T(t)$ is an isomorphism from $L^{2}(\Sigma)\oplus L^{2}(\Sigma)$  to 
 $H^{\12}(\Sigma)\oplus H^{-\12}(\Sigma)$.
 \end{remark}
 \subsubsection{Interaction picture}
 From \eqref{ep3.5} we know that the generator of $\cU_{C}(t,s)$ is diagonal, modulo a smoothing error term.  It follows from standard arguments that $\cU_{C}(t,s)$ is also diagonal, modulo smoothing errors. We  review  this argument, known as the `interaction picture' in the physics literature.

Let $H(t)= H_{0}(t)+ V(t)$ be a time-dependent Hamiltonian, $\cU(\cdot, \cdot)$ and $\cU_{0}(\cdot, \cdot)$ the associated propagators.  We fix $t_1\in \rr$ and set:
\[
\cU(t,s)\eqdef  \cU_{0}(t, t_1)\cV_{t_1}(t,s)\cU_{0}(t_1,s).
\]
(Typically $H_{0}$ does not depend on time and one sets $t_1=0$). It follows that $\cV_{t_1}(\cdot, \cdot)$  is an evolution  group and solves
\[
\begin{cases}
\p_{t}\cV_{t_1}(t,s)= \i \tilde{V}_{t_1}(t)\cV_{t_1}(t,s)\hbox{ for }\tilde{V}_{t_1}(t)= \cU_{0} (t_1,t) V(t)\cU_{0}(t,t_1),\\[2mm]
\cV_{t_1}(s,s)= \one. 
\end{cases}
\]
Note the following covariance property:
\[
\cV_{t_{2}}(t,s)= \cU_{0}(t_{1}, t_{2}) \cV_{t_{1}}(t,s)\cU_{0}(t_{2}, t_{1}), \ t_{1}, t_{2}\in \rr.
\]
\subsubsection{Parametrix for the Cauchy problem}
We apply the above procedure to $C=\tilde C + R_{-\infty}$, fix some $t_1\in I$ and set:
\[
\cU_{C}(t,s)\eqdef  \cU_{\tilde{C}}(t,t_1)\cV_{t_1}(t,s)\cU_{\tilde{C}}(t_1,s),
\]
where $\cV_{t_1}(t,s)$ is the evolution generated by $R_{t_1, -\infty}(t)= \cU_{\tilde{C}}(t_1, t)R_{-\infty}(t)\cU_{\tilde{C}}(t, t_1)$, i.e.
\beq\label{ep3.7}
\begin{cases}
\p_{t}\cV_{t_1}(t,s)= \i  R_{t_1, -\infty}(t)\cV_{t_1}(t,s),\\
\cV_{t_1}(s,s)=\one.
\end{cases}
\eeq
Note that $\tilde{C}(t)$ is diagonal, with entries  satisfying condition (E) in Subsect. \ref{secp1.5}. Therefore by Lemma \ref{ego.1} we know that $R_{t_1,-\infty}(t)\in \cinfb(I;\cW^{-\infty}(\Sigma))$. 
For any $s\in \rr$, the equation \eqref{ep3.7} can be solved in $\cinfb(I^{2}; B(H^{s}(\Sigma)))$ by a convergent series. This implies easily that:
\beq\label{toz}
\begin{array}{l}\cV_{t_1}(t,s)= \one+ \cinfb(I^{2}, \Psi^{-\infty}(\Sigma)),
\\[2mm]
\cU_{C}(t,s)= \cU_{\tilde{C}}(t,s)+ \cinfb(I^{2}, \Psi^{-\infty}(\Sigma)).
\end{array}
\eeq
We summarize this discussion with the following theorem.
\begin{theoreme}\label{newprop1}
Let\beq\label{e4.102b}
\cU_{\tilde{A}}(t,s)\defeq  T(t)\cU_{\tilde{C}}(t,s)T(s)^{-1}.
\eeq
Then $\{\cU_{\tilde{A}}(t,s)\}_{(t,s)\in I^{2}}$ is an evolution group and:
\[
\cU_{A}(t,s)= \cU_{\tilde{A}}(t,s)+ \cinfb(I^{2}, \Psi^{-\infty}(\Sigma)).
\]
It follows that  the group $\{\cU_{\tilde{A}}(t,s)\}_{(t,s)\in I^{2}}$ is a parametrix for the Cauchy problem.
\end{theoreme}
Note that  since $\tilde{C}(t)$ is    diagonal, we have:
\beq\label{ep3.8}
\cU_{\tilde{C}}(t,s)= \mat{\cU_{-b^{-}+r_{0}^{-}}(t,s)}{0}{0}{\cU_{-b^{+}+r_{0}^{+}}(t,s)}.
\eeq

\subsection{Decomposition of the Cauchy evolution}\label{secp2.4}
Basing on the constructions in  Subsect. \ref{sec2.3} it is easy to construct a microlocal decomposition of the evolution $\cU_{A}(t,s)$. 
In fact let 
\beq\label{defdepi}
\pi^{+}= \mat{\one}{0}{0}{0}, \ \ \pi^{-}=\mat{0}{0}{0}{\one}.
\eeq
 We fix a reference time $t_{0}\in I$ for example $t_{0}=0$ and 
set:
\beq\label{split.0}
c^{\pm}(0)\defeq  T(0)\pi^{\pm}T^{-1}(0)=  \mat{\mp(b^{+}- b^{-})^{-1}b^{\mp}}{\pm(b^{+}- b^{-})^{-1}}{\mp b^{+}(b^{+}- b^{-})^{-1}b^{-}}{\pm b^{\pm}(b^{+}- b^{-})^{-1}}(0).
\eeq
We have:
\[
c^{\pm}(0)^{2}= c^{\pm}(0), \ c^{+}(0)+ c^{-}(0)= \one, \ c^{\pm}(0)\in \cinfb(I; \Psi^{\infty}(\Sigma)).
\]
It follows that $c^{+}(0), c^{-}(0)$ is a pair of complementary projections. Moreover from \eqref{symplac}, we obtain that:
\beq\label{ep3.10}
c^{\mp*}(0)qc^{\pm}(0)=0,
\eeq
i.e. the ranges of the projections $c^{\pm}(0)$ are $q-$orthogonal.  
We set:
\beq\label{e2.10}
\cU_{A}^{\pm}(t,s)\defeq  \cU_{A}(t,0)c^{\pm}(0)\cU_{A}(0,s).
\eeq
\begin{definition}\label{split.def}
 A pair $\{\cU_{A}^{\pm}(t, s)\}_{(t,s)\in I^{2}}$ as in \eqref{e2.10} will be called a {\em microlocal decomposition} of the evolution group $\{\cU_{A}(t, s)\}_{(t,s)\in I^{2}}$.
\end{definition}
%old version with $\hat{c}$.
%
%
%In fact let us set
%\beq\label{defdepi}
%\pi^{+}= \mat{\one}{0}{0}{0}, \ \pi^{-}=\mat{0}{0}{0}{\one},
%\eeq
%and:
% \beq\label{split.0}
%\hat{c}^{\pm}(s)\defeq  T(s)\pi^{\pm}T^{-1}(s)=  \mat{\mp(b^{+}- b^{-})^{-1}b^{\mp}}{\pm(b^{+}- b^{-})^{-1}}{\mp b^{+}(b^{+}- b^{-})^{-1}b^{-}}{\pm b^{\pm}(b^{+}- b^{-})^{-1}}(s).
%\eeq
%We have:
%\[
%\hat{c}^{\pm}(s)^{2}= \hat{c}^{\pm}(s), \ \hat{c}^{+}(s)+ \hat{c}^{-}(s)= \one, \ \hat{c}^{\pm}(s)\in \cinfb(I; \Psi^{\infty}(\Sigma)).
%\]
%It follows that $\hat{c}^{+}(s), \hat{c}^{-}(s)$ is a pair of complementary projections. Moreover from \eqref{symplac}, we obtain that:
%\beq\label{ep3.10}
%\hat{c}^{\mp*}(s)q\hat{c}^{\pm}(s)=0,
%\eeq
%i.e. the ranges of the projections $\hat{c}^{\pm}(s)$ are $q-$orthogonal.  
%
%
%
%
%Let us now fix a reference time $T_{0}\in I$ for example $T_{0}=0$ and 
%set:
%\beq\label{e2.10}
%\cU_{A}^{\pm}(t,s)\defeq  \cU_{A}(t,0)\hat{c}^{\pm}(0)\cU_{A}(0,s).
%\eeq
\begin{theoreme}\label{newprop.2}
The following properties are true:
 \[
\bea 
i)&\  \ \cU_{A}^{\pm}(t, s)\cU_{A}^{\pm}(s,t')= \cU_{A}^{\pm}(t, t'),\\[2mm]
ii)&\ \ \cU_{A}^{+}(t,s)+ \cU_{A}^{-}(t,s)= \cU_{A}(t,s),\\[2mm]
iii)&\ \ \cU_{A}^{\pm}(t,s)^{*}q\cU_{A}^{\mp}(t,s)=0,\\[2mm]
iv)&\ \ (\pe_{t}- \i A(t))\cU_{A}^{\pm}(t,s)= \cU_{A}^{\pm}(t,s)(\pe_{s}- \i A(s))=0,\\[2mm]
v)& \ \ \WF(\cU_{A}^{\pm}(t,s))'=\{(X, X')\in T^{*}\Sigma\times T^{*}\Sigma : \  X= \Phi^{\pm}(t,s)(X')\},
\eea 
\]
where $\Phi^{\pm}(t,s): T^{*}\Sigma\to T^{*}\Sigma$
is   the symplectic flow generated by the time-dependent Hamiltonian $\pm(\xi\cdot h_{t}^{-1}(x)\xi)^{\12}$.
\end{theoreme}

\proof $i)$ and $ii)$ follow from the fact that $c^{\pm}(0)$ are complementary projections. $iii)$ follows from  \eqref{ep3.1} and \eqref{ep3.10}. $iv)$ is immediate.  From \eqref{ep3.8} and Prop. \ref{pp4} we obtain that $\cU_{C}(t,0)\pi^{\pm}\cU_{C}(0,s)$ has the  wave front set stated in $v)$.  The result follows then from the fact that 
$\cU_{A}^{\pm}(t,s)= T(t)\cU_{C}(t,0)\pi^{\pm}\cU_{C}(0,s)T^{-1}(s)$. \qed

We now gather a couple of formulae that relate various objects at different times.  The proof is a routine computation that uses the first three statements in Thm. \ref{newprop.2}.

\begin{proposition}\label{split.prop1}
 Let \beq\label{cplus}
  c^{\pm}(t)\defeq\cU_{A}^{\pm}(t,t)=\cU_{A}(t, 0)c^{\pm}(0)\cU_{A}(0,t).
\eeq
Then:
\[
\bea 
&c^{\pm}(t)^{2}= c^{\pm}(t), \ c^{+}(t)+ c^{-}(t)= \one, \\[2mm]
&c^{\pm}(t)= \cU_{A}(t,s)c^{\pm}(s)\cU_{A}(s,t),\\[2mm]
&c^{\mp}(t)qc^{\pm}(t)^{*}=0,\ c^{\pm}(t)\cU_{A}(t,s)c^{\mp}(s)=0,\\[2mm]
&\cU_{A}^{\pm}(t,s)= c^{\pm}(t)\cU_{A}(t,s)c^{\pm}(s)=c^{\pm}(t)\cU_{A}(t,s)=\cU_{A}(t,s)c^{\pm}(s).
\eea 
\]
\end{proposition}

\subsection{The Feynman inverse associated to a microlocal decomposition}\label{feyninv}
In this subsection we work in the setup of Subsect. \ref{secp2.1}
\subsubsection{Distinguished parametrices for the Klein-Gordon operator}\label{turlututu}
In our terminology, a continuous map $G: \coinf(M)\to \cinf(M)$ is a (two-sided) {\em parametrix} of the Klein-Gordon operator $P$ if $PG-\one$ and $\one -GP$ have smooth kernels. In what follows we recall the classification of  parametrices of  $P$  due to Duistermaat and H\"{o}rmander in \cite{DH}.

For $x\in M$ we denote by $V_{x\pm}\subset T_{x}M$ the future/past solid lightcones and by $V_{x\pm}^{*}\subset T_{x}^{*}M$ the dual cones $V_{x\pm}^{*}=\{\xi\in T^{*}_{x}M :\  \xi\cdot v>0, \ \forall v\in V_{x\pm}, \ v\neq 0\}$. We write
\[
\xi\rhd 0\ (\hbox{resp. }\xi\lhd 0)\,\hbox{ if }\,\xi\in V_{x+}^{*} \ (\hbox{resp. } V_{x-}^{*}).
\]
For $X= (x, \xi)\in T^{*}M\backslash\{0\}$ denote $p(X)= \xi\cdot g^{-1}(x)\xi$ the principal symbol of $P$ and 
$\cN= p^{-1}(0)\cap \coo{M}$ the characteristic manifold of $P$. If $H_{p}$ is the Hamiltonian vector field of $p$,
integral curves of $H_{p}$ in $\cN$ are called {\em bicharacteristics}. $\cN$ splits into the {\em upper/lower energy shells}
\[
\cN= \cN^{+}\cup \cN^{-}, \ \ \cN^{\pm}=\cN\cap \{\pm\xi\rhd 0\}.
\]
  For $X_{1}, X_{2} \in\cN$ we write  $X_{1}\sim X_{2}$ if  $X_{1},X_{2}$ lie on the same bicharacteristic. 
For $X_{1}\sim X_{2}$, we write $X_{1}\succ X_{2}$ (resp. $X_{1}\prec X_{2}$) if $X_{1}$ comes
strictly after (before) $X_{2}$ w.r.t. the natural  parameter on the
bicharacteristic through $X_{1}$ and $X_{2}$.   Finally one sets
\[
\cC=\{(X_{1}, X_{2})\in \cN\times \cN : \ X_{1}\sim X_{2}\}, \ \ \Delta= \{(X, X):\ X\in \coo{M}\},
\]
and
\[
\begin{array}{l}
\cC_{\rm ret/adv}= \{(X_{1}, X_{2})\in \cC : \ x_{1}\in J^{\pm}(x_{2})\}, \\[2mm]
\cC_{\rm F}= \{(X_{1}, X_{2})\in \cC : \ X_{1}\prec X_{2}\}, \\[2mm]
\cC_{\rm\overline{F}}= \{(X_{1}, X_{2})\in \cC : \ X_{1}\succ X_{2}\}.
\end{array}
\]
The main results of \cite{DH} relevant to us is the following theorem.
\begin{theoreme}\cite[Thm. 6.5.3]{DH}
For $\sharp= {\rm ret}, {\rm adv}, {\rm F},{\rm\bar{F}}$ there exists a parametrix $G_{\sharp}$ of $P$ such that
\beq\label{duih}
\WF(G_{\sharp})'= \Delta\cup \cC_{\sharp}.
\eeq
Any other parametrix $G$ with $\WF(G)'\subset \Delta\cup \cC_{\sharp}$ equals $G_{\sharp}$ modulo a smooth kernel.
\end{theoreme}
A parametrix satisfying \eqref{duih} for $\sharp= {\rm ret/adv}$ resp. $\sharp= {\rm F}/ {\rm\overline{F}}$ will be called a {\em retarded/advanced}  resp. {\em Feynman/anti-Feynman} parametrix (or inverse if $PG_{\sharp}=\one$ and $G_{\sharp} P=\one$ hold exactly).

\subsubsection{The Feynman inverse associated to a microlocal decomposition}
%It is easy to show that the canonically defined {\em retarded/advanced inverses} $G_{\rm ret/adv}$ (see \ref{secp4.1.2}) are  retarded/advanced parametrices.
%On the other hand there does not exist in general  a canonical Feynman or anti-Feynman inverse for $P$.  This fact is  related to the well-known non-uniqueness of Hadamard states (and non-existence of a canonical one) on a curved space\-time.

We now show how to  associate to the decomposition of the Cauchy evolution constructed in Subsect. \ref{secp2.4} a {\em Feynman inverse} for  the Klein-Gordon operator $P$.
 %We know that to a Hadamard state  we can associate a Feynman inverse (not only a parametrix). This follows from the formula
%\[
%\Lambda^{\pm}= \i (\tilde{E}_{F}- E_{\mp})\hbox{ modulo smoothing}, 
%\]
%hence
%\[
%E_{F}= \i^{-1}\Lambda^{+}+ E_{-}
%\]
%is a true inverse with the correct wave front set.
%
%We want to do the same in our formalism. The idea is as follows: if we start from $b^{\pm}$, and set $u^{\pm}(t,s)= \cU_{b^{\pm}}(t,s)$, then an easy computation shows that
%\[
%\tilde{E}_{F}(t,s)= \left(u^{+}(t,s)\theta(t-s)+ u^{-}(t,s)\theta(s-t)\right)\i^{-1}(b^{+}(s)- b^{-}(s))^{-1}
%\]
%solves
%\[
%(\p^{2}_{t}+ a(t)+ r(t)\p_{t}) \tilde{E}_{F}(t,s)= \delta(t-s) \hbox{ modulo }C^{\infty}(\rr^{2}, \Psi^{-\infty}),
%\]
%hence  gives a Feynman parametrix. The same is true if we replace $b^{\pm}$ by $\tilde{b}^{\pm}$ in Lemma \ref{lemma.1}.   Now from Corollary \ref{corr.1} we see that
%\[
%\cU_{\tilde{b}^{\pm}}(t,s) (b^{+}-b^{-})^{-1}(s)= \pm\pi_{0}\cU_{\tilde{A}}^{\pm}(t,s)\pi_{1}^{*},
%\]
%where $\pi_{i}: L^{2}(\Sigma;\cc^{2})\to L^{2}(\Sigma)$ for $i=0,1$ is the projection on the first or second component. So the idea is to do this replacement in the formula above but now with $\cU_{A}^{\pm}(t,s)$ instead.

In the next theorem, we will use the `time kernel' notation: namely if $A: \coinf(M; \cc^{p})\to \cinf(M; \cc^{q})$ we denote by $A(t,s):\coinf(\Sigma;\cc^{p})\to \cinf(\Sigma;\cc^{q})$ its operator-valued kernel, defined by
\[
Au(t)= \int_{\rr}A(t,s)u(s)ds, \ u\in\coinf(M; \cc^{p}).
\]
We denote by $\pi_{i}: L^{2}(\Sigma;\cc^{2})\to L^{2}(\Sigma)$ for $i=0,1$  the projection on the first or second component and by $\theta(s)$ the Heaviside function.
\begin{theoreme}\label{feyninvth}
 Let $\cU_A^\pm(t,s)$ be a microlocal decomposition and let
 \beq\label{trululu}
 G_{\rm F}(t,s)= \i^{-1}\pi_{0} \left(\cU^{+}_{A}(t,s)\theta(t-s)- \cU_{A}^{-}(t,s)\theta(s-t)\right) \pi_{1}^{*}.
 \eeq
 Then $G_{\rm F}: \coinf(M)\to \cinf(M)$  is continuous and:
 \[
 P\circ G_{\rm F}= G_{\rm F}\circ P= \one.
 \]
 Moreover $\WF(G_{\rm F})'= \Delta\cup \cC_{\rm F}$, hence $G_{\rm F}$ is a Feynman inverse.
  \end{theoreme}
 \proof 
The fact that $G_{\rm F}: \coinf(M)\to \cinf(M)$ is continuous follows from   Thm. \ref{newprop1},  Lemma \ref{ego.1} and the fact that $\coinf(M)\subset \cinfb(\rr; H^{\infty}(\Sigma))\subset \cinf(M)$ continuously.
In the rest of the proof we will use freely the time-kernel notation. We will denote by $\rho$ the map $ \coinf(M)\ni u\mapsto (u, \i^{-1}\p_{t}u)\in \coinf(M; \cc^{2})$, whose kernel is $\delta(t-s)\rho(s)$.

To prove that $P\circ G_{\rm F}= \one$, we  set $R(t,s)= \cU_{A}^{+}(t,s)\theta(t-s)- \cU_{A}^{-}(t,s)\theta(s-t)$. Since $(\pe_{t}- \i A(t))\circ \cU_{A}^{\pm}=0$, we obtain
\[
\bea
&((\pe_{t}-\i A(t))\circ R)(t,s)=\cU_{A}^{+}(t,s)\delta(t-s)+ \cU_{A}^{-}(t,s)\delta(s-t)\\[2mm]
&= (c^{+}(s)+ c^{-}(s))\delta(t-s)= \one_{\Sigma}\delta(t-s),
\eea
\]
hence $(\pe_{t}- \i A(t))\circ R=\one$. This implies that
\[
\pi_{0}\circ(\pe_{t}- \i A(t))\circ R\circ\pi_{1}^{*}=0, \ \ \pi_{0}\circ(\pe_{t}- \i A(t))\circ R\circ\pi_{1}^{*}=\one,
\]
which by an easy computation implies that $P\circ G_{\rm F}= \one$.

%\begin{equation}
%\label{e4.1}
%\bea 
%\p_{t}\pi_{0}\cU_{A}^{(\pm)}(t,s)&= \i \pi_{1}\cU_{A}^{(\pm)}(t,s),\\[2mm]
%\p_{t}\pi_{1}\cU_{A}^{(\pm)}(t,s)&= \i a\pi_{0}\cU_{A}^{(\pm)}(t,s)- r\pi_{1}\cU_{A}^{(\pm)}(t,s).
%\eea 
%\end{equation}
% Therefore
% \[
% \bea 
% \p_{t}G_{\rm F}(t,s)&= \pi_{1} \left(\cU^{+}_{A}(t,s)\theta(t-s)- \cU_{A}^{-}(t,s)\theta(s-t)\right)\pi_{1}^{*}\\[2mm]
% &\phantom{=}+ \i^{-1}\pi_{0}\left(\cU_{A}^{+}(s,s)\delta(t-s)+ \cU_{A}^{-}(s,s)\delta(t-s)\right)\pi_{1}^{*}\\[2mm]
% &=\pi_{1} \left(\cU^{+}_{A}(t,s)\theta(t-s)- \cU_{A}^{-}(t,s)\theta(s-t)\right)\pi_{1}^{*},
% \eea 
%  \]
%  since $\cU_{A}^{\pm}(s,s)= c^{\pm}(s)$, $c^{+}(s)+ c^{-}(s)=1$ and $\pi_{0}\pi_{1}^{*}=0$. Next
%  \[
%  \bea 
%  \p^{2}_{t}G_{\rm F}(t,s)&=(\i a\pi_{0}- r\pi_{1})\left(\cU^{+}_{A}(t,s)\theta(t-s)- \cU_{A}^{-}(t,s)\theta(s-t)\right)\pi_{1}^{*}\\[2mm]
%  &\phantom{=}+\pi_{1}(c^{+}(s)+ c^{-}(s))\delta(t-s)\pi_{1}^{*}\\[2mm]
%  &=-aG_{\rm F}(t,s)- r\p_{t}G_{\rm F}(t,s) +\delta(t-s)\one.
%  \eea 
%  \]
To prove that $G_{\rm F}\circ P= \one$ we note that
\[
\pi_{1}^{*}\circ P= \i (\pe_{t}- \i A(t))\circ \rho, \ \ \theta(\pm(t-s))\circ\pe_{s}= \pe_{s} \circ \theta(\pm(t-s))\pm \delta(t-s).
\]
\newcommand{\moveprec}[1][0pt]{
  \mathrel{\raisebox{#1}{$\prec$}}}
\newcommand{\movesucc}[1][0pt]{
  \mathrel{\raisebox{#1}{$\succ$}}}
Using then that 
  \[
  \cU_{A}^{\pm}\circ (\pe_{t}- \i A(t))=0, \ \ \cU_{A}^{+}(s,s)+ \cU_{A}^{-}(s,s)= c^{+}(s)+ c^{-}(s)= \one,
  \]
    we obtain  that $G_{\rm F}\circ P= \pi_{0}\circ\rho= \one$.  Writing $X= (t, x, \tau, \xi)\in \coo{(\rr\times \Sigma)}$ we have:  
    \[
    X_{1}\genfrac{}{}{0pt}{1}{\moveprec[-4pt]}{\movesucc[1pt]}X_{2} \Leftrightarrow \tau_{i}= \pm(\xi_{i}\cdot h^{-1}(t_{i}, x_{i})\xi_{i})^{\12}, (x_{1}, \xi_{1})= \phi^{\pm}(t_{1}, t_{2})(x_{2},\xi_{2}). 
    \]
    Using Thm. \ref{newprop.2} {\it v)} this easily implies that $\WF(G_{\rm F})'= \Delta\cup \cC_{\rm F}$.  \qed
\section {Hadamard states}\init\label{secp4}\init
In this section  we associate to a microlocal decomposition as in Def. \ref{split.def} a unique {\em pure Hadamard state} $\omega$. The Cauchy surface two-point functions (see Def. \ref{defcs2pt}) are (matrices of) pseudodifferential operators on $\Sigma$.  We give the relation between the space\-time two-point functions of $\omega$ and the operators $\cU_{A}^{\pm}(\cdot, \cdot)$ in Def. \ref{split.def}.

 {We say that} a state is {\em regular}  if its Cauchy surface two-point functions are (matrices) of pseudodifferential operators {(in the sense of the calculus on manifolds of bounded geometry)}. We show that any pure regular Hadamard state is actually associated to a microlocal decomposition.
\subsection{Klein-Gordon fields}\label{secp4.1}
We start by reviewing classical material about quasi-free states for Klein-Gordon fields, see e.g. \cite{DG,KM,HW}. We use the complex formalism, based on charged (i.e., complex) fields $\psi, \psi^{*}$, which turns out to be more convenient for our analysis. 
\subsubsection{Bosonic quasi-free states}\label{secp4.1.1}
Let $\cV$ be a complex vector space, $\cV^{*}$ its anti-dual and let us denote $L_{\rm h}(\cV, \cV^{*})$ the space of hermitian sesquilinear forms on $\cV$.
  A pair $(\cV,q)$ consisting of a complex vector space $\cV$ and a non-degenerate hermitian form $q$ on $\cV$ will be called a {\em phase space}.
  We denote by $U(\cV, q)$ the pseudo-unitary group for $(\cV, q)$.
  
As outlined in the introduction, given a phase space $(\cV,q)$ one can  define the 
{\em   CCR $*$-algebra} ${\rm CCR}(\cV,q)$ (see e.g. \cite[Sect. 8.3.1]{DG}) \footnote{See \cite{GW1} for  the transition between real and complex vector space terminology.}. The  (complex) field operators $\cV\ni v\mapsto \psi(v), \psi^{*}(v)$,   which generate ${\rm CCR}(\cV,q)$, are anti-linear, resp. linear in $v$ and  satisfy the canonical commutation relations
\[
[\psi(v), \psi(w)]= [\psi^{*}(v), \psi^{*}(w)]=0,  \ \ [\psi(v), \psi^{*}(w)]=  \bar{v} q w \one, \ \ v, w\in \cV.
\]
The \emph{complex covariances}  $\Lambda^\pm\in L_{\rm h}(\cV,\cV^*)$ of a state $\omega$ on ${\rm CCR}(\cV,q)$ are defined in terms of the abstract field operators  by
\beq\label{eq:lambda}
\bar{v}\cdot\Lambda^+ w \defeq \omega\big(\psi(v)\psi^*(w)\big), \quad \bar{v}\cdot\Lambda^- w \defeq \omega\big(\psi^*(w)\psi(v)\big), \quad v,w\in \cV
\eeq
Note that  $\Lambda^{\pm}\geq 0$ and   $\Lambda^+ - \Lambda^- = q$ by the canonical commutation relations.
Conversely if   $\Lambda^\pm$  are Hermitian forms on $\cV$ such that  
\begin{equation}
\label{positon}
\Lambda^+ - \Lambda^- = q, \ \ \Lambda^\pm\geq 0,
\end{equation}
then there is a unique  quasi-free state $\omega$ such that (\ref{eq:lambda}) holds, see e.g. \cite[Sect. 17.1]{DG}. \medskip

{In order to discuss purity of quasi-free states in terms of their two-point functions, one needs to work in a $C^{*}$-algebraic framework instead. 

If $\cV_{\rr}$ is $\cV$ considered as a real vector space and $\sigma= \i^{-1}q$, then $(\cV_{\rr}, {\rm Re}\,\sigma)$ is a real symplectic space. We denote by ${\rm W}(\cV, q)$ the  {\em Weyl $C^{*}$-algebra} over $(\cV_{\rr}, {\rm Re}\,\sigma)$, see e.g. \cite[Sect. 8.5.3]{DG}, whose generators are denoted by $W(v)$. 
We still denote by $\omega$ the quasi-free state on ${\rm W}(\cV, q)$ defined by
\[
\omega(W(v))= \e^{- \12v \eta v}, \hbox{ for }\eta= \Re(\Lambda^{\pm}\mp \12 q),
\]
see \cite[Sect. 2.3]{GW1}. By definition  $\omega$ is {\em pure} if it is pure as a state on the $C^{*}$-algebra ${\rm W}(\cV, q)$.

Note that \eqref{positon} implies that $\Ker(\Lambda^{+}+ \Lambda^{-})=\{0\}$, hence $\|v\|_{\omega}^{2}\defeq \bar{v}\Lambda^{+}v+ \bar{v}\Lambda^{-}v$ is a Hilbert norm on $\cV$.  Denoting by  $\cV^{\rm cpl}$ the completion of $\cV$ for $\|\!\cdot\! \|_{\omega}$, the hermitian forms $q, \Lambda^{\pm}$ extend uniquely  to $q^{\rm cpl}, \Lambda^{\pm, {\rm cpl}}$ on $\cV^{\rm cpl}$, and $\omega$ uniquely extends to a state $\omega^{\rm cpl}$ on ${\rm CCR}(\cV^{\rm cpl}, q^{\rm cpl})$ or ${\rm W}(\cV^{\rm cpl}, q^{\rm cpl})$. Note that $q^{\rm cpl}$ may be degenerate.

If $\cV_{1}\subset \cV^{\rm cpl}$ with $\cV\subset \cV_{1}$ densely for $\|\!\cdot\!\|_{\omega}$, then we also obtain unique 
objects $q_{1}, \Lambda^{\pm}_{1}, \omega_{1}$ that extend $q, \Lambda^{\pm}, \omega$.

In the proposition below,  we give a characterization of pure quasi-free states. Note  that the characterization given in \cite[Prop. 2.7]{GW1} was incorrect, unless $\cV= \cV^{\rm cpl}$.
\begin{proposition}\label{puritypure}
The state $\omega$ is pure on ${\rm CCR}(\cV, q)$ iff there exists $\cV_{1}\subset \cV^{\rm cpl}$ with $\cV\subset \cV_{1}$ densely for $\|\!\cdot\!\|_{\omega}$  and projections $c^{\pm}_{1}\in L(\cV_{1})$ such that
\begin{equation}
\label{turlututututu}
c^{+}_{1}+ c^{-}_{1}=\one, \ c_{1}^{+*}q_{1}c_{1}^{-}= 0, \ \Lambda_{1}^{\pm}= \pm q_{1}\circ c_{1}^{\pm}.
\end{equation}
\end{proposition}
The proof is given in Appendix \ref{subsecapp2}.}

% then there is a unique  quasi-free state $\omega$ such that (\ref{eq:lambda}) holds, see e.g. \cite[Sect. 17.1]{DG}. One can associate to $\omega$ the pair of operators $c^{\pm}\in L(\cV)$:
%\beq\label{defodefo}
%c^{\pm}\defeq  \pm q^{-1}\circ \Lambda^{\pm}.
%\eeq
%The properties \eqref{positon} become then:
%\beq\label{defidefi}
%c^{+}+ c^{-}=\one, \ c^{\pm*}q= q c^{\pm}, \ \pm q c^{\pm}\geq 0.
%\eeq
%The following characterization of {\em pure} quasi-free states is well known, see e.g. \cite[Sect. 17.1]{DG}, \cite[Prop. 2.7]{GW1}.
%\begin{proposition}\label{purity}
% The following are equivalent:
% \ben
% \item the state $\omega$ is pure,
% \item\label{it:pure1} 
% $c^{+}+ c^{-}=\one,  \ (c^{\pm})^{2}= c^{\pm}, \ c^{\pm*}q= q c^{\pm}, \ \pm q c^{\pm}\geq 0$,
%\item\label{it:pure2}  $ \eta\defeq  \Lambda^{+}+ \Lambda^{-}\geq 0, \  \eta q^{-1}\eta= q$.
%\een
%\end{proposition}
%
%{The identities in \eqref{it:pure1}, \eqref{it:pure2} are to be understood as to hold on the completion of $\cV$ with respect to the scalar product defined by  $\Lambda^{+}+ \Lambda^{-}$ (note that this was not stated precisely in \cite[Prop. 2.3 \& 2.7]{GW1}). In practice it is often convenient to check \eqref{it:pure1} or \eqref{it:pure2} on a dense subset.} 

 \subsubsection{Phase spaces for Klein-Gordon fields}\label{secp4.1.2}
Let $(M, g)$ be a globally hyperbolic space\-time and 
$P=- \nabla^{a}\nabla_{a}+ V(x)$, for $V\in \cinf(M, \rr)$ a Klein-Gordon operator on $(M, g)$.   More generally $P$ can be any formally selfadjoint second order differential operator, whose principal symbol $\sigma_{\rm pr}(P)$ equals  $\xi\cdot g^{-1}(x)\xi$.

We denote by $G_{\rm ret/adv}$ the retarded/advanced inverses for $P$ and by $G\defeq  G_{\rm ret}- G_{\rm adv}$,  the Pauli-Jordan commutator.  We set
\beq\label{pscal.e1}
(u|v)_{M}\defeq  \int_{M}\bar{u}v \,d\vol_{g},  \ u, v\in\coinf(M).
\eeq
%We denote by $\cinf_{\rm sc}(M)$ the space of smooth {\em space-compact} functions on $M$.

The classical phase space associated to $P$ is $(\cV,Q)$, where
\beq\label{defo}
\cV\defeq \frac{\coinf(M)}{P\coinf(M)}, \quad \overline{[u]} \cdot Q[ v]\defeq \i(u| G v)_{M}.
\eeq
Let $\Sigma$ be  a Cauchy hypersurface,
\[
\rho_{\Sigma}u\defeq  \begin{pmatrix}u\traa{\Sigma} \\ \i^{-1}\partial_n u\traa{\Sigma}\end{pmatrix},
\]
where $n$ is the future unit normal to $\Sigma$ and $\cV_{\Sigma}= \coinf(\Sigma;\cc^{2})$. We equip $\cV_{\Sigma}$ with the scalar product
\beq\label{pscal.e3}
(f|g)_{\Sigma}\defeq \int_{\Sigma}\overline{f}_{0}g_{0}+ \overline{f}_{1}g_{1}d\sigma_{\Sigma},
\eeq
Then $\rho_{\Sigma}G: \cV\to \cV_{\Sigma}$ is bijective, it makes thus sense to define $G_{\Sigma}: \cV_{\Sigma}=\coinf(\Sigma;\cc^{2})\to \cinf(\Sigma;\cc^{2})$ by the identity:
\[
G\eqdef (\rho_{\Sigma}G)^{*}G_{\Sigma}\rho_{\Sigma}G,
\]
where the adjoint is taken with respect to the scalar products \eqref{pscal.e1}, \eqref{pscal.e3}. 
Finally we set
\beq\label{pscal.e5}
\bar{f}q_{\Sigma}g\defeq  \i(f| G_{\Sigma}g)_{\Sigma},
\eeq
so that the map:
\[
\rho_{\Sigma}G: (\cV, Q)\to (\cV_{\Sigma}, q_{\Sigma})
\]
is pseudo-unitary.  One can use equivalently
either of the  above phase spaces.   By a computation that uses Stoke's theorem, one has concretely (see e.g. \cite{DG})
\begin{equation}
\label{pscal.e4}
q_{\Sigma}=\mat{0}{1}{1}{0}.
\end{equation}
By the definition of $G_{\Sigma}$,
\beq
\label{eq:idU}
\one=G^*\rho_{\Sigma}^{*} G_{\Sigma}\rho_{\Sigma} \mbox{\ \ on \ } G\coinf(M).
\eeq
This also implies $\rho_{\Sigma}=\rho_{\Sigma} G^*\rho_{\Sigma}^* G_{\Sigma}\rho_{\Sigma}$ on $G\coinf(M)$. On the other hand, denoting $C^\infty_{\rm sc}(M)$ the space of \emph{space-compact} smooth functions (i.e. smooth functions whose restriction to $\Sigma$ have compact support), it is well-known that $G\coinf(M)$ is exactly $\Ker P|_{\cinf_{\rm sc}(M)}$, see e.g. \cite{BGP}. Furthermore, since the Cauchy problem 
\beq\label{eq:cauh}
\bec
Pu=0,\\
\rho_{\Sigma}u=f.
\eec
\eeq
is well-posed in $u\in C^\infty_{\rm sc}(M)$ for any $f\in \coinf(\Sigma;\cc^{2})$, the map $\rho_\Sigma:\Ker P|_{\cinf_{\rm sc}(M)}\to\coinf(\Sigma;\cc^{2})$ is bijective and therefore
\beq\label{eq:idU2}
\one=\rho_{\Sigma} G^*\rho_{\Sigma}^* G_{\Sigma} \mbox{\ \ on \ } \coinf(\Sigma;\cc^{2}).
\eeq 
%The sign is OK

\subsubsection{Cauchy evolution operator}\label{secp4.1.3}

It is useful to introduce the {\em Cauchy evolution operator}:
\beq
\cU_{\Sigma}\defeq G^* \rho_{\Sigma}^* G_{\Sigma}.
\eeq
By (\ref{eq:idU}) and (\ref{eq:idU2}), it satisfies $\rho_{\Sigma} \cU_{\Sigma} =\one$ on $\coinf(\Sigma;\cc^{2})$ and $\cU_{\Sigma} \rho_{\Sigma} = \one$ on $\Ker P|_{\cinf_{\rm sc}(M)}$. Moreover, since $G^*=-G$ we get $P \cU_{\Sigma}=0$ hence for $f\in \coinf(\Sigma;\cc^{2})$, $u= \cU_{\Sigma}f$ is the unique solution in $\cinf_{\rm sc}(M)$ of the Cauchy problem \eqref{eq:cauh}.

\subsubsection{Spacetime two-point functions}\label{secp4.1.4}

 We use the phase space defined in (\ref{defo}). Let us  introduce the assumptions:
\begin{equation}
\label{eq:titu}
\begin{array}{rl}
i)&\Lambda^{\pm}: \coinf(M)\to\cinf(M)\\[2mm]
ii)&\Lambda^{\pm}\geq 0 \hbox{ for }(\cdot| \cdot)_{M} \hbox{ on }\coinf(M),\\[2mm]
iii)& \Lambda^{+}-\Lambda^{-}= \i G,\\[2mm]
iv)& P\Lambda^{\pm} = \Lambda^{\pm}P=0.
\end{array}
\end{equation}

Note that (\ref{eq:titu})  implies that $\Lambda^{\pm}: \cE'(M)\to \cD'(M)$.
Let us set  with a slight abuse of notation:
\[
\overline{u}\cdot\Lambda^{\pm}v\defeq  (u|\Lambda^{\pm}v), \ u, v\in \coinf(M).
\]
If  (\ref{eq:titu}) hold,   then $\Lambda^{\pm}$  define a pair of complex pseudo-covariances on the phase space $(\cV, q)$ defined in (\ref{defo}), hence define a unique quasi-free-state  on ${\rm CCR}(\cV, Q)$.
\begin{definition}
 A pair  of maps  $\Lambda^{\pm}: \coinf(M)\to\cinf(M)$  satisfying (\ref{eq:titu})   will be called a pair of {\em  space\-time two-point functions}. 
\end{definition}
\subsubsection{Hadamard condition}\label{secp4.1.5}
By the Schwartz kernel theorem, we can also identify $\Lambda^\pm$ with a pair of distributions $\Lambda^\pm(x, x')\in \cD'(M\times M)$, and  one is especially interested in the subclass of \emph{Hadamard states}, subject to a condition on the \textit{wave front set} of $\Lambda^\pm(x, x')$.  We recall that the sets $\cN^{\pm}$ were defined in  \ref{turlututu}.

\begin{definition}\label{def:hadamard}A pair of  two-point functions $\Lambda^\pm$  satisfying (\ref{eq:titu}) is  \emph{Hadamard} if \beq\label{hadamard}\tag{Had}
\wf(\Lambda^\pm)'\subset \cN^\pm\times\cN^\pm.
\eeq
\end{definition}
{This form of the Hadamard condition is taken from \cite{SV,hollands}, see also \cite{W2} for a review on the equivalent formulations.} The original formulation in terms of wave front sets is due to Radzikowski \cite{R}, who showed its equivalence with older definitions \cite{KW}.

\subsubsection{Cauchy surface two-point functions}\label{secp4.1.6}
We will need a version of two-point functions acting on Cauchy data of $P$ instead of  test functions on $M$. 

%\begin{lemma}\label{lem:sobolev}
% The operator $\rho_{\Sigma} G$ extends continuously to a surjection
% \[
%\rho_{\Sigma} G: \cE'(M)\to\cE'(\Sigma; \cc^{2})
%\]
%with $\Ker \rho_{\Sigma} G|_{\cE'}= \Ran P|_{\cE'}$.
%\end{lemma}
%\proof To show that $\rho_{\Sigma} G: \cE'(M)\to \cE'(\Sigma; \cc^{2})$ is well-defined and continuous, it suffices to use the well-known fact that 
%\beq\label{eq:WFofG}
%{\rm WF}'(G)\subset \cN\times \cN
%\eeq
%and the rules for composition of distributional kernels in terms of the wave front set (see \cite{H1}).
%The fact that $\rho_{\Sigma} G: \cE'(M)\to \cE'(\Sigma; \cc^{2})$ follows then from the support properties of $G$. To prove the surjectivity it suffices to show that the identity \eqref{eq:idU2}
%extends to $\cE'(\Sigma; \cc^{2})$ which follows from \eqref{pscal.e4}.

%The fact that $\Ker \rho G|_{ \cE'}= \Ker G|_{ \cE'}= \Ran P|_{\cE'}$ follows by the same proof as before. \qed

Let us introduce the assumptions:
\begin{equation}
\label{eq:titucauchy}
\begin{array}{rl}
i)&\lambda^{\pm}_{\Sigma}: \coinf(\Sigma; \cc^{2})\to\cinf(\Sigma; \cc^{2}), \\[2mm]
ii)&\lambda_{\Sigma}^{\pm}\geq 0 \hbox{ for }(\cdot| \cdot)_{\Sigma},\\[2mm]
iii)& \lambda_{\Sigma}^{+}- \lambda_{\Sigma}^{-}= \i G_{\Sigma}.
\end{array}
\end{equation}
\begin{definition}\label{defcs2pt}
 A pair of maps $\lambda_{\Sigma}^{\pm}$ satisfying (\ref{eq:titucauchy}) will be called a pair of {\em Cauchy surface two-point functions}.
\end{definition}

\begin{proposition}\label{minusu}
The maps:
 \[
 \bea\lambda_{\Sigma}^{\pm}\mapsto \Lambda^{\pm}\defeq  (\rho_{\Sigma} G)^{*}\lambda_{\Sigma}^{\pm}(\rho_{\Sigma} G),\\[2mm]
 \Lambda^{\pm}\mapsto\lambda_{\Sigma}^{\pm}\defeq (\rho_{\Sigma}^{*}G_{\Sigma})^{*} \Lambda^{\pm} (\rho_{\Sigma}^{*}G_{\Sigma})
 \eea
\]
are bijective and inverse from one another. Furthermore, $\lambda_{\Sigma}^{\pm}$ are Cauchy surface two-point functions iff $\Lambda^\pm$ are two-point functions.
\end{proposition}

Prop. \ref{minusu} is proved in \cite{GW2} in a slightly more general context. 

% \proof  
%(1): let $\lambda^{\pm}_{\Sigma}$ satisfy (\ref{eq:titucauchy}). Clearly $\Lambda^{\pm}$ is well defined as a map from $\coinf(M)$ to $\cinf(M)$. If $u\in \coinf(M)$, then $f^{\pm}\defeq\lambda^{\pm}_{\Sigma}\rho_{\Sigma} G u\in \cinf(\Sigma; \cc^{2})$, hence ${\rm WF} (\rho_{\Sigma}^{*}f^{\pm})\subset N^{*}_{\Sigma}M$, the {\em conormal bundle} to $\Sigma$ in $M$. We use now (\ref{eq:WFofG}), the fact that $\Sigma$ is non-characteristic i.e. $N^{*}_{\Sigma}M\cap \cN= \emptyset$ and standard arguments with wave front sets (see \cite{H1}) to obtain that $\Lambda^{\pm} u= -G \rho^{*}f^{\pm}\in \cinf(M)$.  The other conditions in (\ref{eq:titu}) are straightforward.
%
%(2): let $\Lambda^{\pm}$ satisfies (\ref{eq:titu}).  Since $\Lambda^{\pm}P=0$, we  have ${\rm WF}'(\Lambda^{\pm})\subset T^{*}M\times \cN$ which implies that $\Lambda^{\pm}\circ  (\rho^{*}_{\Sigma}G_{\Sigma}): \coinf(\Sigma; \cc^{2})\to\cinf(M)$. Next we use \eqref{pscal.e4} hence $G_{\Sigma}: \cinf(\Sigma; \cc^{2})\to \cinf(\Sigma; \cc^{2})$ to obtain that $\lambda^{\pm}_{\Sigma}: \coinf(\Sigma; \cc^{2})\to \cinf(\Sigma; \cc^{2})$. The other conditions in (\ref{eq:titucauchy}) are straightforward.
 
%The fact that the two maps are inverse from each other  follows from $\rho_{\Sigma} \cU_{\Sigma}=\one$, see \ref{secp4.1.3}.  The last statement about positivity is obvious. \qed

\subsection{Reduction to the model case}\label{secp2.2}
In this subsection we  consider a Klein-Gordon operator $P$ on $(M, g)$ in \eqref{kgkg} satisfying hypotheses (H), (M)  introduced in Subsect. \ref{secp3.3}. 
We show that the construction of Hadamard states for $P$ can be reduced to the case of a model operator $\tilde{P}$  on $I\times \Sigma$ as introduced  in Subsect. \ref{secp2.1}.

We use the notation in Subsect. \ref{secp3.3}.  We equip $\tilde{M}= I\times \Sigma$  with the Lorentzian metric $\tig=  - dt^{2}+ h_{t}(y)dy^{2}$. We recall that  $U= \chi(\tilde{M})$  is an open neighborhood of $\Sigma$ in $M$. We equip $\coinf(U)$  and $\coinf(\tilde{M})$ with their canonical scalar products $(\cdot |\cdot)_{M}$ and $(\cdot |\cdot)_{\tilde{M}}$.

\begin{proposition}\label{reduc-prop}
Let us set $W: \coinf(U)\ni u\mapsto c^{(d-1)/2}u\circ \chi\in \coinf(\tilde{M})$. Then the following holds:
\ben
\item There exists $a(t, y, \pe_{y})$ satisfying the conditions in Subsect. \ref{secp2.1} for $k_{0}= h_{0}$,  such that
 \[
 \tilde{P}\defeq (W^{-1})^{*}P W^{-1}= \pe^{2}_{t}+ r(t,y)\pe_{t}+ a(t, y, \pe_{y}). 
 \]
 \item if $\tilde{G}$ is  the Pauli-Jordan commutator for $\tilde{P}$ one has $\tilde{G}= W G W^{*}$.
 \item Let $\tilde{\Lambda}^{\pm}$ be  the  space\-time two-point functions of a Hadamard state $\tilde\omega$ for $\tilde{P}$ on $\tilde{M}$. Then there exists a unique Hadamard state $\omega$ for $P$ on $M$  with two-point functions $\Lambda^{\pm}$ such that 
 \[
 \tilde{\Lambda}^{\pm}= W \Lambda^{\pm}W^{*}.
 \]
 Moreover $\omega$ is pure iff $\tilde{\omega}$ is pure.
 \een 
\end{proposition}
\proof Without loss of generality we  may assume that $\chi= {\rm Id}$.  Let us first prove (1).
 We set $\tih_{t}= c^{2}h_{t}$ so that $g= - c^{2}dt^{2}+ \tih_{t}dx^{2}$. We have:
\[
\bea
 P= &-|g|^{-\12}\pe_{\mu}|g|^{\12}g^{\mu\nu}\pe_{\nu}+ V\\[2mm]
 &= c^{-1}|\tih|^{-\12}\pe_{t}c^{-1}|\tih|^{\12} \pe_{t}- c^{-1}|\tih|^{-\12}\pe_{i}c\tih^{ij}|\tih|^{\12}\pe_{j}+V.
\eea
\]
A routine computation shows that:
\[
\bea
&c^{-1}|\tih|^{-\12} \pe_{t} c^{-1} |\tih|^{\12}\pe_{t}\\
&= c^{-1}\left( \pe_{t}^{2}+ c^{-1}|\tih|^{-\12}\p_{t}( c|\tih|^{\12})\pe_{t}+ |\tih|^{-\12}\p_{t}(|\tih|^{\12}\p_{t}\ln c)\right)c^{-1}\\
&\eqdef c^{-1}P_{0}(t, y, \pe_{t})c^{-1},
\eea
\]
\[
\bea
&c^{-1}|\tih|^{-\12}\pe_{i}c\tih^{ij}|\tih|^{\12}\pe_{j}+V\\
&=c^{-1}\left(|\tih|^{-\12}\pe_{i} c\tih^{ij}|\tih|^{\12}\pe_{j}c+ c^{2}V\right)c^{-1}\\
&=c^{-1}\left(|\tih|^{-\12} \pe_{i}h^{ij} |\tih|^{\12}\pe_{j}+ |\tih|^{-\12}\pe_{i}|\tih|^{\12}h^{ij}\p_{j}\ln c+ c^{2}V\right)c^{-1}\\
&\eqdef c^{-1}P_{\Sigma}(t, y ,\pe_{y})c^{-1}.
\eea
\]
Using that $|\tih|^{\12}= c^{d}|h|^{\12}$, we can rewrite these two operators as follows:
\[
\bea
P_{0}(t, y, \pe_{t})
&=\pe_{t}^{2}+ (|h|^{-\12}\p_{t}|h|^{\12}+ (d+1) \p_{t}\ln c)\pe_{t}\\[2mm]
&\phantom{=}+ |h|^{-\12}\p_{t}|h|^{\12}\p_{t}\ln c+ d (\p_{t}\ln c)^{2}+ \p_{t}^{2}\ln c,\\[2mm]
P_{\Sigma}(t, y, \pe_{y})&=|h|^{-\12}\pe_{i}h^{ij}|h|^{\12}\pe_{j}+ d \p_{i}\ln c
h^{ij}\pe_{j}\\[2mm]
&\phantom{=}+ \pe_{i}h^{ij}\p_{j}\ln c- \p_{i}|h|^{-\12}|h|^{\12}h^{ij}\p_{j}\ln c
+ d \p_{i}\ln ch^{ij}\p_{j}\ln c+ c^{2}V.
\eea
\]
Let now $U$ be the operator of multiplication by $c^{-(d+1)/2}$. Since 
\[
U^{-1}\pe_{t}U= \pe_{t}- \12 (d+1)\p_{t}\ln c, \ U^{-1}\pe_{i}U= \pe_{i}- \12(d+1)\p_{i}\ln c,
\]
we obtain:
\[
 cU^{-1}PUc= \pe_{t}^{2}+ r(t, y)\pe_{t}+ a(t, y, \pe_{y}).
\]
By hypotheses (H), (M) we have $r\in \cinfb(I; \BT^{0}_{0}(\Sigma, h_{0}))$, $a\in \cinfb(I; \Diff^{2}(\Sigma, h_{0}))$, with principal symbol: 
\[
\sigma_{\rm pr}(a)(t, y, \eta)=\eta \cdot h_{t}^{-1}(y)\eta.
\]
If $S: \coinf(U)\ni u\mapsto  c^{(d+3)/2}u\in \coinf(\tilde{M})$ we check that $W^{*}= S^{-1}$, hence $cU^{-1}PUc=SP W^{-1}= \tilde{P}$.
Since $\tilde{P}$ is selfadjoint for the scalar product $(\cdot | \cdot)_{\tilde{M}}$ it follows using  $d \vol_{\tilde{g}}= |h_{t}|^{\12}dtdy$ that 
\[
r= |h_{t}|^{-\12}\p_{t}|h_{t}|^{\12}, \ a(t, y, \pe_{y})= a^{*}(t, y, \pe_{y}).
\]
This completes the proof of (1).  From the uniqueness of   retarded/advanced inverses we obtain that $\tilde{G}_{\rm ret/adv}= WG_{\rm ret/adv}W^{*}$, which proves (2).  

To prove (3) we use two well-known arguments: the first one is the {\em time-slice property}, which  means that   $\cV= \frac{\coinf(U)}{P \cinf(U)}$, i.e. we can replace $M$ by $U$ in \eqref{defo}, since $U$ is a neighborhood of a Cauchy surface.   In other words, a pair of two-point functions $\Lambda^{\pm}\in \cD'(U\times U)$ satisfying \eqref{eq:titu} over $U\times U$ uniquely extends to $\Lambda^{\pm}\in \cD'(M\times M)$ satisfying \eqref{eq:titu} over $M\times M$. 

The second follows from a result based on H\"{o}rmander's propagation of singularities theorem, see \cite{R,SV}: if $\Lambda^{\pm}$ satisfy (Had) over $U\times U$, then they satisfy (Had) globally, using that $P\Lambda^{\pm}= \Lambda^{\pm}P=0$. The proof is complete. \qed 

%\begin{remark}
%This should  be removed, I write it just to fix ideas.  Since  $g= -c^{2}dt^{2}+ \tih_{t}dx^{2}$, $\tih_{t}= c^{2}h_{t}$, the normal vector field is $c^{-1}\p_{t}$, the induced metric on $\Sigma_{t}$ is $|\tih_{t}|^{\12}dx$.  The charge is:
% \[
% \bar{\psi}\cdot q \psi= \int_{\Sigma_{t}}(\overline{c^{-1}D_{t}\psi}\psi+ \overline{\psi}c^{-1}D_{t}\psi) |\tih_{t}|^{\12}dx= \int_{\Sigma_{t}}(\overline{D_{t}\psi}\psi+ \overline{\psi}D_{t}\psi )c^{d-1}|h_{t}|^{\12}dx.
% \]
% If we set $W\psi= \tilde{\psi}$, so that $P\psi=0$ iff $\tilde{P}\tilde{\psi}=0$, we have $\psi= W^{-1}\tilde{\psi}= c^{-(d-1)/2}\tilde{\psi}$ so
% \[
%  \bar{\psi}\cdot q \psi= \bar{\tilde{\psi}}\tilde{q}\tilde{\psi}=\int_{\Sigma_{t}}(\overline{D_{t}\tilde\psi}\tilde\psi+ \overline{\tilde\psi}D_{t}\tilde\psi )|h_{t}|^{\12}dx.
% \]
%\end{remark}

\subsection{The pure Hadamard state associated to a microlocal decomposition}\label{secp4.2}
In this subsection we consider   the  model Klein-Gordon operator $\tilde{P}$  obtained from Prop. \ref{reduc-prop}. To simplify notation, we denote $\tilde{P}$ by $P$, $\tilde{M}= I\times \Sigma$ by $M$.
We will associate to a microlocal decomposition for $P$ a unique {\em pure Hadamard state}.  First we need to introduce some more notation. 

The level sets $\Sigma_{t}= \{t\}\times \Sigma$ are all Cauchy hypersurfaces. The various objects
associated to the Cauchy surface $\Sigma_{t}$, like  $\rho_{\Sigma_{t}}$, $G_{\Sigma_{t}}$, $\lambda^{\pm}_{\Sigma_{t}}$, $\cU_{\Sigma_{t}}$ will be simply denoted by $\rho(t)$, $G(t)$, $\lambda^{\pm}(t)$, $\cU(t)$, etc.

\begin{lemma}\label{l6.6}
 One has:
 \[
\begin{array}{rl}
i)&\lambda^{\pm}_{\omega}(t)= \cU_{A}(s,t)^{*}\lambda^{\pm}_{\omega}(s)\cU_{A}(s,t),\\[2mm]
ii)&c_{\omega}^{\pm}(t)= \cU_{A}(t,s)c_{\omega}^{\pm}(s)\cU_{A}(s,t),\\[2mm]
iii)&q= \cU_{A}(s,t)^{*}q \cU_{A}(s,t),\\[2mm]
iv)& \cU_{A}(t,s)= \rho(t)G^{*}\rho(s)^{*}G(s).
\end{array}
\]
\end{lemma}
\proof It suffices to use the various identities in \ref{secp4.1.2}, \ref{secp4.1.3} and the fact that $\cU_{A}(t,s)= \rho(t)\cU(s)$. \qed

\begin{theoreme}\label{thp4.1}
 Let $\{\cU_{A}^{\pm}(\cdot, \cdot)\}_{(t,s)\in I^{2}}$ be a microlocal decomposition  of the evolution $\cU_{A}$ as in Subsect. \ref{secp2.4} and let $\lambda^{\pm}(t)\defeq  \pm q\circ c^{\pm}(t)$, where $c^{\pm}(t)$ are defined in Prop. \ref{split.prop1}. Then $\lambda^{\pm}(t)$ are the Cauchy surface two-point functions of a pure Hadamard state. 
 One has:
 \beq\label{keyform}
 \bea
 \lambda^{\pm}(t)&= \cU_{A}(0,t)^{*}T^{-1}(0)^{*}\pi^{\pm}T^{-1}(0)\cU_{A}(0,t)\\[2mm]
 &=  T^{-1}(t)^{*}\cU_{C}(0,t)^{*}\pi^{\pm}\cU_{C}(0,t)T^{-1}(t)\\[2mm]
 &=T^{-1}(t)^{*}\pi^{\pm}T^{-1}(t)+ \cinfb(I; \Psi^{-\infty}(\Sigma)).
 \eea
 \eeq
 where $\pi^{\pm}$ are defined in \eqref{defdepi} and $T(t)$ in \eqref{defdeT}.
 \end{theoreme}

\proof  Let us first prove \eqref{keyform}. From the definition of $c^{\pm}(t)$ (see \eqref{split.0}, \eqref{cplus}), we have:
 \[
 \bea
 \lambda^{\pm}(t)&=\pm
q\cU_{A}(t,0) T(0)\pi^{\pm}T^{-1}(0)\cU_{A}(0,t)\\[2mm]
&=\pm\cU_{A}(0,t)^{*}q T(0)\pi^{\pm}T^{-1}(0)\cU_{A}(0,t)\\[2mm]
&=\pm\cU_{A}(0,t)^{*}T^{-1}(0)^{*}\hat{q}\pi^{\pm}T^{-1}(0)\cU_{A}(0,t)\\[2mm]
&=\cU_{A}(0,t)^{*}T^{-1}(0)^{*}\pi^{\pm}T^{-1}(0)\cU_{A}(0,t),
\eea
 \]
 where we used successively \eqref{ep3.1}, \eqref{symplac} and the fact that $\pm \hat{q}\pi^{\pm}= \pi^{\pm}$.
The second line in \eqref{keyform} follows then from \eqref{taz}. From \eqref{staro}, \eqref{toz} we obtain then
\[
\hat{q}= \cU_{C}(0,t)^{*}\hat{q}\cU_{C}(0,t)= \cU_{\tilde{C}}(0,t)^{*}\hat{q}\cU_{\tilde{C}}(0,t)+ \cinfb(I, \Psi^{-\infty}(\Sigma)).
\]
Since $\hat{q}$ and $\cU_{\tilde{C}}(0,t)$ are diagonal this implies that 
\[
\pi^{\pm}= \cU_{\tilde{C}}(0,t)^{*}\pi^{\pm}\cU_{\tilde{C}}(0,t)+  \cinfb(I, \Psi^{-\infty}(\Sigma)),
\]
which using once more \eqref{toz} gives:
\[
\pi^{\pm}= \cU_{C}(0,t)^{*}\pi^{\pm}\cU_{C}(0,t)+  \cinfb(I, \Psi^{-\infty}(\Sigma)),
\]
which completes the proof of \eqref{keyform}.

To check that $\lambda^{\pm}(t)$ are the Cauchy surface two-point functions of a pure Ha\-da\-mard state we work 
with the Cauchy surface $\Sigma_{0}$. By Prop. \ref{split.prop1} we know that $c^{+}(0)+ c^{-}(0)=\one$  hence condition \eqref{eq:titucauchy} {\it iii)} is satisfied. Condition {\it i)} follows from the fact that $c^{\pm}(0)\in \Psi^{\infty}(\Sigma)$.  The positivity condition {\it iii)}  follows from \eqref{keyform}. 
To check the Hadamard condition one can use the arguments in \cite{GW2}, which we recall for the sake of self-containedness. We have by \eqref{e2.10}:
\[
\cU(0) c^{\pm}(0)= \cU_{A}^{\pm}(\cdot, 0)c^{\pm}(0)\hbox{ on }\cE'(\Sigma; \cc^{2}),
\]
hence
\[
\Lambda^{\pm}= \pm \i\cU(0)c^{\pm}(0)\rho(0)G= \pm \i\cU_{A}^{\pm}(\cdot, 0)\rho(0)G, \hbox{ on }\cE'(M). 
\]
From Thm. \ref{newprop.2} this implies that  $\wf'(\Lambda^{\pm})\subset \cN^{\pm}\times \cN$. Since $\Lambda^{\pm}=(\Lambda^{\pm})^*$ we have also $\wf'(\Lambda^\pm)\subset \cN^\pm\times \cN^\pm$.  This shows that the state $\omega$ is Hadamard. The fact that $\omega$ is pure follows from the fact that $c^{\pm}(0)$ are projections. \qed

 We recall  from Subsects. \ref{sec2.3.4}, \eqref{secp2.4} that a  microlocal decomposition $\cU_{A}^{\pm}(\cdot, \cdot)$ is uniquely obtained from a generator $b(t)$ constructed in Thm. \ref{th2.1} as an approximate solution of a Riccati equation.
Consequently to any such $b(t)$ corresponds a unique pure Hadamard state. 

\begin{definition}\label{bodlo.def1}
 The pure Hadamard state associated to a generator $b(t)$ in Thm. \ref{thp4.1} will be denoted by $\omega_{b}$.
\end{definition}

It is now easy to find the relationship between the space\-time two-point functions $\Lambda^{\pm}$ of $\omega$ and the operators $\cU_{A}(t,s)$.  As in Subsect. \ref{feyninv} we denote by $A(t,s)$ the time kernel of an operator $A$.
\begin{theoreme}\label{michal's-identities}
Let $\Lambda^{\pm}$ be the space\-time two-point functions of the state $\omega$ constructed in Thm. \ref{thp4.1}. Then:
\begin{equation}
\label{e3.9}
\cU_{A}^{\pm}(t,s)= \pm\begin{pmatrix} \i\p_s\Lambda^\pm(t,s)  & \Lambda^\pm(t,s) \\ \p_t\p_s\Lambda^\pm(t,s)   &  \i^{-1}\p_t\Lambda^\pm(t,s)   \end{pmatrix}.
\end{equation} 
Consequently we have:
\beq\label{trouloulou}
\Lambda^{\pm}(t,s)= \pm \pi_{0}\cU^{\pm}_{A}(t,s)\pi_{1}^{*},
\eeq
where $\pi_{i}$ are defined in Subsect. \ref{feyninv}.
\end{theoreme}
\proof Using \eqref{e2.10} and the identities in Lemma \ref{l6.6}, Prop. \ref{minusu} we obtain:
\[
\bea 
\cU_{A}^{\pm}(t,s)&=\cU_{A}(t,0)c^{\pm}(0) \cU_{A}(0,s)\\[2mm]
&= \rho(t)G^{*}\rho(0)^{*}G(0)c^{\pm}(0)\rho(0)G^{*}\rho(s)^{*}G(s)\\[2mm]
&= \pm \i \rho(t)G^{*}\rho(0)^{*}\lambda^{\pm}(0)\rho(0)G^{*}\rho(s)^{*}G(s)\\[2mm]
&=\pm\i \rho(t)\Lambda^{\pm}\rho(s)^{*}G(s)= \pm \rho(t)\Lambda^{\pm}\rho(s)^{*}q.
\eea 
\]
Using $\rho(s)^{*}f= f^{0}\otimes \delta(s)- \i f^{1}\otimes \delta'(s)$, this yields:
\[\cU_{A}^{\pm}(t,s)= \pm\begin{pmatrix} \i\p_s\Lambda^\pm(t,s)  & \Lambda^\pm(t,s) \\ \p_t\p_s\Lambda^\pm(t,s)   &  \i^{-1}\p_t\Lambda^\pm(t,s)   \end{pmatrix},
\]
which completes the proof. \qed

In Subsect. \ref{feyninv} we associated to a microlocal decomposition a canonical Feynman inverse $G_{\rm F}$, see Thm. \ref{feyninvth}. On the other hand, it is well-known (see e.g. \cite{R} or \cite[Thm. 3.4.4]{W2} for the complex case) that if $\Lambda^{\pm}$ are the space\-time two-point functions of a Hadamard state $\omega$, then the operator
\[
\i^{-1}\Lambda^{+}+ G_{\rm adv}= \i^{-1}\Lambda^{-}+ G_{\rm ret}
\]
is a Feynman inverse of $P$.  It is easy to show that if $\omega$ is the  state in Thm. \ref{thp4.1} then both Feynman inverses are the same.

\begin{proposition}\label{feyninvprop}
 Let $G_{\rm F}$ and $\Lambda^{\pm}$ the Feynman inverse  and space\-time two-point functions associated to the microlocal decomposition $\{\cU_{A}^{\pm}(\cdot, \cdot)\}_{(t,s)\in I^{2}}$. Then 
 \[
G_{\rm F}= \i^{-1}\Lambda^{+}+ G_{\rm adv}= \i^{-1}\Lambda^{-}+ G_{\rm ret}
\]
\end{proposition}
\proof  Arguing as in the proof of Thm. \ref{feyninvth} we see that
\[
G_{\rm ret}(t,s)= \i^{-1}\pi_{0}\cU_{A}(t,s)\pi_{1}^{*}\theta(t-s), \ G_{\rm adv}(t,s)= -\i^{-1}\pi_{0}\cU_{A}(t,s)\pi_{1}^{*}\theta(s-t).
\]
Then the proposition follows from the identities \eqref{trululu}, \eqref{trouloulou}. \qed

\subsection{Regular Hadamard states}\label{secp4.3}

\begin{definition}
  A state $\omega$ is {\em regular} if $\lambda_{\omega}^{\pm}(t)\in \cinf(\rr, \Psi^{\infty}(\Sigma; M_{2}(\cc)))$.
\end{definition}
In other words regular states have  Cauchy surface two-point functions equal to matrices of pseudodifferential operators. The following lemma shows that it suffices to check the pseudodifferential property for one time $t$.
\begin{lemma}
 $\omega$ is regular iff  there exists $s\in I$ such that $\lambda_{\omega}^{\pm}(s)\in \Psi^{\infty}(\Sigma; M_{2}(\cc))$.
\end{lemma}
\proof   Assume that $\lambda_{\omega}^{\pm}(s)\in \Psi^{\infty}(\Sigma; M_{2}(\cc))$ for some $s\in I$.  Then $\lambda_{\omega}^{\pm}(t)$ is given by  Lemma \ref{l6.6} {\it i)}.  By Thm. \ref{newprop1} 
 we can replace  $\cU_{A}(s,t)$  by $\cU_{\tilde{A}}(s,t)$ and then by $\cU_{\tilde{C}}(s,t)$, which has a diagonal generator, see \eqref{ep3.6}. Then we apply Egorov's Theorem, Thm. \ref{pth1}. \qed
 
 Let  now $\omega$ be the  Hadamard state associated to a microlocal decomposition as in Thm. \ref{thp4.1}, and $\omega_{1}$ another regular state. We denote by $\Lambda^{\pm}$, $\Lambda_{1}^{\pm}$, $\lambda^{\pm}(t)$, $\lambda_{1}^{\pm}(t)$ their respective space\-time and Cauchy surface two-point functions. 

\begin{proposition}  A regular state $\omega_{1}$ is Hadamard iff:
  \beq\label{poopoo}
  \lambda_{1}^{\pm}(t)- \lambda^{\pm}(t)\in \cinf(\rr; \Psi^{-\infty}(\Sigma; M_{2}(\cc))).
  \eeq
\end{proposition}
\proof  It is well known (see e.g. \cite{R} or \cite[Thm. 3.4.4]{W2} for the complex case) that  $\Lambda^{\pm}- \Lambda_{1}^{\pm}$ are smoothing  operators on $M$, hence $\lambda^{\pm}(t)- \lambda_{1}^{\pm}(t)$ are smoothing operators on $\Sigma$.  By Lemma \ref{lemmatoche} this yields  the $\Rightarrow$ implication. The $\Leftarrow$ implication is immediate. \qed

  \subsubsection{Bogoliubov transformations}
 We work in the setup of \ref{secp4.1.1}. It is well known, see e.g. \cite[Thm. 11.20]{DG}, that if $\omega, \omega_{1}$ are two pure quasi-free states on ${\rm CCR}(\cV, q)$,  then there exists $u\in U(\cV, q)$ such that 
 \[
 \lambda_{1}^{\pm}= u^{*}\lambda^{\pm}u.
 \]
 We now examine the form of the operator $u$ if $\omega$ is a pure Hadamard state associated to a microlocal decomposition as in Thm. \ref{thp4.1} and $\omega_{1}$ is another pure, regular Hadamard state. In the proposition below, we fix the reference time $t=0$.  The operator $b(0)\in \Psi^{1}(\Sigma)$ entering in the definition of $\lambda^{\pm}(0)$ (see formulas \eqref{keyform} and \eqref{split.0}) will be simply denoted by $b$.

\begin{proposition}\label{boldrick}
 Let $\lambda_{1}^{\pm}(0)$ be the $t=0$ two-point functions of a pure regular Hadamard state $\omega_{1}$. Then there exists $a\in \Psi^{-\infty}(\Sigma)$ such that:
 \beq\label{boldo.e1}
 \lambda_{1}^{+}= T^{-1}(0)^{*}U^{*}\pi^{+}U T^{-1}(0), \hbox{ for }U=  \mat{(\one + aa^{*})^{\12}}{a}{a^{*}}{(\one + a^{*}a)^{\12}}.
 \eeq
 \end{proposition}
\proof  Let us set 
\[\eta_{1}= \lambda_{1}^{+}+ \lambda_{1}^{-},  \ \ \hat{\eta}_{1}= T^{-1}(0)^{*}\eta_{1}T^{-1}(0).
\]
 Since $\omega_{1}$ is pure, we deduce from Prop. \ref{puritypure} and identity \eqref{symplac} that $\hat{\eta}_{1}$ satisfies:
\begin{equation}
\label{turu}
i) \ \hat{\eta}_{1}\geq 0, \ \ ii) \ \hat{\eta}_{1}\hat{q}\hat{\eta}_{1}= \hat{q},
\end{equation}
where we recall that $\hat{q}= \mat{1}{0}{0}{-1}$. We write $\hat{\eta}_{1}$ as $\mat{a}{b}{b^{*}}{c}$, where  using \eqref{poopoo} and the fact that $\hat{\eta}=\one$ we know that $b, \one - a, \one - c\in \Psi^{-\infty}(\Sigma)$. Now   (\ref{turu}) is equivalent to:
\[
\begin{array}{rl}
i')& a\geq0, \ c\geq 0, \ | (u| bv)|\leq  (u| au)^{\12}(v| cv)^{\12}, \ u, v\in\coinf(\Sigma),\\[2mm]
ii')& a^{2}= \one + bb^{*}, \ c^{2}= \one + b^{*}b, \ ab-bc=0.
\end{array}
\]
Since $a,c\geq 0$ by {\it i')}, the first two equations of {\it ii')} yield 
\[
a= (\one + bb^{*})^{\12}, \ c= (\one + b^{*}b)^{\12}.
\]
The third equation of {\it ii')} then holds using the identity 
\begin{equation}
\label{tara}
bf(b^{*}b)= f(bb^{*})b,\ f\hbox{ any Borel function}. 
\end{equation}
 The second condition in {\it i')} is equivalent to $\| (\one + bb^{*})^{\12}b(\one + b^{*}b)^{\12}\|\leq 1$, which holds using again (\ref{tara}).
This implies that $\hat{\lambda}_{1}\defeq  T^{-1}(0)^{*}\lambda_{1}^{+}T^{-1}(0)$ equals
\[
\hat{\lambda}_{1}=\12\mat{(\one + bb^{*})^{\12}+ \one}{b}{b^{*}}{(\one + b^{*}b)^{\12}-\one}.
\]
Let now 
\[
a\defeq  \frac{b}{\sqrt{2}}((\one + b^{*}b)^{\12}+ \one)^{\12}\in\Psi^{-\infty}(\Sigma).
\]
Using (\ref{tara}) we obtain by an easy computation that
\[
\one + a^{*}a= \12((\one + b^{*}b)^{\12}+ \one), \ \one + aa^{*}=  \12((\one + bb^{*})^{\12}+ \one), \ b= 2a(\one + a^{*}a)^{\12}.
\]
Hence
\[
\hat{\lambda}_{1}= \mat{\one + aa^{*}}{a(\one + a^{*}a)^{\12}}{(\one + a^{*}a)^{\12}a^{*}}{a^{*}a}= U^{*}\pi^{+}U,
\]
for $U$ as in the proposition. 
 \qed

 The following theorem shows that any pure regular Hadamard state is actually associated to a microlocal decomposition. 
\begin{theoreme}\label{boldo.th1}
Let $\omega_{1}$ be a pure regular Hadamard state. Then there exists  a generator $b_{1}(t)$ as in Thm. \ref{th2.1} such that $\omega_{1}= \omega_{b_{1}}$.
\end{theoreme}
 Before proving Thm. \ref{boldo.th1} we need one more lemma.
 
\begin{lemma}\label{boldo.l1}
 Let $a\in\Psi^{-\infty}(\Sigma)$ and set $r(a)\defeq (\one + aa^{*})^{\12}- a$. Then $r(a)$ is boundedly invertible with $r(a)^{-1}\in \one + \Psi^{-\infty}(\Sigma)$.
\end{lemma}
\proof By the polar decomposition theorem, we have $a= u|a|= |a^{*}|u$, where $u$ is a partial isometry. Moreover,
$r(a)= (\one+ aa^{*})^{\12}(\one - (\one+ aa^{*})^{-\12}a)$. To prove invertibility it suffices to notice that $(\one+ aa^{*})^{-\12}a= (\one+ | a^{*}|^{2})^{-\12}|a^{*}| u$ has norm $<1$, which is easily checked by using the self-adjoint functional calculus and the fact that $a^{*}$ is bounded. The fact that $r(a)^{-1}-\one\in\Psi^{-\infty}(\Sigma)$ follows by the argument used already to prove Lemma \ref{lp.1}. \qed

\noindent {\bf Proof of Thm. \ref{boldo.th1}.} From Prop. \ref{boldrick}, we know that  there exists $a\in \Psi^{-\infty}(\Sigma)$ such that \eqref{boldo.e1} holds. 
Let us first try to find some $b_{1}\in\Psi^{1}(\Sigma)$ such that
\begin{equation}
\label{boldo.e2}
\lambda_{1}^{+}= (T_{1}^{-1})^{*}\pi^{+}T_{1}^{-1},
\end{equation}
where $T_{1}$ is defined as in \eqref{defdeT} with $b=b(0)$ replaced by $b_{1}$. The proof is divided in several steps.

{\it Step 1:} 
 we first solve \eqref{boldo.e2}. Let $r(a)$ be as in Lemma \ref{boldo.l1} and set 
\beq
\label{e2d}
z\defeq  r(a)(b+ b^{*})^{-\12}\in \Psi^{-\12}(\Sigma).
\eeq 
Note  that $z^{-1}\in \Psi^{\12}(\Sigma)$. We claim  that 
\beq\label{e2bis}
b_{1}\defeq b + (b+ b^{*})^{\12}a^{*}z^{*-1}= b+ \Psi^{-\infty}(\Sigma)
\eeq
solves (\ref{boldo.e2}). Indeed, if $V_{1}=\mat{\alpha_{1}}{\beta_{1}}{\gamma_{1}}{\delta_{1}}$ the equation 
\[
V^{*}\pi^{+}V=V_{1}^{*}\pi^{+}V_{1},
\]
is equivalent to the system
\beq\label{e2f}
\left\{
\bea 
i)& \ \alpha_{1}^{*}\alpha_{1}= \alpha^{*}\alpha\\[2mm]
ii)& \ \alpha_{1}^{*}\beta_{1}= \alpha^{*}\beta,\\[2mm]
iii)& \ \beta_{1}^{*}\beta_{1}= \beta^{*}\beta.
\eea \right.
\eeq
If  $V= UT(b)$ and $V_{1}= T(b_{1})$ we have:
\[
\bea
\alpha_{1}&= (b_{1}+ b_{1}^{*})^{-\12}b_{1}^{*},\ 
\beta_{1}= (b_{1}+ b_{1}^{*})^{-\12},\\[2mm]
\alpha&= (1+ aa^{*})^{\12}(b + b^{*})^{-\12}b^{*}+ a(b+ b^{*})^{-\12}b,\\[2mm]
\beta&= (1+ aa^{*})^{\12}(b+ b^{*})^{-\12}- a (b+ b^{*})^{-\12}.
\eea
 \]
Using  the operator $z$ introduced in \eqref{e2d} we see that
\begin{equation}
\label{e2e}
\alpha= zb^{*}+ a (b+ b^{*})^{\12}, \ \beta= z.
\end{equation}
Note also that:
 \begin{equation}
\label{e3c}
\bea 
&r(a)r(a)^{*}+ r(a)a^{*}+ a r(a)^{*}\\[2mm]
&=(r(a)+a)(r(a)^{*}+a^{*})- aa^{*}=1.\eea 
\end{equation}
Hence for  $b_{1}$ given by \eqref{e2bis} we have:
\[
\bea 
b_{1}+ b_{1}^{*}&=(b+ b^{*})+ (b + b^{*})^{\12}a^{*}z^{*-1}+ z^{-1}a(b+ b^{*})^{-1}\\[2mm]
&=(b+ b^{*})^{\12}\left(1 + a^{*}r(a)^{*-1}+ r(a)^{-1}a\right)(b+ b^{*})^{\12}\\[2mm]
&=z^{-1}\left (r(a)r(a)^{*}+ z(b+ b^{*})^{\12}a^{*}+ a (b+ b^{*})^{\12}z^{*}\right)z^{*-1}\\[2mm]
&= z^{-1}\left(r(a)r(a)^{*}+ r(a)a^{*}+ a r(a)^{*}\right) z^{*-1}= z^{-1}z^{*-1},
\eea 
\]
by \eqref{e3c}, hence:
\[
 \beta_{1}^{*}\beta_{1}= (b_{1}+ b_{1}^{*})^{-1}= z^{*}z=\beta^{*}\beta,
\]
hence \eqref{e2f} $iii)$ is satisfied.
 We also obtain
\[
\bea 
\alpha_{1}^{*}\beta_{1}&=b_{1}(b_{1}+ b_{1}^{*})^{-1}=b_{1}z^{*}z \\[2mm]
&=b z^{*}z+ (b+ b^{*})^{\12}a^{*}z =\alpha^{*}\beta,
\eea 
\]
hence \eqref{e2f} $ii)$ is satisfied.  Finally we have
\[
\bea 
\alpha_{1}^{*}\alpha_{1}&=  b_{1}(b_{1}+ b_{1}^{*})^{-1}b_{1}^{*}= b_{1} z^{*}z b_{1}^{*}\\[2mm]
&= (b z^{*}+(b+ b^{*})^{\12}a^{*})( z b^{*}+ a(b+ b^{*})^{\12})= \alpha^{*}\alpha,
\eea 
\]
by \eqref{e2e}.  Therefore \eqref{e2f} $i)$ is satisfied  and $b_{1}$ solves \eqref{boldo.e2}.

{\it Step 2}: we check  that $b_{1}= b_{1}(0)$ satisfies the properties {\it i)}, {\it ii)} and {\it iii)} in Thm. \ref{th2.1} at $t=0$. First of all  $b_{1}= b + \Psi^{-\infty}(\Sigma)$ hence {\it i)} is satisfied. We claim that
\beq\label{e3.8}
b_{1}+ b_{1}^{*}\sim b+ b^{*},
\eeq
(see Subsect. \ref{sec0.not} for notation), 
which implies properties $ii)$, $iii)$ at $t=0$. 
In fact  we have:
\[
\bea 
b_{1}+ b_{1}^{*}&= (b+ b^{*})^{\12}\left(1+ a^{*}r(a)^{-1*}+ r(a)^{-1}a\right)(b+ b^{*})^{\12}\\[2mm]
&=(b+ b^{*})^{\12}r(a)^{-1}\left(r(a)r(a)^{*}+ r(a)a^{*}+ a r(a)^{*}\right) r(a)^{-1*}(b+ b^{*})^{\12}\\[2mm]
&= (b+ b^{*})^{\12}r(a)^{-1}\left((r(a)+a)(r(a)^{*}+a^{*})- aa^{*}\right) r(a)^{-1*}(b+ b^{*})^{\12}\\[2mm]
&= (b+ b^{*})^{\12}r(a)^{-1}r(a)^{-1*}(b+ b^{*})^{\12},
\eea 
\]
which implies \eqref{e3.8} since $r(a)$ is boundedly invertible by Lemma \ref{boldo.l1}. 

{\it Step 3}: we now extend $b_{1}$ into $b_{1}(t)$. We set
\[
b_{1}(t)= b(t)+ r_{-\infty}(t),
\]
where $r_{-\infty}\in \cinf(\rr, \Psi^{-\infty}(\Sigma))$ is chosen such that  $r_{-\infty}(0)= b_{1}(0)- b(0)$ and properties $i)$, $ii)$, $iii)$ are satisfied for all $t\in I$. Then $iv)$ is automatically satisfied also.  This completes the proof of the theorem. 
\qed

\appendix
\section{}\init\label{secapp1}
\subsection{Computations for Kerr-de Sitter}\label{subsecapp1}
We recall  now an identity of the kind which is often used in the literature (see e.g. \cite[Lemma 2.2.1]{O2} for the Kerr metric).
\begin{remark}
The identity in Lemma \ref{limo}  allows to check that the Kerr-de Sitter metrics are smooth Lorentzian metrics on $M$  despite the fact  the forms $d\varphi$, $d\theta$ are  singular at the poles of $\mathbb{S}^{2}$. In fact the forms $\sin 2 \theta d\theta$ and $\sin^{2}\theta d\varphi$ are smooth  on $\mathbb{S}^{2}$, since they equal $x dx + y dy$ and $xdy- ydx$ in Cartesian coordinates near the poles, and the standard metric on $\mathbb{S}^{2}$ equals $d\theta^{2}+ \sin^{2}\theta d\varphi^{2}$.

\end{remark}
 
\begin{lemma}\label{limo}
Let 
\[
d\omega^{2}= d \theta^{2}+ \frac{1+ \alpha \cos^{2}\theta}{1+\alpha}\sin^{2}\theta d\varphi^{2}
\]
Then:
\ben
\item  $d\omega^{2}$ is a smooth Riemannian metric on $\mathbb{S}^{2}$.

\item One has:
 \[
 g_{\theta\theta}d\theta^{2}+ g_{\varphi\varphi}d\varphi^{2}= \frac{\rho^{2}}{\Delta_{\theta}} d\omega^{2}+\left(\frac{2M\vara^{2}r}{(1+ \alpha)^{2}\rho^{2}}+ \frac{\vara^{2}}{1+\alpha}\right)(\sin^{2}\theta d\varphi)^{2}.
 \]
 \een
 \end{lemma}
\proof 
$d\omega^{2}$ is clearly positive definite. We have 
\[
d\omega^{2}= (d \theta^{2}+ \sin^{2}\theta d\varphi^{2})- \frac{\alpha}{1+\alpha} (\sin^{2}\theta d\varphi)^{2}.
\]
The first term is the standard metric on $\mathbb{S}^{2}$, the second term is smooth   since $\sin^{2}\theta d\varphi$ is a smooth $1-$form on $\mathbb{S}^{2}$. Therefore $d\omega^{2}$ is smooth, which proves (1).

A routine computation shows that:
\[
\sigma^{2}= (r^{2}+\vara^{2})(1+ \alpha)\rho^{2}+ 2M\vara^{2}r\sin^{2}\theta.
\]
Using this identity in the definition of $g_{\varphi\varphi}$  (see \eqref{kds.e1}) we easily obtain (2). \qed

\subsubsection{Some classes of functions}
  The map $I\ni r\mapsto s(r)\in \rr$ is bijective.   Setting: 
 \[
 \begin{array}{rl}
({\rm KdS})& \kappa_{h/c}\defeq  \mp\frac{\p_{r}\Delta_{r}(r_{h/c})}{(1+ \alpha)(r_{h/c}^{2}+ \vara^{2})},\\[2mm]
({\rm K})&\kappa_{h}\defeq  \mp\frac{\p_{r}\Delta_{r}(r_{h})}{(1+ \alpha)(r_{h}^{2}+ \vara^{2})},
 \end{array}
  \]
 which are related to the surface gravities at  
 the Kerr-de Sitter resp. Kerr case, one has:
\[
\begin{array}{rl}
({\rm KdS})&(r- r_{h/c})\sim \e^{-\kappa_{h/c}|s|}, \ \p^{\alpha}_{s}(r- r_{h/c})\in O(\e^{-\kappa_{h/c}|s|}),\hbox{ for }s\to\mp\infty, \\[2mm]
({\rm K})&\bec
 (r- r_{h})\sim \e^{-\kappa_{h}|s|}, \ \p^{\alpha}_{s}(r- r_{h})\in O(\e^{-\kappa_{h}|s|}),\hbox{ for }s\to-\infty,\\
 r\sim s, \ \p^{\alpha}_{s}r\in O(\langle s\rangle^{1- \alpha}), \hbox{ for }s\to +\infty. 
\eec
\end{array}
\]
\begin{definition}\label{defkds}
 We set:
 \[
 \bea
 T^{0}_{{\rm KdS}}&=\{f\in \cinf(]r_{h}, r_{c}[\times \mathbb{S}^{2})\, : \, \p^{\alpha}_{r}\p^{\beta}_{\omega}f\in O(1)\},\\[2mm]
 T^{0, 0}_{{\rm K}}&=\{f\in \cinf(]r_{h}, +\infty[)\, : \, \p^{\alpha}_{r}\p^{\beta}_{\omega}f\in O(\langle r\rangle^{-\alpha})\}\\[2mm]
 T^{p}_{{\rm KdS}}&=(r-r_{h})^{p}(r-r_{c})^{p}T^{0}_{{\rm KdS}}, \ p\in \zz,\\[2mm]
 T^{m,p}_{{\rm K}}&=\langle r\rangle^{m-p}(r-r_{h})^{p}T^{0, 0}_{{\rm K}}, m\in \rr, \ p\in \zz.
 \eea
  \]
 \end{definition}
The following are the images of the above spaces under the change of variable $r\mapsto s(r)$.
\begin{definition}\label{defkds2}
 We set:
 \[
 \bea
 S^{p}_{\rm KdS}&= \{f\in \cinf(\rr\times \mathbb{S}^{2}) : \, \p^{\alpha}_{s}\p_{\omega}^{\beta}f\in O(\e^{p\kappa_{h/c}|s|}), \ \pm s<0\},  \ p\in\zz^{*},\\[2mm]
 S^{0}_{\rm KdS}&= \{f\in \cinf(\rr\times \mathbb{S}^{2}) : \, f\hbox{ bounded }, \ \p_{s}f\in S^{-1}_{\rm KdS} \},\\[2mm]
 S^{m,p}_{\rm K}&=\{f\in \cinf(\rr\times \mathbb{S}^{2}) : \,\p^{\alpha}_{s}\p_{\omega}^{\beta}f\in O(\e^{p\kappa_{h}|s|}), \, s<0, \, \p^{\alpha}_{s}\p_{\omega}^{\beta}f\in O(\langle s\rangle^{m- \alpha}), \, s>0\}, \ p\in\zz^{*},\\[2mm]
 S^{m, 0}_{\rm K}&= \{f\in  \cinf(\rr\times \mathbb{S}^{2}) : \,\p^{\beta}_{\omega}f\in O(\langle s\rangle^{m}), \, \p_{s}f\in S^{m-1, -1}_{\rm K} \}.
 \eea
  \]
 \end{definition}
\begin{definition}
 A function $f\in S^{p}_{\rm KdS}$, resp. $f\in S^{m,p}_{\rm K}$ is {\em elliptic} if  $f(s, \omega)\neq 0$ on $\rr\times\mathbb{S}^{2}$ and $f^{-1}\in S^{-p}_{\rm KdS}$, resp. $f^{-1}\in S^{-m, -p}_{\rm K}$. 
\end{definition}
The following result is easy to prove (see \cite[Sect. 9.3]{Ha}).
\begin{lemma}\label{kds.lem1}
\ben 
\item  $S^{p_{1}}_{\rm KdS}\times S^{p_{2}}_{\rm KdS}\subset S^{p_{1}+ p_{2}}_{\rm KdS}$ and $S^{m_{1}, p_{1}}_{\rm K}\times S^{m_{2}, p_{2}}_{\rm K}\subset S^{m_{1}+ m_{2}, p_{1}+ p_{2}}_{\rm K}$,
\item Set $\tilde{f}(s, \omega)= f(r(s), \omega)$ for $f\in\cinf(I\times\mathbb{S}^{2})$. Then if 
$f\in T^{p}_{\rm KdS}$, resp. $f\in T^{m,p}_{\rm K}$  one has $\tilde{f}\in  S^{-p}_{\rm KdS}$, resp. $\tilde{f}\in S^{m,-p}_{\rm K}$.
\een
\end{lemma}
From Lemma \ref{kds.lem1} we obtain easily the following lemma.
\begin{lemma}\label{kds.lem2}
 One has:
 \[
\begin{array}{rl}
i)&\rho^{2}, \ r^{2}+ \vara^{2} \hbox{ are elliptic in }S^{0}_{\rm KdS}, \hbox{ resp. in }S^{2,0}_{\rm K},\\[2mm]
ii)&\sigma^{2}\hbox{ is elliptic in }S^{0}_{\rm KdS}, \hbox{ resp. in }S^{4,0}_{\rm K},\\[2mm]
iii)&\Delta_{\theta}\hbox{ is elliptic in } S^{0}_{\rm KdS}, \hbox{ resp. in }S^{0,0}_{\rm K},\\[2mm]
iv)&\Delta_{r}\hbox{ is elliptic in }S^{-1}_{\rm KdS},\hbox{ resp. in }S^{2, 1}_{\rm K}\\[2mm]
v)&F(s)\defeq  (1+ \alpha)^{2}\frac{(r^{2}+ \vara^{2})^{2}}{\Delta_{r}}\hbox {is elliptic in }S^{1}_{\rm KdS}, \hbox{ resp. in }S^{2, 1}_{\rm K},\\[2mm]
vi)&G(s, \theta)\defeq \frac{\sigma^{2}}{(r^{2}+ \vara^{2})^{2}\Delta_{\theta}},\hbox { is  elliptic in }S^{0}_{\rm KdS}, \hbox{ resp. in }S^{0,0}_{\rm K},\\[2mm]
vii)&\tilde{c}^{2}=\frac{\Delta_{r}\Delta_{\theta}\rho^{2}}{(1+ \alpha)^{2}\sigma^{2}}\in S^{-1}_{\rm KdS}\hbox{ resp. }\in S^{0, -1}_{\rm K}.
\end{array}
\]
\end{lemma}
 \proof Statements {\it i)}, \ldots, {\it iv)} are routine computations, using Lemma \ref{kds.lem1} (2). The remaining statements follow then from Lemma \ref{kds.lem1} (1). \qed
 
 In the next lemma we estimate the function $R$ defined in \ref{sssecchange}.
\begin{lemma}\label{kds.lem3}
 Let $R= g_{t\varphi}g_{\varphi\varphi}^{-1}$ and set
 \[
R_{r}(s, \theta)= \p_{r}R(r(s), \theta), \ R_{\theta}(s, \theta)\defeq  (\sin 2\theta)^{-1}\p_{\theta}R(r(s, \theta)).
\]
Then:
\begin{equation}
\label{kds.e8}
\begin{array}{l}
R_{r}\in S^{0}_{\rm KdS}, \hbox{ resp. }\in S^{0, -3}_{\rm K}, \\[2mm]
R_{\theta}\in S^{-1}_{\rm KdS}, \hbox{ resp. }\in S^{-1, -2}_{\rm K}.
\end{array}
\end{equation}
\end{lemma}
 \proof We have
 \[
 \bea
R(r, \theta)&= -\frac{\vara}{\sigma^{2}}\left(\Delta_{r}- (r^{2}+ \vara^{2})\Delta_{\theta}\right)\\[2mm]
&=\frac{\vara}{r^{2}+ \vara^{2}}\left(1- \frac{\Delta_{r}}{(r^{2}+\vara^{2})\Delta_{\theta}}\right)\left(1-\frac{\vara^{2}\Delta_{r}\sin^{2}\theta}{(r^{2}+\vara^{2})^{2}\Delta_{\theta}}\right)^{-1},
\eea
\]
hence
\[
R= \frac{\vara}{r^{2}+\vara^{2}}(1+ R_{1}(r, \theta)), \ R_{1}\in T^{1}_{\rm KdS}, \hbox{ resp. }R_{1}\in T^{1, -1}_{\rm K}.
\]
It follows that (using that $R$ depends on $\theta$ only through $\sin^{2}\theta$):
\[
\begin{array}{l}
R_{r}\defeq \p_{r}R\in T^{0}_{\rm Kds}, \hbox{ resp. }\in T^{0, -3}_{\rm K}, \\[2mm]
R_{\theta}\defeq (\sin 2\theta)^{-1}\p_{\theta}R\in T^{1}_{\rm KdS}, \hbox{ resp. }\in T^{1, -2}_{\rm K}.
\end{array}
\]
Passing to the variable $s$ using  Lemma \ref{kds.lem1} we obtain \eqref{kds.e8}. \qed

{ 
\subsection{Proof of Prop. \ref{puritypure}}\label{subsecapp2}

Let us recall that in our notations, $\cV^{\rm cpl}$ is the completion of $\cV$ with respect to the Hilbert norm $\|v\|_{\omega}^{2}\defeq \bar{v}\Lambda^{+}v+ \bar{v}\Lambda^{-}v$, and the superscript `${\rm cpl}$' is also used to denote canonical extensions of various objects on $\cV$ to $\cV^{\rm cpl}$.  

If $\cV= \cV^{\rm cpl}$, then  the real symplectic space $(\cV, \Re \,\sigma)$ is complete for the Euclidean norm $(v\eta v)^{\12}$, $\eta= \Re(\Lambda^{\pm}\mp \12 q)$. In that situation, we know from \cite[Thm. 17.13]{DG} that $\omega$ is pure  iff  $(2\eta^{\rm cpl}, \Re\sigma^{\rm cpl})$ is K\"{a}hler, i.e. there exists  an anti-involution $\ii_{1}\in Sp(\cV^{\rm cpl}, \Re\, \sigma^{\rm cpl})$  such that $2\eta^{\rm cpl}= \Re\,\sigma^{\rm cpl}\ii_{1}$.   This is known to be equivalent to the existence of projections $c^{\pm}$ satisfying
\begin{equation}
\label{turlutututututu}
c^{+}+ c^{-}=\one, \ c^{+*}q c^{-}= 0, \ \Lambda^{\pm}= \pm q \circ c^{\pm},
\end{equation}
as requested, see \cite[Prop. 2.7]{GW1}. 

Let us now treat the general case.
We recall from  \cite[Thm. 2.3.19]{BR} that a state $\omega$ on a $C^{*}$-algebra $\mathfrak{A}$ is pure iff its GNS representation $(\cH_{\omega}, \pi_{\omega})$ is irreducible, i.e. iff $\cH_{\omega}$ does not contain non-trivial closed subspaces invariant under $\pi_{\omega}(\mathfrak{A})$.

For  $\cV_{1}$ as in the statement of the proposition, we set  
$\mathfrak{A}_{(1)}= {\rm W}(\cV_{(1)}, q_{(1)})$, and we let $(\cH_{(1)}, \pi_{(1)}, \Omega_{(1)})$ be the GNS triple for $(\mathfrak{A}_{(1)}, \omega_{(1)})$.  Using that $\cV$ is dense in $\cV_{1}$ for $\|\!\cdot\!\|_{\omega}$, we first obtain that $\cH= \cH_{1}$, $\Omega= \Omega_{1}$ and $\pi_{1}|_{\mathfrak{A}}= \pi$. 

We also easily obtain that $\pi(\mathfrak{A})$ is strongly dense in $\mathfrak{A}_{1}$. In fact, if $A= \sum_{1}^{N}\lambda_{i}\pi_{1}(W(v_{i}))\in \pi_{1}(\mathfrak{A}_{1})$ and $v_{i, n}\in \cV$ with $v_{i, n}\to v_{i}$ for $\|\!\cdot\!\|_{\omega}$, we obtain that $A_{n}= \sum_{1}^{N}\lambda_{i}\pi(W(v_{i,n}))$ is bounded by $\sum_{1}^{N}|\lambda_{i}|$ and converges strongly to $A$ on the dense subspace $\pi(\mathfrak{A})\Omega$, hence on $\cH$.

From this fact we see that a closed subspace $\cK\subset \cH$ is invariant under $\pi(\mathfrak{A})$ iff it is invariant under $\pi_{1}(\mathfrak{A}_{1})$, hence $\omega$ is pure iff  its extension $\omega_{1}$  to $\mathfrak{A}_{1}$ is pure.

Therefore, the $\Rightarrow$ direction is shown simply by taking $\cV_{1}= \cV^{\rm cpl}$.  Conversely, if on a space $\cV_{1}$ as in the statement of the proposition, there exist projections $c_{1}^{\pm}$ satisfying \eqref{turlutututututu} (with $c_1^\pm,\Lambda^{\pm}_1$ in place of $c^\pm,\Lambda^{\pm}$), then  an easy computation shows that as identities on $L(\cV_{1}, \cV_{1}^{*})$, one has
\[
c_{1}^{\pm*}\lambda_{1}^{\pm}c_{1}^{\pm}= \lambda_{1}^{\pm}, \ \ c_{1}^{\pm*}\lambda_{1}^{\mp}c_{1}^{\pm}=0,
\]
hence $c_{1}^{\pm}$ are bounded for $\|\!\cdot\!\|_{\omega}$. Therefore they extend to projections on $\cV^{\rm cpl}$ satisfying \eqref{turlutututututu}. This implies that $\omega^{\rm cpl}$ is pure, hence $\omega$ is pure. \qed }

\subsection*{Acknowledgments} The authors would like to thank Victor Nistor and Jean-Philippe Nicolas for useful conversations. M.\,W. gratefully acknowledges the France-Stanford Center for Interdisciplinary Studies for financial support and the Department of Mathematics of Stanford University for its kind hospitality. The authors also wish to thank the Erwin Schr\"odinger Institute in Vienna for its hospitality during the program ``Modern theory of wave equations''.

\end{document}